\documentclass[sigconf]{acmart}
\usepackage{multirow, makecell}
\usepackage{pifont}
\usepackage{subfig}
\usepackage{graphicx}
\usepackage{rotating}

\AtBeginDocument{%
  }

\newcommand{\ourmethod}{\textit{CoWear}}

\copyrightyear{2026}
\acmYear{2026}
\setcopyright{cc}
\setcctype{by-nc-nd}
\acmConference[CHI '26]{Proceedings of the 2026 CHI Conference on Human Factors in Computing Systems}{April 13--17, 2026}{Barcelona, Spain}
\acmBooktitle{Proceedings of the 2026 CHI Conference on Human Factors in Computing Systems (CHI '26), April 13--17, 2026, Barcelona, Spain}
\acmPrice{}
\acmDOI{10.1145/3772318.3791951}
\acmISBN{979-8-4007-2278-3/2026/04}

\begin{document}
\title{From Invisible to Actionable: Augmented Reality Interactions with Indoor CO\textsubscript{2}}

\author{Prasenjit Karmakar}
\email{prasenjitkarmakar52282@gmail.com}
\affiliation{%
	\institution{IIT Kharagpur}
	\country{India}
}

\author{Manjeet Yadav}
\email{p25cs0011@iitj.ac.in}
\affiliation{%
	\institution{IIT Jodhpur}
	\country{India}
}

\author{Swayanshu Rout}
\email{22it3057@rgipt.ac.in}
\affiliation{%
	\institution{Rajiv Gandhi Institute of Petroleum Technology}
	\country{India}
}

\author{Swadhin Pradhan}
\email{swapradh@cisco.com}
\affiliation{%
	\institution{Cisco Systems Inc}
	\country{USA}
}

\author{Sandip Chakraborty}
\email{sandipc@cse.iitkgp.ac.in}
\affiliation{%
	\institution{IIT Kharagpur}
	\country{India}
}

\begin{abstract}
Indoor carbon dioxide (CO\textsubscript{2}) can rapidly accumulate to form invisible pollution \textit{hotspots}, posing significant health risks due to its odorless and colorless nature. Despite growing interest in wearable or stationary sensors for pollutant detection, effectively visualizing CO\textsubscript{2} levels and engaging individuals remains an ongoing challenge. In this paper, we develop a portable wrist-sized pollution sensor that detects CO\textsubscript{2} in real time at any indoor location and reveals \textit{CO\textsubscript{2} bubbles} by highlighting sudden spikes. In order to promote better ventilation habits and user awareness, we also develop a smartphone-based augmented reality (AR) game for users to locate and disperse these high-CO\textsubscript{2} zones. A user study with $35$ participants demonstrated increased engagement and heightened understanding of CO\textsubscript{2}’s health impacts. Our system's usability evaluations yielded a median score of $1.88$, indicating its strong practicality.
\end{abstract}



\begin{teaserfigure}
  \centering 
  \includegraphics[width=0.95\textwidth]{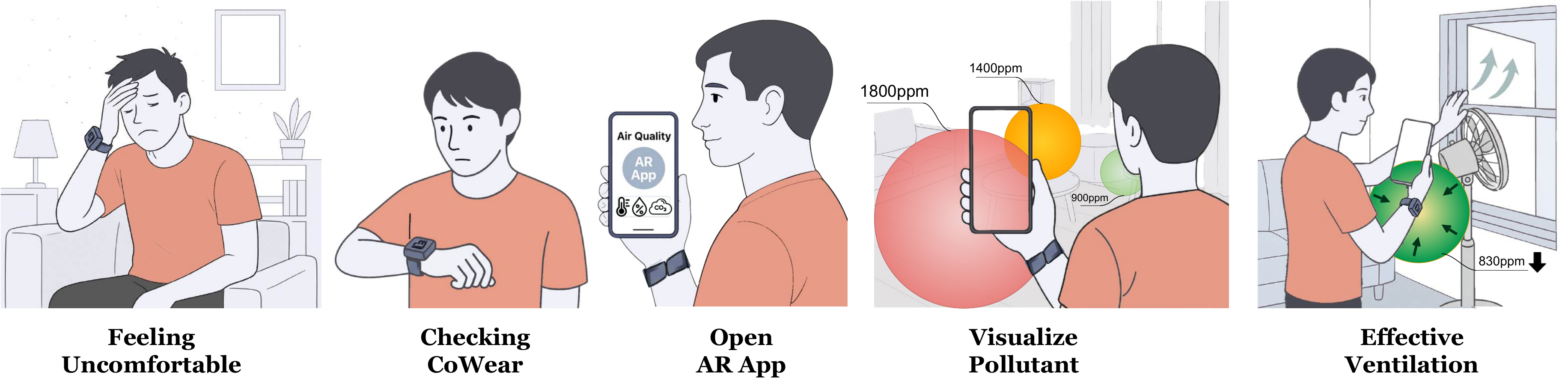}
  \caption{Our system enables actionable awareness of indoor air pollution using augmented reality (AR). (1) When feeling uncomfortable indoors, the user (2) checks the CoWear wrist-wearable sensor to monitor real-time air quality. (3) By launching the AR app, the user can (4) visualize invisible CO\textsubscript{2} pollution as color-coded, spatial bubbles overlaid in their environment. (5) Guided by these visualizations, the user employs effective ventilation strategies, such as directing airflow toward high-CO\textsubscript{2} zones, and immediately observes reductions in pollutant concentration for healthier indoor air.}
  \Description{The figure shows a step-by-step illustration of how the system works. First, a user indoors begins to feel uncomfortable. Next, the user checks the CoWear wrist-wearable sensor to track real-time air quality. By opening the AR app on a smartphone, the user can see color-coded floating bubbles representing CO2 levels around them. Green bubbles indicate safe levels, while yellow and red bubbles represent increasingly unhealthy concentrations. The user then applies ventilation strategies, such as opening a window or using a fan, and observes the bubbles shrinking and changing color as the CO2 concentration reduces.}
  \label{fig:teaser_cowear}
\end{teaserfigure}

\maketitle

\section{Introduction}
\label{sec:intro}

\textbf{Motivation.}
Air pollution remains one of the most pressing environmental and public health challenges of our era. While outdoor pollution has received significant attention in HCI through work on sensing, visualization, and community engagement~\cite{li2023visual,schurholz2019myaqi,hsu2017community,hsu2020smell,gupta2022human,liu2020making,liu2018third}, the indoor context is comparatively underexplored, despite evidence that indoor air quality can be equally or more harmful~\cite{epa_indoor}. Pollutants such as VOCs, CO, CO\textsubscript{2}, and other indoor gases impair respiratory and cognitive function~\cite{patel2018physiology}. Yet, most remain invisible, odorless, and unperceived by occupants. Given that people spend up to $90\%$ of their time indoors~\cite{morales2021air,whoindoor}, raising awareness and enabling actionable strategies for indoor air quality is a crucial challenge for HCI.

Despite the well-documented health effects of indoor pollutants, most occupants are unaware of and lack the means to monitor their immediate exposure. Traditional sensing approaches typically rely on static monitors placed in fixed locations. However, these monitors are difficult to deploy at dense spatial scales and often fail to provide information that is personally meaningful to the occupant. This limitation creates a disconnect between \textit{sensing} and \textit{actionable awareness}. In contrast, wearables offer a promising alternative by moving the sensing closer to the individual, allowing user-centric real-time measurements of personal exposure~\cite{jiang2011maqs,maag2018w}. 

When combined with Augmented Reality (AR), wearable data can be visualized \textit{in situ}, anchored to the user's immediate surroundings, and embodied perspective~\cite{assor2024augmented,mathews2021air}. AR on mobile devices has emerged as a powerful medium to enhance human perception of the environment and support new forms of interaction~\cite{cao2023mobile}. Prior research demonstrates the broad applicability of AR: from enabling interactions with 3D depth maps~\cite{du2020depthlab}, supporting digital literacy among older adults~\cite{jin2024exploring}, and facilitating shopping~\cite{ahn2015supporting,an2021arshoe}, to assisting navigation in unfamiliar smart spaces~\cite{clark2022articulate}. More broadly, AR has been used to promote environmental sustainability through immersive and situated awareness experiences~\cite{cosio2023virtual,prandi2019augmenting,silva2022understanding,ceccarini2022escapecampus}.
By integrating invisible environmental data with everyday perception, abstract sensor readings can be transformed into situated and interactive visualizations. Coupling these pathways enables users not only to monitor continuously, but also to interpret invisible pollutants and take action.

\textbf{Objective.}  
In this paper, we focus on carbon dioxide (CO\textsubscript{2}), a pollutant often overlooked in everyday discourse but associated with significant health risks. Elevated concentrations above $1000$ ppm can impair cognition, while levels exceeding $2000$ ppm cause headaches, nausea, and reduced attention~\cite{ramalho2015association,azuma2018effects,tsai2012office}. Prolonged exposure at such concentrations can also contribute to \textit{hypercapnia}, a condition of excessive CO\textsubscript{2} in the bloodstream that is particularly dangerous for older adults and individuals with respiratory conditions such as chronic obstructive pulmonary disease (COPD), obesity-hypoventilation syndrome, or asthma, who already have compromised gas exchange~\cite{csoma2022hypercapnia,wei2018aecopd}. Even healthy individuals may experience dizziness, sleepiness, or reduced work performance when exposed to sustained indoor CO\textsubscript{2} levels above recommended thresholds~\cite{allen2015cognitive,bowen2020review,fan2023short}.

Indoors, CO\textsubscript{2} accumulates due to human respiration, cooking and combustion, with accumulation patterns shaped by room size, ventilation, and occupancy. This means that common environments such as \textit{office spaces, conference rooms, classrooms, or basements} can easily reach concentrations associated with headaches, drowsiness, and impaired decision-making if fresh air supply is insufficient~\cite{allen2015cognitive,fan2023short}. \textit{Kitchens} and \textit{indoor gyms} are also high-risk, as cooking and heavy breathing rapidly generate additional CO\textsubscript{2} in confined spaces. Industrial or recreational environments such as \textit{breweries and wineries}, where fermentation produces large amounts of CO\textsubscript{2}, pose occupational risks of acute hypercapnia and even unconsciousness if ventilation is inadequate~\cite{permentier2017carbon}. Air conditioning is not always effective in mitigating these build-ups~\cite{zhong2021complexity,snow2019performance}, leaving users dependent on informed ventilation strategies. Consequently, in this paper, we examine how AR interactions can render CO\textsubscript{2} perceptible and contextualized within a user's immediate environment, and how such visualizations can be made actionable -- nudging individuals to intervene and experiment with mitigation strategies in real time. By doing so, this work extends AR from environmental awareness to situated actionability, bridging sensing, visualization, and interaction in everyday health contexts.

\textbf{Proposed Solution.}  
We begin with a mixed-mode prestudy survey of $140$ participants to assess awareness and behavioral practices around indoor CO\textsubscript{2}. We find that while many participants acknowledge CO\textsubscript{2} as harmful, few have measured it personally, and most lack knowledge of safe thresholds or effective ventilation strategies. Echoing prior studies on localized ``CO\textsubscript{2} bubbles'' around occupants~\cite{enriquez2014quasi,ghahramani2019personal,pantelic2020personal}, our pilot experiments confirm that CO\textsubscript{2} accumulates unevenly and can persist in corners unless actively dispersed. To address this gap, we design and implement \ourmethod{}, a wrist-worn CO\textsubscript{2} sensor integrated with a smartphone AR application. The system visualizes localized CO\textsubscript{2} concentrations as color- and size-coded AR bubbles that users can place, observe, and interact with (Figure~\ref{fig:teaser_cowear}). Through these interactions, users can both monitor their personal exposure and experiment with actions, such as opening windows or directing airflow, to mitigate CO\textsubscript{2} hotspots. We evaluate \ourmethod{} with $35$ participants across semi-controlled and in-the-wild contexts. Through AR-mediated visualization, we demonstrate that indoor air quality can be improved not only as a result of improving awareness, but also as a result of informing timely and targeted ventilation interventions. In addition, the system also supports play-like, interactive interactions, highlighting the broader potential of AR for situated environmental management.  

\textbf{Contributions.} 
In this paper, we contribute to the HCI community by advancing the development of interaction techniques that allow invisible pollutants to be perceived, situated, and made actionable. Specifically, our work offers: 
\begin{enumerate}
    \item \textbf{Human survey to assess awareness about personalized CO\textsubscript{2} exposure.}  
    We present a mixed-mode pre-study survey with 140 participants, combining online and offline responses. The survey reveals significant gaps in public understanding of CO\textsubscript{2} as an indoor pollutant, misconceptions about ventilation strategies, and a strong demand for more intuitive visualization methods. It is these insights that underpin the design requirements for our system and underscore the role of HCI in bridging awareness gaps.  

    \item \textbf{Design of a wearable (\textit{CoWear}) with a smartphone AR interface for personalized CO\textsubscript{2} visualization.}  
    We introduce \textit{CoWear}, a low-cost wrist-worn prototype that continuously monitors personal CO\textsubscript{2} exposure, integrated with a smartphone-based AR application. Through the use of dynamic, color-coded AR bubbles, our system becomes a multi-modal HCI interface that translates sensor data into situated, embodied experiences that facilitate environmental sensing and interaction.  

    \item \textbf{Developing a single-player AR game for pollution awareness and mitigation of CO\textsubscript{2} bubbles.}  
    Extending the AR interface, we design and evaluate a game mechanic where users actively interact with CO\textsubscript{2} bubbles through ventilation strategies (e.g., opening windows, directing airflow). The game not only enhances user engagement but also demonstrates how playful interaction modalities can motivate behavior change for healthier indoor environments. We validate this approach with $35$ participants across semi-controlled and in-the-wild settings, demonstrating effectiveness and usability.  
\end{enumerate}
\section{Related Work}
\label{sec:relatedwork}
We review related work in two areas: (i) studies on the impact of CO\textsubscript{2} as an indoor pollutant, and (ii) research on pollution measurement, visualization, and interactive methods for environmental awareness.

\subsection{CO\textsubscript{2} as an Indoor Pollutant}
Several studies have identified that CO\textsubscript{2} can negatively impact decision-making at concentrations as low as $1000$ ppm~\cite{satish2012co2,allen2016associations}. However, both home and bedroom environments can contribute to significant CO\textsubscript{2} exposure~\cite{karmakar2024exploring}, with individuals spending nearly one-third of their life sleeping~\cite{azuma2018effects} and more than $60$\% of their time in homes. In particular, for bedroom, CO\textsubscript{2} levels may exceed $2500$ ppm when doors and windows are closed~\cite{strom2016effects,gall2016real}, although lower CO\textsubscript{2} concentrations in bedrooms improves the sleep quality~\cite{karmakar2024exploring}. Notably, air conditioning (AC) systems also influence personal CO\textsubscript{2} exposure, as CO\textsubscript{2} tends to concentrate at lower heights due to its density. Sedentary office work can similarly result in higher CO\textsubscript{2} levels, as static air and limited occupant movement create conditions where individuals may re-inhale their exhaled CO\textsubscript{2}~\cite{ghahramani2019personal}. Studies have shown that CO\textsubscript{2} bubbles from personal respiration can average around $1200$ ppm, compared to $650$ ppm in surrounding indoor air under normal ventilation conditions~\cite{ghahramani2019personal}. Thus, a comprehensive understanding of personal CO\textsubscript{2} exposure requires measurements near the inhalation zone.

\subsection{Pollution Measurements and Visualization}
 Several studies~\cite{masic2020evaluation,zheng2018field} have shown the utility of static electrochemical and BAM (Beta Attenuation Monitor) sensors to measure the concentration of indoor air pollutants (i.e., CO, CO\textsubscript{2}, VOC, PM\textsubscript{2.5}, etc.). In addition, dashboards~\cite{lee2022online,hsu2017community} and alert systems~\cite{schurholz2019myaqi} from the measurements to provide actionable recommendations to the user. Pollution data visualization is traditionally achieved through time-series plots and 2D representations overlaid on satellite imagery. For instance, Chen \textit{et al.}~\cite{chen2019visualization} introduced a method for visualizing air quality data collected from fixed reference stations across China using Google Earth. While such representations are adequate for data visualization, they often fall short in terms of user engagement, as they are presented through static dashboards or mobile applications~\cite{moore2018managing,jiang2011maqs}. 

Recent studies have highlighted advancements in visualization techniques, such as dynamic and interactive dashboards or user-driven suggestions, which can significantly enhance user engagement with data. For instance, Lindrup \textit{et al.}~\cite{lindrup2023carbon} demonstrated how physical representations of data could improve understanding of the environmental impacts associated with food consumption. \cite{kim2020awareness} reported user experience and awareness of indoor air quality based on data visualization in the in-built display of the sensor. \cite{li2023visual} designed a data visualization platform to extract possible spatiotemporal transport patterns from large-scale pollutant transport trajectories. \cite{margariti2024evaluating} deploy a modular, shape-changing, configurable display for climatic awareness in the workplace. Another example, \textit{AirSense}~\cite{fang2016airsense}, can automatically identify pollution events, pinpoint the sources of pollution, and offer valuable suggestions for improving indoor air quality. Unlike outdoors, indoor pollutants can be distributed non-uniformly and persist only in specific spatiotemporal instances, making static sensor deployment challenging. Several studies~\cite{zhong2020hilo,fang2016airsense} have proposed mobile handheld or wearable sensor solutions to better approximate the spatiotemporal pollution distribution of indoor environments.

\subsection{Immersive Visualization for Environmental Awareness and Actionability}
\label{sec:related_immersive}
Notably, AR is rapidly becoming the new medium through which users can understand and interact with their surroundings innovatively and meaningfully by enhancing their perception of the environment and objects. Several studies have utilized AR visualization to improve awareness and decision-making. For instance, a game-based learning tool, EscapeCampus~\cite{ceccarini2022escapecampus}, simulates an AR escape room to increase students' (11-16 years old) awareness of the 17 sustainable development goals of the United Nations.  Sa{\ss}mannshausen \textit{et al.}~\cite{sassmannshausen2021citizen} created an AR visualization for citizens to contribute design ideas for urban planning, where building projects and environment changes are visualized. Studies~\cite{roddiger2018armart, alvarez2020store} also helped customers choose by utilizing AR to overlay nutrition facts and comparative data with online and in-store products. Jin \textit{et al.}~\cite{jin2024exploring} develops an AR support tool for learning smartphone applications, particularly in improving digital literacy among older adults.

Moreover, AR visualization has been proven effective in persuading users to take proactive actions. For instance, Schaper \textit{et al.}~\cite{schaper2022addressing} developed a persuasive mobile AR app that teaches the recycling process of each type of waste. Assor \textit{et al.}~\cite{assor2024augmented} introduced ARwavs, an AR waste accumulation visualization representing waste data embedded in users’ familiar environment. ARwavs yields stronger emotional responses than non-immersive waste accumulation visualizations and plain numbers. Mittmann \textit{et al.}~\cite{mittmann2022lina} introduced an AR smartphone-based collaborative game that facilitates improved social interactions among students. Recent studies have utilized AR to visualize air quality data. For instance,~\cite{mathews2021air} visualized outdoor air pollutants in an AR scene with real-time data from nearby air quality monitoring sites. Katsiokalis \textit{et al.}~\cite{katsiokalis2023gonature} overlayed real-time data with AR to provide citizen awareness of outdoor air and noise pollution data. Chae \textit{et al.}~\cite{chae2022virtual} explored mobile AR systems to visualize air conditioner airflow and temperature changes. However, a few studies have investigated AR applications for indoor pollution visualization and interaction.

\subsection{Interaction Design for Environmental Awareness and Actionability}
\label{sec:related_interaction}
Our research is also connected to the growing work on games and gamification. Recent literature has seen a surge in publications exploring the use of games across various domains~\cite{bauer2023improving, albar2024playful, ishoj2021axo, prophet2018cultivating}. For example, the gamified application AXO~\cite{ishoj2021axo} was designed to teach preteens about recycling by engaging them in sorting and identifying common household waste to protect a chain of islands. Simultaneously, Prophet \textit{et al.}~\cite{prophet2018cultivating} developed an application where air quality is symbolized by the growth of a virtual tree, with users interacting via augmented reality to care for the tree based on local air quality. Similarly, Feldpausch-Parker \textit{et al.}~\cite{feldpausch2013adventures} created an educational game to enhance students’ understanding of the impacts of CO\textsubscript{2} emissions, while Albar \textit{et al.}~\cite{albar2024playful} emphasized game-based learning to foster environmental sustainability awareness among children. Similarly, \textit{CityOnStats}~\cite{teles2020game} employs a 3D game-like environment where users can explore and interpret air quality data through various visual cues, catering to diverse user preferences. Furthermore, the work by Relvas \textit{et al.}~\cite{relvas2024serious} underscores the importance of raising awareness about air pollution. Their study introduces ``\textit{Problems in the Air,}" a Unity-based game designed to educate players about air pollution through interactive gameplay.

\subsection{Open Challenges and Hypothesis}
Although prior work has established CO\textsubscript{2} as a significant indoor pollutant and developed methods for measuring individual exposure, these efforts largely stop at data collection. Invisible phenomena such as personal CO\textsubscript{2} ``bubbles'' remain difficult for occupants to perceive or interpret, leaving a critical gap between sensor readings and everyday experience. In contrast, visualization and gamification techniques have been effectively used to foster awareness of environmental issues such as outdoor air pollution. Yet, we still lack an understanding of how best to bridge \emph{invisible environmental data} with everyday perception through interactive means. We approach this challenge by expanding the role of AR beyond environmental awareness toward situated actionability. Physical properties of CO\textsubscript{2} (e.g., diffusion, accumulation, persistence) lend themselves to embodied representation in AR. Our aim is to connect abstract sensing data, situated visualization, and interaction into a coherent pipeline that enables users not only to perceive their personal CO\textsubscript{2} environments but also to act upon them in everyday contexts~\cite{assor2024augmented,albar2024playful,ceccarini2022escapecampus}. Accordingly, we hypothesize:

\begin{enumerate}
    \item[\textbf{H1.}] \textbf{Lack of CO\textsubscript{2} Pollution Awareness.} Occupants may be aware of CO\textsubscript{2}'s general health impacts but often lack knowledge of concrete remedies. We hypothesize that making CO\textsubscript{2} exposure \emph{visible and situated} at a personal scale can improve awareness of effective preventive measures.

    \item[\textbf{H2.}] \textbf{From Awareness to Actionability.} Beyond raising awareness, we hypothesize that AR visualizations of CO\textsubscript{2} exposure will nudge occupants toward concrete actions, such as engaging with ventilation systems, windows, or other devices, to reduce concentrations. Prior research demonstrates that AR can scaffold in-situ decision making and proactive behaviors~\cite{schaper2022addressing,assor2024augmented,mittmann2022lina}, particularly when immersive data visualizations are tightly coupled to everyday contexts~\cite{krings2020development,roddiger2018armart}.
\end{enumerate}

\section{Prestudy Survey and Motivational Experiment}
\label{sec:motivation}
Hypothesis \textbf{H1} aims to determine the perception and awareness of communities on indoor pollution in general and CO\textsubscript{2} exposure in particular. For this purpose, we designed a pre-study survey questionnaire with the following objectives.
\begin{enumerate}
    \item [\textbf{O1.}] How experienced are the participants in measuring pollutants, like CO\textsubscript{2}, at a personal scale, like at their homes?  
    \item [\textbf{O2.}] Are participants aware about the CO\textsubscript{2} standards and its impact? 
    \item [\textbf{O3.}] What is participants' perception of ventilation's role in indoor pollution, particularly CO\textsubscript{2} exposure? 
    \item [\textbf{O4.}] What is the participants' perception about improving their awareness level for breathing healthy air indoors? 
\end{enumerate}

\subsection{Survey Methodology}
We have conducted a mixed-mode pre-study survey with two groups of participants -- offline and online participants. Using a one-sided one-sample proportion test for awareness and experience with null-hypothesis proportion ($p_0$) $0.40$, and expected true proportion ($p_1$) $0.25$ (i.e., small-to-medium effect, Cohen’s $h=0.32$), with statistical significance ($\alpha$) $0.05$ and $80\%$ power, the required sample size is N $\approx$ $61$ respondents (normal approximation). In the pre-study survey, we collected a total of $N = 140$ responses. The offline participants ($N = 40$) were recruited from the university campus and further chose to volunteer for the entire study at its different stages (system performance and testing, as we discuss later in Section~\ref{sec:res}). The online responders ($N = 100$) are more diverse. For online participation, we have broadcast the survey questionnaire through social media, public forums, and emails to targeted groups/communities. The survey questionnaire contained four significant sections -- (1) general demographic information including age, gender, location, profession, family income, house type, number of members in the family, history in the family about pollution-related diseases, etc., (2) general awareness on indoor pollution ($7$ questions), (3) understanding about the type, spread, and harmfulness of indoor pollutants ($7$ questions), and (4) perception on counter-measure to reduce indoor pollution ($10$ questions with a rating $1$--$5$). The offline survey participants were mainly university students and faculties, aged between $20$--$38$, with $82.5\%$ male and $17.5\%$ female. We have received online survey responses from $30$ different cities over $4$ countries -- Germany, India, the UK, and the USA. The online participants were aged between $17$--$60$, with $75\%$ of males and $25\%$ of females. Most of the online participants were from within the $20$-$30$ age group, representing $55.4\%$ of the total participants; the $ 30$-$40$ age group accounted for $24.1\%$, while $14.4\%$ participation was between $40$-$50$ years old. The remaining participants ($6.1\%$) were more than $50$ years of age. Among these online participants, $46\%$ were from urban areas, $34\%$ were from suburban areas, and $20\%$ were from rural backgrounds.

\subsection{Observations}
\label{sec:observe}

\begin{figure*}[]
	\captionsetup[subfigure]{}
	\begin{center}
             \subfloat[CO\textsubscript{2} is harmful\label{fig:mot_co2_harm}]{
			\includegraphics[width=0.37\linewidth,keepaspectratio]{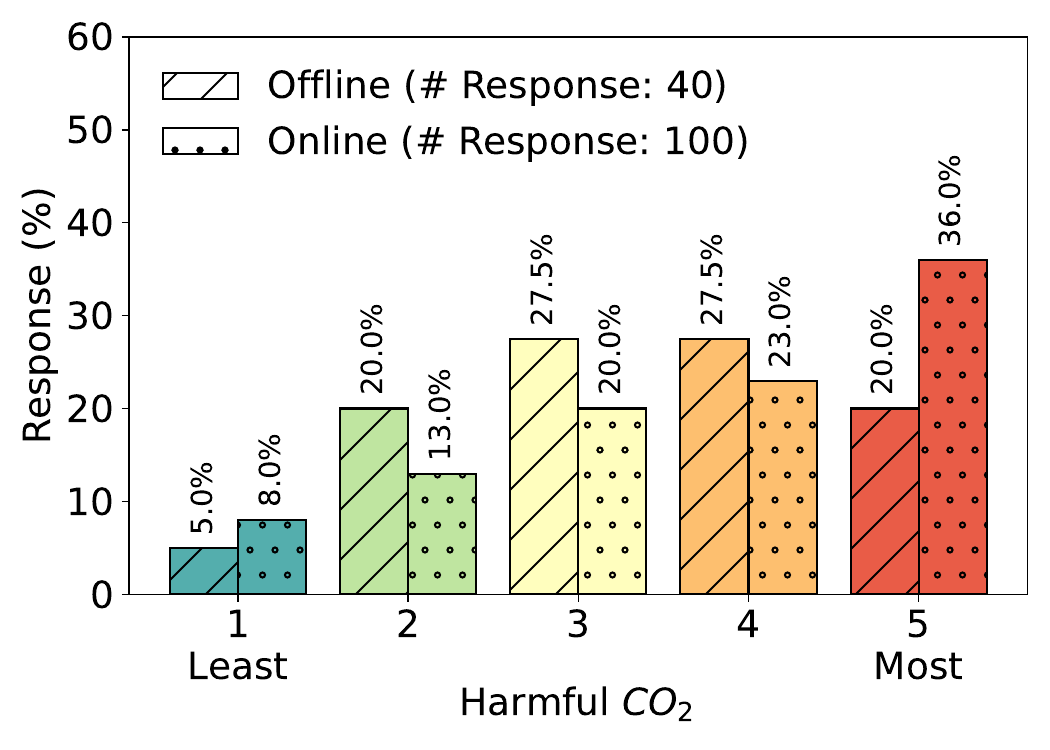}
		}
            \subfloat[Measured Pollutants once\label{fig:mot_measured}]{
			\includegraphics[width=0.29\linewidth,keepaspectratio]{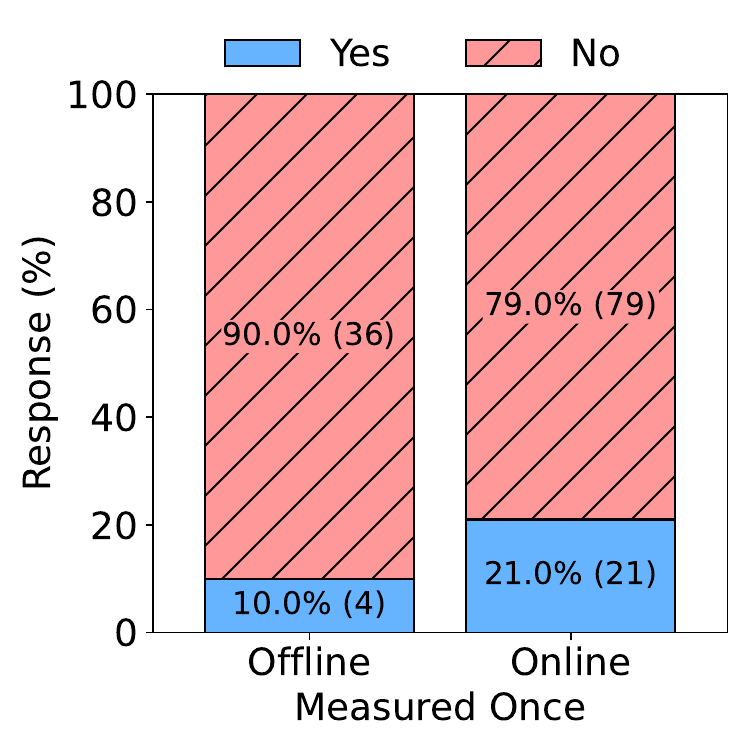}
		}
            \subfloat[400ppm is permissible in Indoors\label{fig:standard_400}]{
			\includegraphics[width=0.29\linewidth,keepaspectratio]{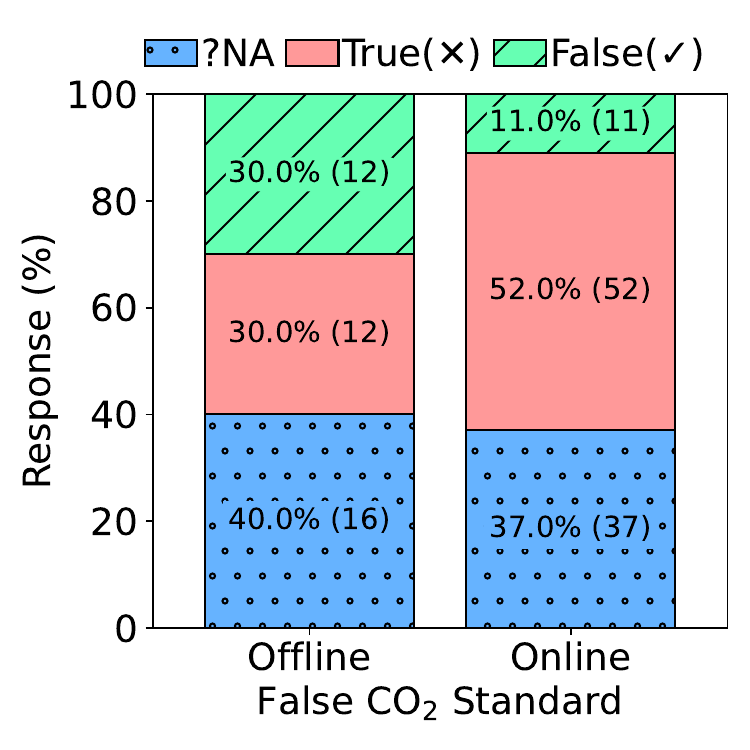}
		}
	\end{center}
	\caption{The responders of the prestudy (a) agree that CO\textsubscript{2} is harmful, but (b) most of them have never measured indoor pollutants. (c) Awareness about indoor pollutants -- response to a false statement about indoor CO\textsubscript{2} standards.}
    \Description{Three chart report responses to a prestudy survey: (a) bar chart shows that most participants agreed that CO2 is harmful, (b)  stacked bar chart indicates that most participants have never measured pollutants indoors, (c) another stacked bar chart shows low awareness about true CO2 standards in indoor environments, based on responses to a false statement.}
	\label{fig:co2_harm_measure}
\end{figure*}

\subsubsection{Experience with Pollution Measurement}
As per \figurename~\ref{fig:mot_co2_harm}, around half of responders (i.e., $47.5\%$ in offline, and $59\%$ online) think that CO\textsubscript{2} is harmful. However, when it comes to measuring these pollutants in their personal indoor spaces, only $10\%$ offline and $21\%$ online responders have experience with pollution monitors and measurements, as shown in \figurename~\ref{fig:mot_measured}. Overall proportion ($\hat{p}$) $0.178$ (95\% CI: $0.115-0.242$, z-statistic $-5.35$, p$<.01$) indicates that most of the responders who think CO\textsubscript{2} is harmful have never actually tried to measure their exposure level.

\subsubsection{Awareness and Ventilation Strategy}
We observe that the majority of the responders (i.e., $87.5\%$ offline and $85\%$ online) are aware that indoor gatherings increase CO\textsubscript{2}. However, as depicted in \figurename~\ref{fig:standard_400}, we observe that most of the responders are unfamiliar with permissible CO\textsubscript{2} limits in indoors (i.e., only \textit{30\%} in offline and \textit{11\%} in online were correct) with overall proportion ($\hat{p}$) $0.164$ (95\% CI: $0.103-0.225$, z-statistic $-5.69$, p$<.01$). Notably, a significant number of survey participants were not able to answer the facts related to indoor pollutants like CO\textsubscript{2} and the strategy to mitigate them from indoor spaces under different situations. For example, most of the participants, both from the offline and online survey, believed that split AC can ventilate pollutants, contrary to the observations made in existing studies~\cite{zhong2021complexity,snow2019performance}. To improve energy efficiency, the split AC recirculates the cold indoor air rather than injecting outside fresh air. Thus, the inside polluted air is not ventilated unless an embedded ventilation system with the AC exists. 

To analyze this further, we asked how the participants would ventilate the CO\textsubscript{2} in a family gathering. Interestingly, we observed that $27.5\%$ offline and $33\%$ online responders proposed an ineffective solution (i.e., turn on split AC or ceiling fan). Further, $57.5\%$ offline and $31\%$ online responders get confused and propose a solution that might be effective for reducing CO\textsubscript{2} but would waste resources (i.e., turn on AC and open the windows at the same time). $7.5\%$ offline and $23\%$ online responders could choose an effective method (i.e., turn on the ceiling fan and open the windows while keeping the AC off); however, opening up windows can increase other pollutants like particulate matter (e.g., PM\textsubscript{2.5}) if the outdoor is heavily polluted (e.g., for cities like Delhi, Hotan, Dhaka, etc.). Only $7.5\%$ offline and $13\%$ online responders proposed a safe and effective approach (using electric window ventilation or an exhaust fan). Overall proportion ($\hat{p}$) $0.114$ (95\% CI: $0.061-0.167$, z-statistic $-6.9$, p$<.01$) of effective and safe ventilation approach in the survey indicates that due to less understanding of the indoor pollution dynamics, most responders may not be able to mitigate indoor pollutants like CO\textsubscript{2} in an effecient way.

\begin{figure*}[]
	\captionsetup[subfigure]{}
	\begin{center}
            \subfloat[Ventilation in Gathering\label{fig:gather_action}]{
			\includegraphics[width=0.29\linewidth,keepaspectratio]{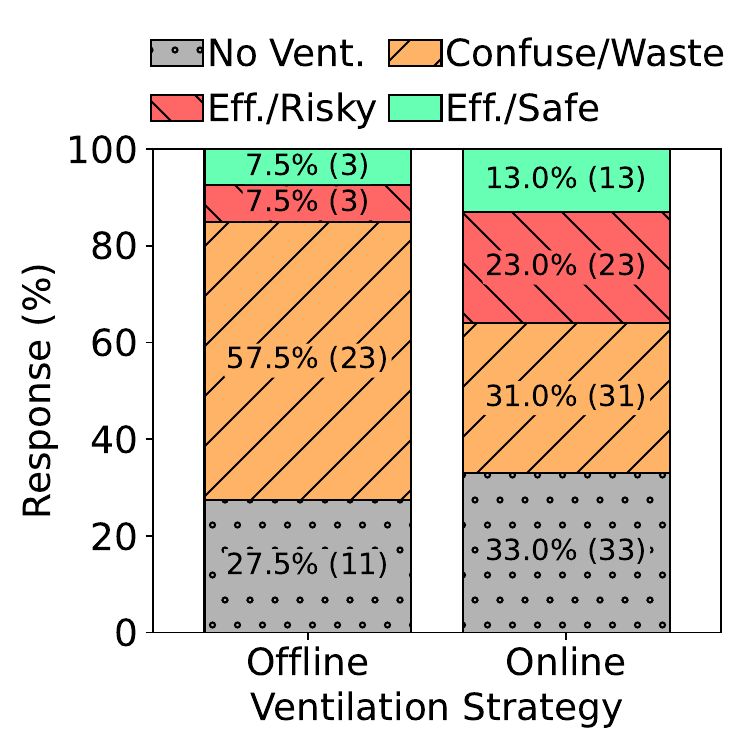}
		}
             \subfloat[Unaware of Indoor pollution\label{fig:unaware}]{
			\includegraphics[width=0.33\linewidth,keepaspectratio]{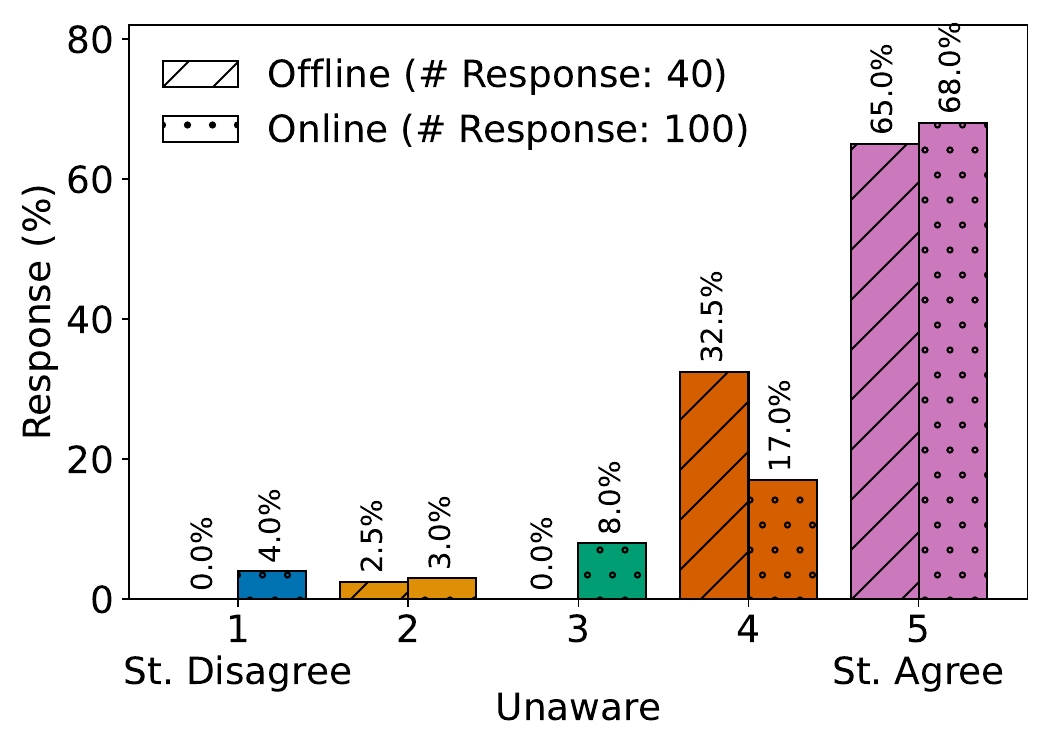}
		}
            \subfloat[Can reduce pollutants if I can see it\label{fig:canreduce}]{
			\includegraphics[width=0.33\linewidth,keepaspectratio]{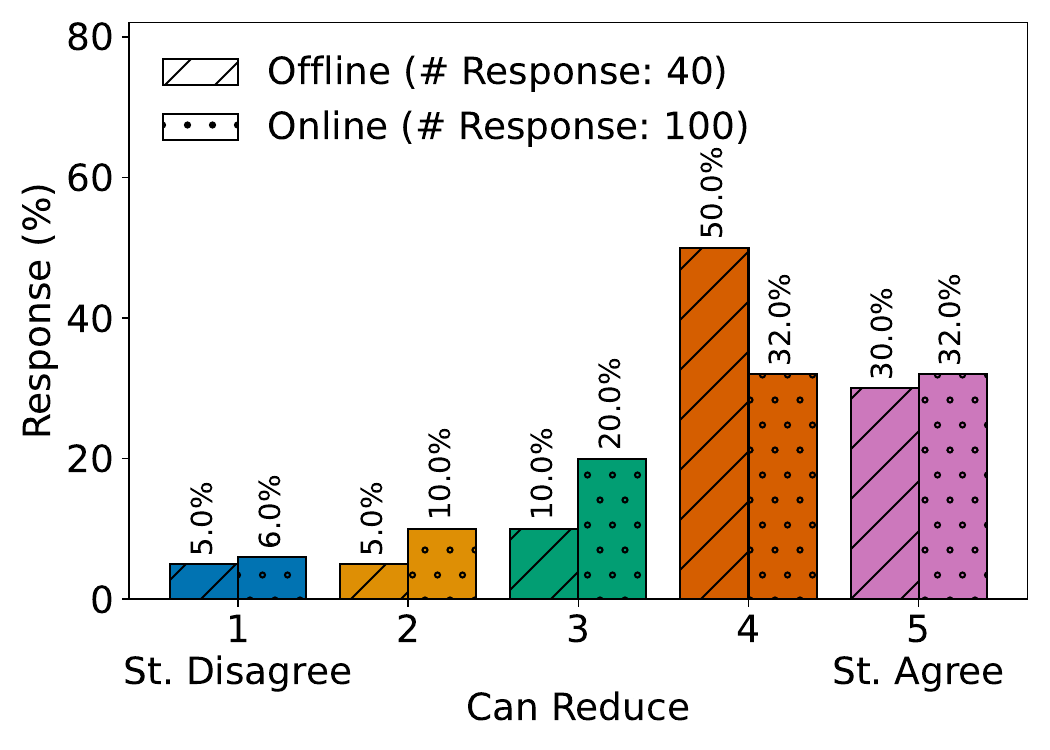}
		}
	\end{center}
	\caption{(a) Effectiveness of different ventilation methods in a family gathering, as perceived by the responders. (b,c) Perception of the responders on (b) the fact that the general population is unaware of indoor pollution and (c) their confidence in reducing the pollution through a visualization method. The survey indicates the need for effective pollution visualization, like an AR-based solution.}
    \Description{The figure consists of three charts summarizing prestudy survey responses: (a) stacked bar chart shows how respondents perceived the effectiveness of different ventilation methods during a family gathering, (b) bar chart highlights that most respondents believe the general population is unaware of indoor pollution, (c) Another bar chart shows that a majority feel confident they could reduce indoor pollution if they could see it, suggesting the importance of visualization tools.}
	\label{fig:unaware_cabreduce}
\end{figure*}

\subsubsection{Can Visualization Help?}
As shown in \figurename~\ref{fig:unaware}, most of the respondents in the mixed-mode survey agree that there is less awareness about indoor pollutants among the general population, whereas only $0.06$ (95\% CI: $0.019-0.101$, z-statistic $-7.95$, p$<.01$) of the overall proportion ($\hat{p}$) disagrees. Moreover, $80\%$ of offline and $64\%$ of online responders think that they can reduce the pollutants if they can see them, as shown in \figurename~\ref{fig:canreduce}, indicating the importance of a visualization method for indoor pollutants to improve awareness among the general population, fostering adequate ventilation, healthier, and happier indoors. Based on this prestudy that validates \textbf{H1}, we next discuss our observations from the pilot experiment to understand the indoor CO\textsubscript{2} dynamics to design effective data-driven visualization.

\subsection{How do we \textit{actually} manage CO\textsubscript{2} in Indoors?}
\label{sec:pilot_lab}
\begin{figure*}[]
	\captionsetup[subfigure]{}
	\begin{center}
            \subfloat[3 occupants \label{fig:lowoccupency}]{
			\includegraphics[width=0.316\linewidth,keepaspectratio]{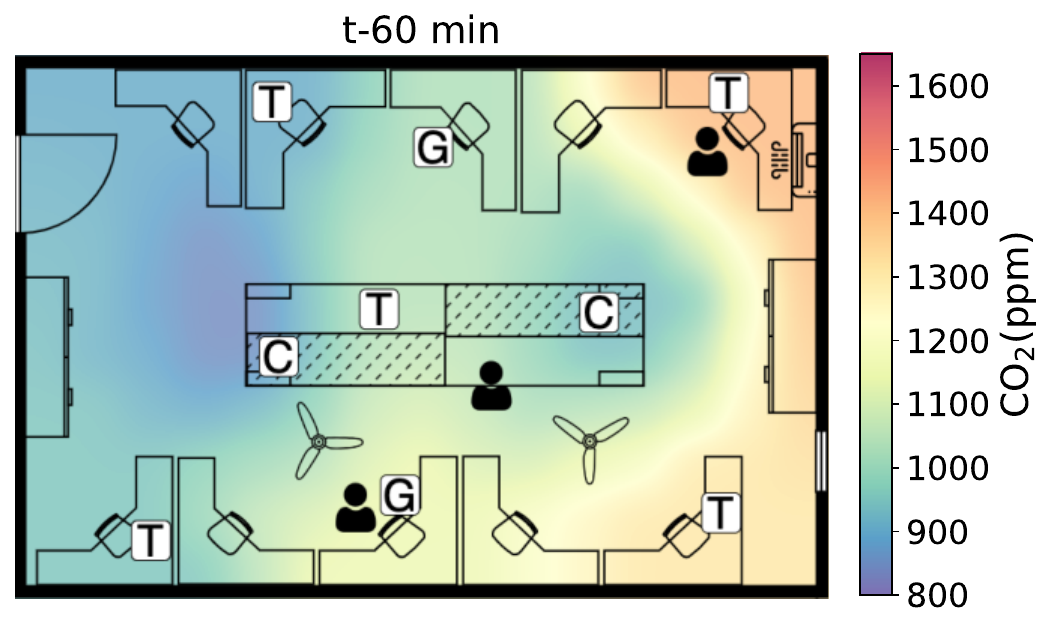}
		}
            \subfloat[4 occupants\label{fig:midoccupancy}]{
			\includegraphics[width=0.316\linewidth,keepaspectratio]{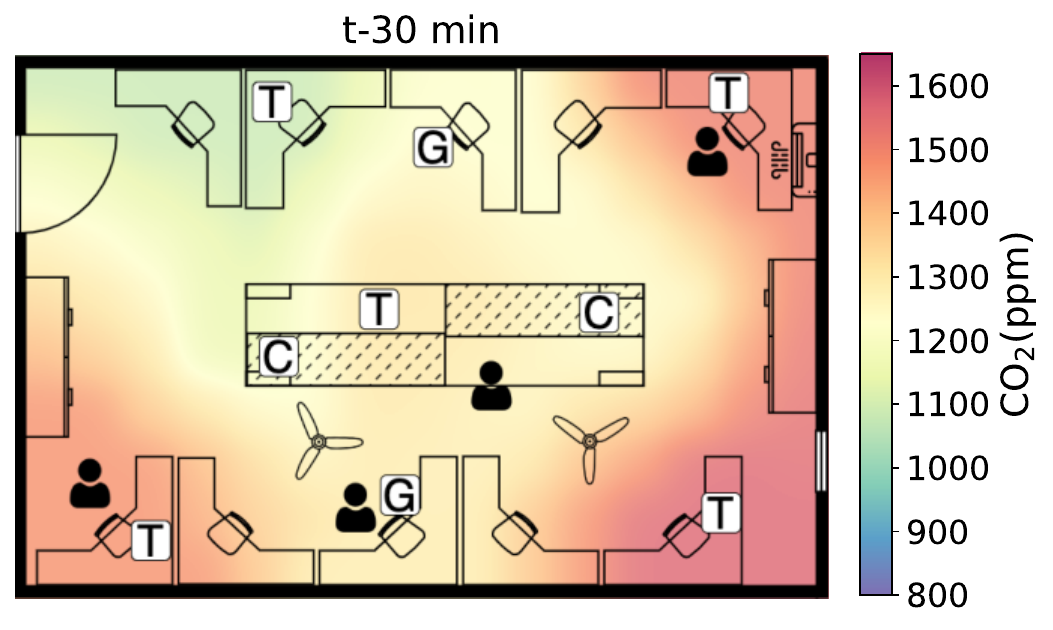}
		}
            \subfloat[7 occupants, window ventilator on\label{fig:highoccupancy}]{
			\includegraphics[width=0.316\linewidth,keepaspectratio]{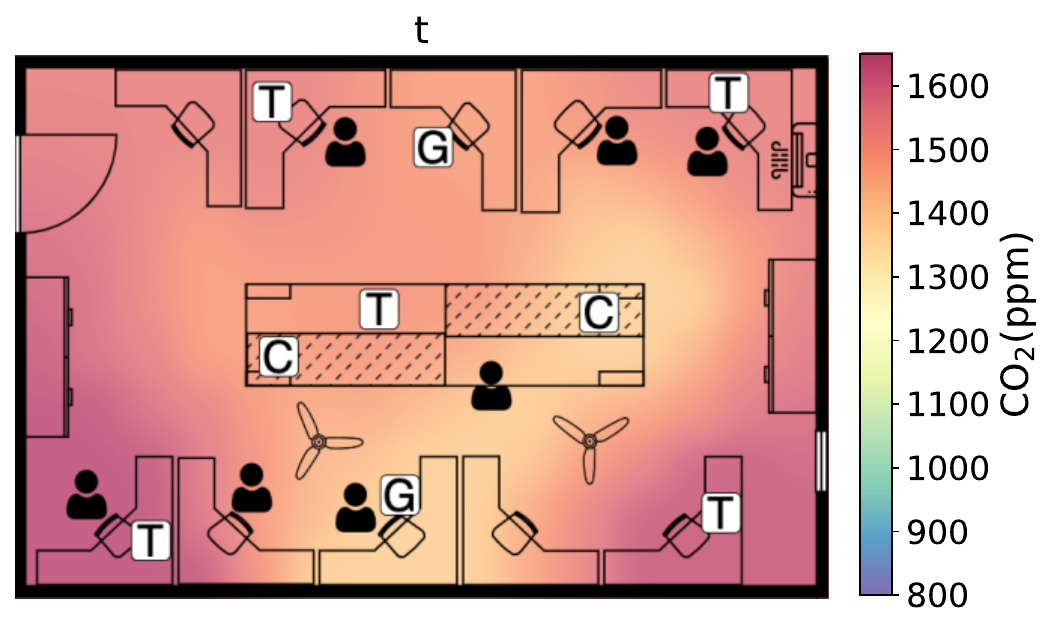}
		} \
             \subfloat[Room empty, window ventilator on \label{fig:ventmid}]{
			\includegraphics[width=0.316\linewidth,keepaspectratio]{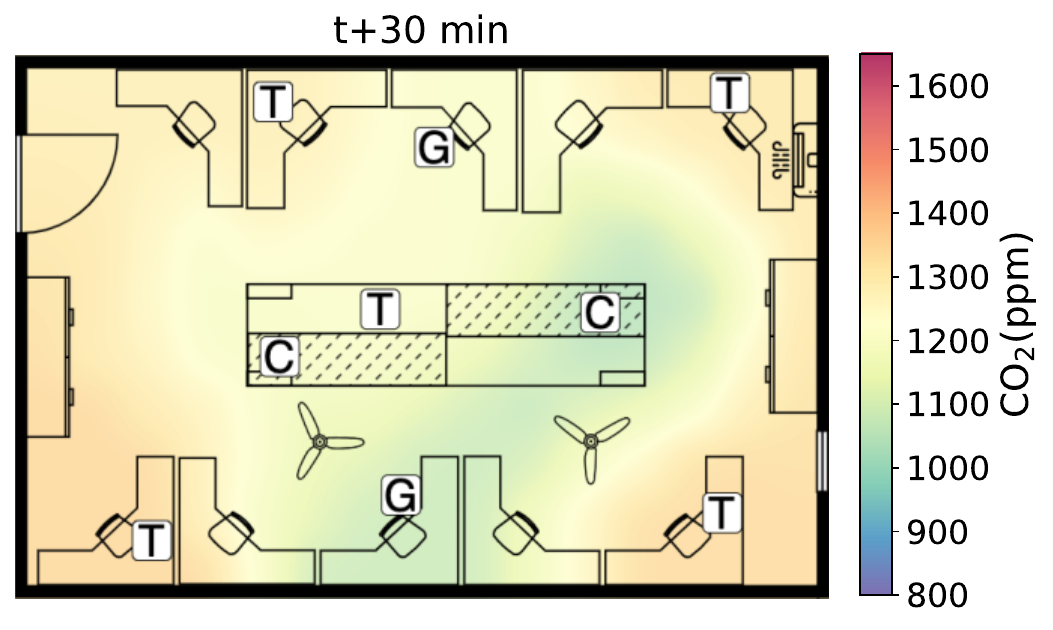}
		}
            \subfloat[Room empty, window ventilator on\label{fig:ventmax}]{
			\includegraphics[width=0.316\linewidth,keepaspectratio]{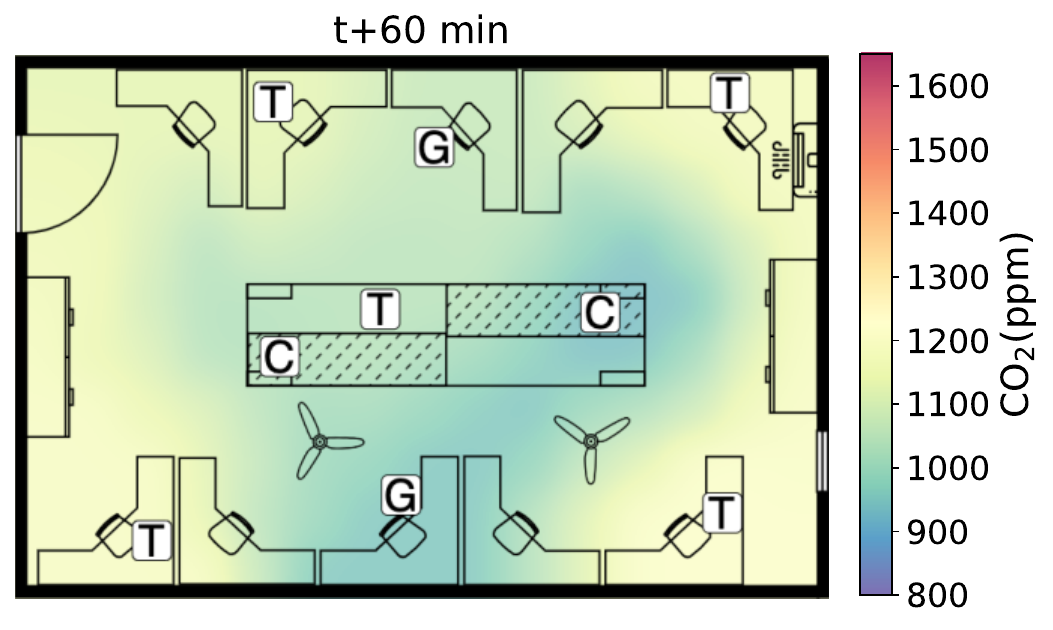}
		}
            \subfloat[Pedestal fan on in bottom-right corner\label{fig:fanaction}]{
			\includegraphics[width=0.316\linewidth,keepaspectratio]{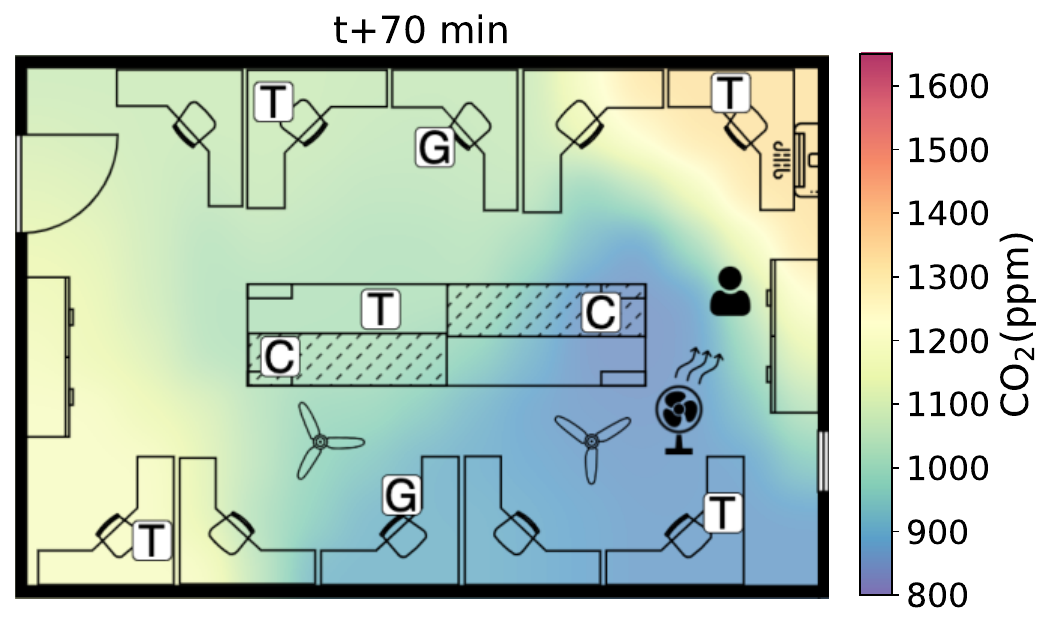}
		}
	\end{center}
	\caption{CO\textsubscript{2} distribution at various locations of a room at different heights. \textbf{[G]} represents ground height, \textbf{[T]} represents table height, and \textbf{[C]} represents ceiling height. The CO\textsubscript{2} concentration increases over time with the occupancy level of the room without ventilation (i.e., window ventilator is turned on at time $t$) - (a) three-person occupancy at $t$-60 minutes, (b) four-person occupancy at $t$-30 minutes, (c) seven-person occupancy at $t$, maximal CO\textsubscript{2} concentration of $1635$ ppm in bottom-left corner, window ventilator is turned on for ventilation. Occupants leave the room. CO\textsubscript{2} distribution when the room is - (d) ventilated for 30 minutes, (e) ventilated for 60 minutes. We observe that CO\textsubscript{2} accumulates and gets trapped in corners of the room at source height (i.e., occupants, table height). Lastly, (f) Turned on the stand fan from the bottom-right corner towards the top-right corner, reducing accumulated CO\textsubscript{2} in the bottom-right corner. Thus, targeted airflow can reduce trapping of CO\textsubscript{2} in specific areas of the room (e.g., corners).}
    \Description{A series of six heatmaps displays CO2 distribution in a room under different occupancy and ventilation conditions. Charts (a)–(c) show increasing CO2 concentration with three, four, and seven people in the room, peaking at 1635 ppm in a room corner. Charts (d) and (e) show how CO2 decreases after 30 and 60 minutes of window ventilation, though some CO2 remains trapped in corners. Chart (f) shows that turning on a pedestal fan helps disperse trapped CO2 in one corner. The figure demonstrates how occupancy and airflow shape indoor CO2 dynamics.}
	\label{fig:co2_manage}
\end{figure*}

We conducted a pilot study to investigate key physical behaviors of CO\textsubscript{2}, including spreading, trapping, and lingering, to inform data-driven visualization in personal indoor spaces. The study took place in a $15 \times 10\,m^2$ office containing $10$ cubicle workstations (cubicle height: $5\,m$), two ceiling fans, two pedestal fans, a window ventilator, and a split AC. We deployed pollution sensors at three heights across $9$ spatial locations (Figure~\ref{fig:co2_manage}). During the experiment, the ceiling fans and split AC remained off; only the pedestal fan and window ventilator were selectively operated.

As occupancy increased, localized CO\textsubscript{2} accumulation formed around active workstations (Figure~\ref{fig:lowoccupency}), with concentrations rising proportionally to the number of occupants (Figure~\ref{fig:midoccupancy}). CO\textsubscript{2} levels were consistently highest at table height, compared to floor or ceiling measurements. Once all sensors exceeded $1400$~ppm, the window ventilator was activated (Figure~\ref{fig:highoccupancy}). After occupants left, CO\textsubscript{2} levels decreased, particularly at ceiling height and central room regions; however, persistent elevated concentrations remained in corner cubicles at table height, forming localized \textit{personal CO\textsubscript{2} bubbles}~\cite{pantelic2020personal,ghahramani2019personal} exceeding $1200$~ppm (Figure~\ref{fig:ventmax}). Activating the pedestal fan near one such bubble significantly reduced local accumulation (Figure~\ref{fig:fanaction}). These CO\textsubscript{2} bubbles grow quasi-statically~\cite{enriquez2014quasi} and may remain trapped in corners unless dispersed by directed airflow.

\subsection{Personified, Situated Visualization and Actionability}
\label{sec:AR_wearable_motiv}
Central to our approach is the concept of \textit{personified CO\textsubscript{2} bubbles}, representing an individual's localized exposure zone. Capturing these highly dynamic and spatially heterogeneous bubbles typically requires dense networks of static CO\textsubscript{2} sensors, as demonstrated in our pilot study (see Fig.~\ref{fig:co2_manage}). However, such infrastructure is costly, immobile, and poorly suited to evolving environments or personal routines. In contrast, a wrist-worn CO\textsubscript{2} sensor travels with the user, enabling high-resolution, personalized exposure monitoring at lower cost and with greater flexibility. Real-time wearable data allows AR-based visualization of evolving CO\textsubscript{2} bubbles at meaningful physical locations and heights, transforming abstract sensor values into situated, perceptible risks. This spatial anchoring supports rapid identification of hotspots and encourages exploratory mitigation actions (e.g., directing airflow, opening windows). Immediate visual feedback, such as bubble shrinkage following ventilation, not only improves comprehension but also reinforces confidence and sustained engagement. We argue that AR serves as a bridge between \textit{risk location}, \textit{affected user}, and \textit{effective action}. The following sections detail the wearable prototype and AR platform.

\section{Prototype Design and Study Procedure}
\label{sec:method}

\begin{figure}
        \centering
        \includegraphics[width=0.7\linewidth,keepaspectratio]{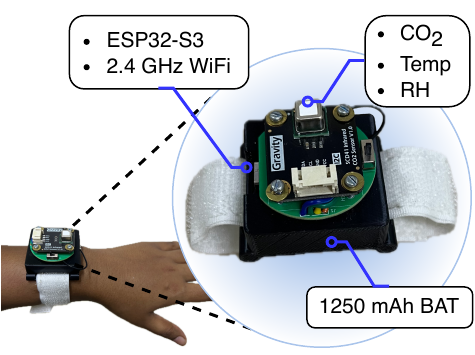}
	\caption{\ourmethod{} wrist-wearable.}
    \Description{This figure shows the CoWear wrist-wearable prototype. It is a compact rectangular device with embedded CO2, temperature, and humidity sensors, a rechargeable battery, and wireless connectivity. The design allows it to be worn on the wrist, making it portable and personal for continuous air quality monitoring.}
	\label{fig:cowear}
\end{figure}

On the basis of the pre-study survey and the pilot experiments, we design a visualization and interaction methodology to mitigate high concentrations of CO\textsubscript{2} (i.e., bubbles). We consider the following design goals to build the prototype. 
\begin{itemize}
    \item [\textbf{D1.}] To design a personal wrist-wearable prototype that can effectively measure CO\textsubscript{2} concentration in real-time at a personal indoor space. 
    \item [\textbf{D2.}] To design a smartphone-based AR gaming application with two sub-objectives: (1) visualizing the CO\textsubscript{2} bubbles in a personal space while displaying their severity in real-time, (2) interacting with the CO\textsubscript{2} bubbles through directed actions, like turning on the window ventilator, opening the window, directing airflow towards the bubble, etc., to reduce their severity. We discuss these in detail below.
\end{itemize}

\subsection{\ourmethod{} Wrist-wearable Sensor}
We developed a wrist-wearable module named \ourmethod{} to sense personalized pollution exposure of the user. \ourmethod{} is equipped with \textit{Temperature}, \textit{Humidity}, and \textit{Carbon Dioxide} (CO\textsubscript{2}) sensors~\cite{SENSIRION_co2_sensor} along with a \textit{1250 mAh battery}, enabling day-long power backup with only two hours of recharging time. Notably, wrist-wearable works best because of the following reasons: (1) the participants can move their hands to check the varying CO\textsubscript{2} exposure around them, (2) wrist-wearables are lightweight and provide a comfortable and robust solution compared to other commercial wearable pollution monitors available in the market like Atmotube Pro\footnote{\url{https://atmotube.com/atmotube-pro} (Accessed: \today)}, a neck-wearable or AirSniffler\footnote{\url{https://www.airsniffler.com/} (Accessed: \today)} that needs to be carried explicitly. For developing \ourmethod{}, we use the ESP32-S3 chip as the on-device processing unit that packs a dual-core Xtensa 32-bit LX7 microcontroller with 2.4 GHz Wi-Fi (802.11 b/g/n) capabilities. The connectivity board is a two-layer printed circuit board (FR4 material). This wearable uses a 3D printed shell (PLA+ material) to package the sensors and battery. \tablename~\ref{tab:ovl_spec} details the overall specifications of the \ourmethod{} wrist-wearable. The microcontroller periodically measures CO\textsubscript{2} at a $5$-second interval. The measurements are transmitted to client devices over HTTPS GET queries via the wireless channel with a latency of $58.76$ ($\pm5.32$) ms. Next, we describe the smartphone AR app that visually grounds real-time pollution data over the indoor space.

\begin{figure*}
    \centering
    \includegraphics[width=0.95\linewidth]{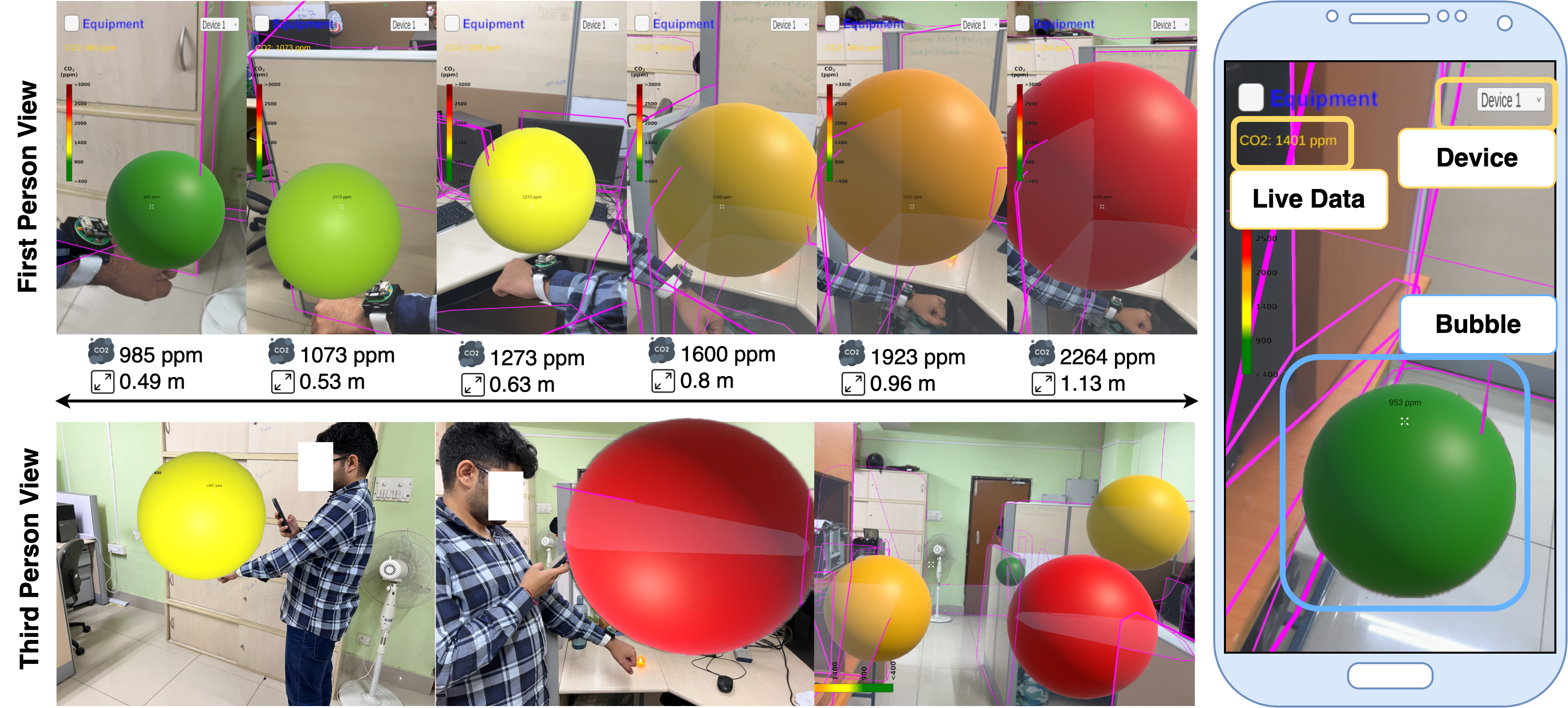}
    \caption{Augmented reality application - green bubble represents less CO\textsubscript{2} concentration, relatively larger yellow bubble represents moderate CO\textsubscript{2} concentration, and the largest red bubble represents more than $2000$ ppm CO\textsubscript{2} concentration. The bubbles' color and diameter vary according to the sensor readings (i.e., $400$ ppm $\Rightarrow$ green, $0.2$m bubble, and $3000$ ppm $\Rightarrow$ red, $1.5$m bubble).}
    \Description{A visualization from the AR application shows floating bubbles of different colors and sizes. A small green bubble represents a low CO2 concentration. A larger yellow bubble indicates moderate CO2 levels. The largest red bubble indicates dangerous CO2 concentrations. The figure illustrates how the AR system encodes pollution severity through color and bubble size.}
    \label{fig:app_scenario}
\end{figure*}

\subsection{Smartphone Augmented Reality (AR) Interactions}

\begin{table*}[]
\centering
\scriptsize
\caption{Overall specifications of \ourmethod{} wrist-wearable. Typical conditions represent $25^{\circ}$C, $50\%$ RH, $1013$ mbar ambient pressure.}
\Description{Overall specifications of CoWear wrist-wearable. The table reports operational details of the CO2 sensor and the system specifications of the microcontroller used to develop CoWear.}
\label{tab:ovl_spec}
\begin{tabular}{|l|l|l|l|l|l|l|l|c|l|} 
\cline{1-2}\cline{4-10}
\multicolumn{2}{|c|}{\textbf{System Specification}}                                                                                             & \multirow{10}{*}{} & \multicolumn{2}{c|}{\multirow{3}{*}{\textbf{Sensors}}}     & \multicolumn{5}{c|}{\textbf{Operational Details}}                                                                                                                                                                                                                                                                                                                                         \\ 
\cline{1-2}\cline{6-10}
\multirow{2}{*}{Microcontroller} & \multirow{2}{*}{\begin{tabular}[c]{@{}l@{}}Xtensa®32-bit LX7\\Clock 80\textasciitilde{}240 MHz\end{tabular}} &                    & \multicolumn{2}{c|}{}                                     & \multicolumn{1}{c|}{\multirow{2}{*}{\textbf{Range}}}                                       & \multicolumn{1}{c|}{\multirow{2}{*}{\textbf{Repeatability}}} & \multicolumn{2}{c|}{\textbf{Response Time}}                                                                                               & \multicolumn{1}{c|}{\multirow{2}{*}{\textbf{Yearly Drift}}}                       \\ 
\cline{8-9}
                                &                                                                                                              &                    & \multicolumn{2}{c|}{}                                     & \multicolumn{1}{c|}{}                                                                      & \multicolumn{1}{c|}{}                                        & \multicolumn{1}{c|}{\textbf{Preheat}} & \textbf{Poll}                                                                                 & \multicolumn{1}{c|}{}                                                             \\ 
\cline{1-2}\cline{4-10}
PSRAM+Flash                     & 8MB+8MB                                                                                                      &                    & \multirow{2}{*}{CO\textsubscript{2}}         & Condition                  & 400 ppm – 5000 ppm                                                                         & Typical                                                      & \multirow{2}{*}{60 sec}                & \multirow{7}{*}{\rotatebox[origin=c]{90}{\begin{tabular}[c]{@{}c@{}}5 sec\\(periodic samples)\end{tabular}}} & \multirow{2}{*}{\begin{tabular}[c]{@{}l@{}}$\pm$ 5 ppm +\\0.5 \% value\end{tabular}}  \\ 
\cline{1-2}\cline{5-7}
Connectivity                    & Wi-Fi 2.4 GHz                                                                                                &                    &                              & Accuracy                   & $\pm$ 40 ppm + 5\% value                                                                       & $\pm$ 10 ppm                                                     &                                        &                                                                                                  &                                                                                   \\ 
\cline{1-2}\cline{4-8}\cline{10-10}
COM Latency (ms)                & 58.76 ($\pm$5.32)                                                                                                &                    & \multirow{3}{*}{Humidity}    & \multirow{2}{*}{Condition} & \multirow{2}{*}{\begin{tabular}[c]{@{}l@{}}0\% RH – 95\% RH\\-10 $^{\circ}$C – 60 $^{\circ}$C\end{tabular}} & \multirow{2}{*}{Typical}                                     & \multirow{3}{*}{90 sec}                &                                                                                                  & \multirow{3}{*}{ 0.25\% RH}                                                       \\ 
\cline{1-2}
Avg Power (W)                   & 0.244 @3.7V                                                                                                    &                    &                              &                            &                                                                                            &                                                              &                                        &                                                                                                  &                                                                                   \\ 
\cline{1-2}\cline{5-7}
Battery (mAh)                   & 1250                                                                                                         &                    &                              & Accuracy                   & $\pm$ 6\% RH                                                                                   & $\pm$ 0.4\% RH                                                   &                                        &                                                                                                  &                                                                                   \\ 
\cline{1-2}\cline{4-8}\cline{10-10}
Dimensions (mm$^3$)                & 42$\times$49$\times$18                                                                                                     &                    & \multirow{2}{*}{Temperature} & Condition                  & - 10 $^{\circ}$C – 60 $^{\circ}$C                                                                              & Typical                                                      & \multirow{2}{*}{120 sec}               &                                                                                                  & \multirow{2}{*}{ 0.03 $^{\circ}$C}                                                         \\ 
\cline{1-2}\cline{5-7}
Weight (g)                      & 50                                                                                                           &                    &                              & Accuracy                   & $\pm$ 0.8 $^{\circ}$C                                                                                   & $\pm$ 0.1 $^{\circ}$C                                                      &                                        &                                                                                                  &                                                                                   \\
\cline{1-2}\cline{4-10}
\end{tabular}
\end{table*}

We developed an AR application that visualizes indoor CO\textsubscript{2} concentration with virtual bubble objects based on our observations from Section~\ref{sec:pilot_lab}. The app has three key features: spatial anchoring, personified pollution visualization, and real-time AR interactions with pollutants. The app helps to understand indoor pollution hotspots as follows.

\subsubsection{3D Spatial Anchoring}
The app creates a relative 3D coordinate system for the indoor environment using the Unity3D Plane Manager library. By detecting planar objects and walls in the environment, the app provides spatial anchoring for tracking virtual objects' location and size, regardless of the smartphone's location. To place objects at any location, the user must scan the entire indoor space at the start of the app. Although this is an essential step, modern smartphone cameras (e.g., Apple iPhone 13 Pro in this study) allow us to scan an entire space in minimal time.

\subsubsection{Personified Pollution Visualization with \ourmethod{}}
The AR app is coupled with \ourmethod{} wrist-wearable -- the wearable measures user-centric CO\textsubscript{2} exposure at any location of the indoor space. The AR app visually anchors the CO\textsubscript{2} data by allowing the user to spawn representative AR bubbles that vary in terms of color and diameter with the sensor readings at any particular location. A smaller, more greenish bubble represents a lower CO\textsubscript{2} concentration. Typical outdoor $400$ ppm, CO\textsubscript{2} is represented as a green $0.2$m bubble. Whereas increased CO\textsubscript{2} reading leads to yellow and then red bubbles in a continuous spectrum. An unhealthy high CO\textsubscript{2} reading of $3000$ ppm indoors is represented as a red, $1.5$m bubble. The bubbles placed in different pollution scenarios are shown in \figurename~\ref{fig:app_scenario}. The users must align their hand to co-localize the \ourmethod{} wrist-wearable and the AR bubble such that the bubble represents the local pollution concentration. The user must stay near the AR bubble ($\leq1$m) to notice the change in the bubble's color and size with the accumulation or ventilation of the pollutants.

\subsubsection{AR Interactions with the Pollutants}
\label{sec:interact}
CO\textsubscript{2} sources, such as cooking, candles, and indoor gatherings, form CO\textsubscript{2} bubbles in the indoor environment. Candles produce CO\textsubscript{2} steadily, creating CO\textsubscript{2} bubbles around it. Cooking food generates a significant amount of pollutants due to the fire. Baking soda-based food items (i.e., cakes and fried foods) produce CO\textsubscript{2} even when heated in an induction oven. Additionally, small indoor gatherings cause significant CO\textsubscript{2} accumulation in that area due to the respiratory emission of the occupants, as shown in our pilot experiment on how to manage indoor CO\textsubscript{2} in section~\ref{sec:pilot_lab}. Subsequently, CO\textsubscript{2} bubbles get trapped at various corners of indoor spaces unless they are removed through external airflow. The user places the representative CO\textsubscript{2} bubbles using the AR app and reduces the bubbles with these tools and available ventilation equipment, such as ceiling fans, pedestal fans, open windows, and window ventilators. For instance, the user may use the hand fan to direct airflow towards the window ventilator or the opened window, observing a gradual shrinkage and color shift in the AR bubble with a reduction in CO\textsubscript{2} concentration in the indoor location. With these AR bubbles, users can identify areas of accumulation or potential pollution sources. In addition, bubble shrinkage can be monitored to confirm effective ventilation of the indoor space.

\begin{figure}[]
	\captionsetup[subfigure]{}
	\begin{center}
             \subfloat[Office\label{fig:office}]{
			\includegraphics[width=0.45\columnwidth,keepaspectratio]{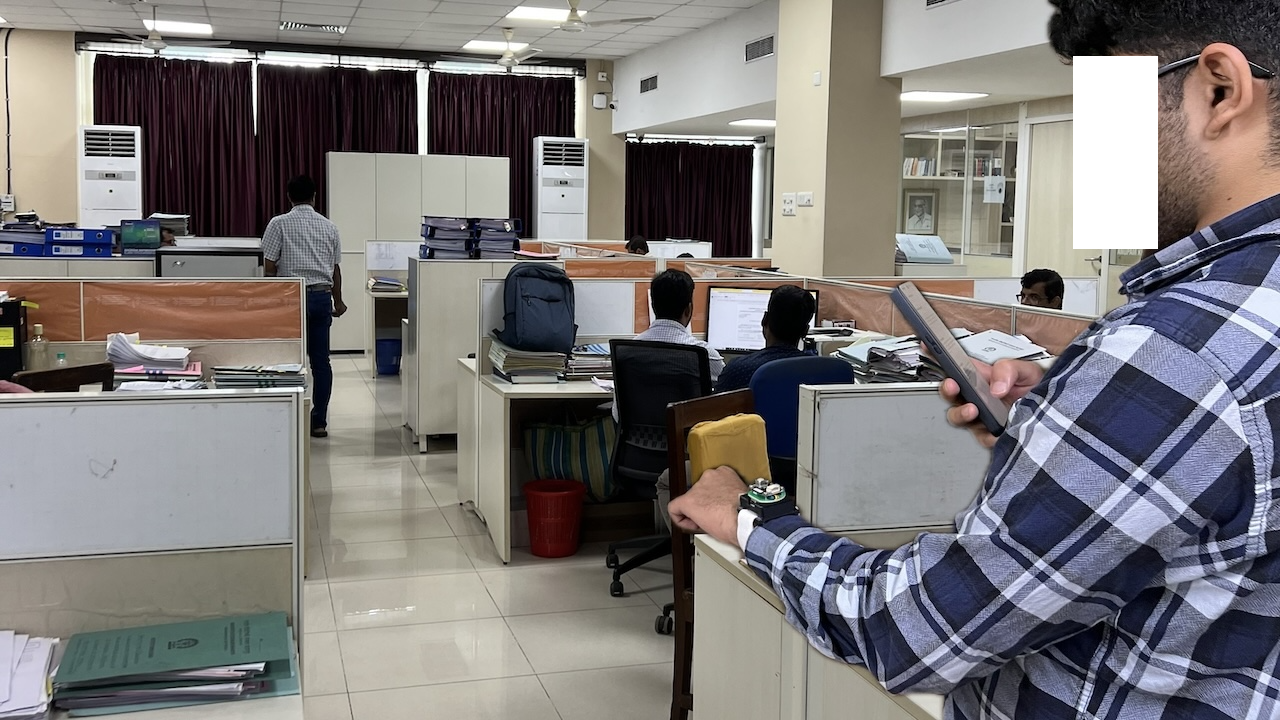}
		}
            \subfloat[Home\label{fig:home}]{
			\includegraphics[width=0.45\columnwidth,keepaspectratio]{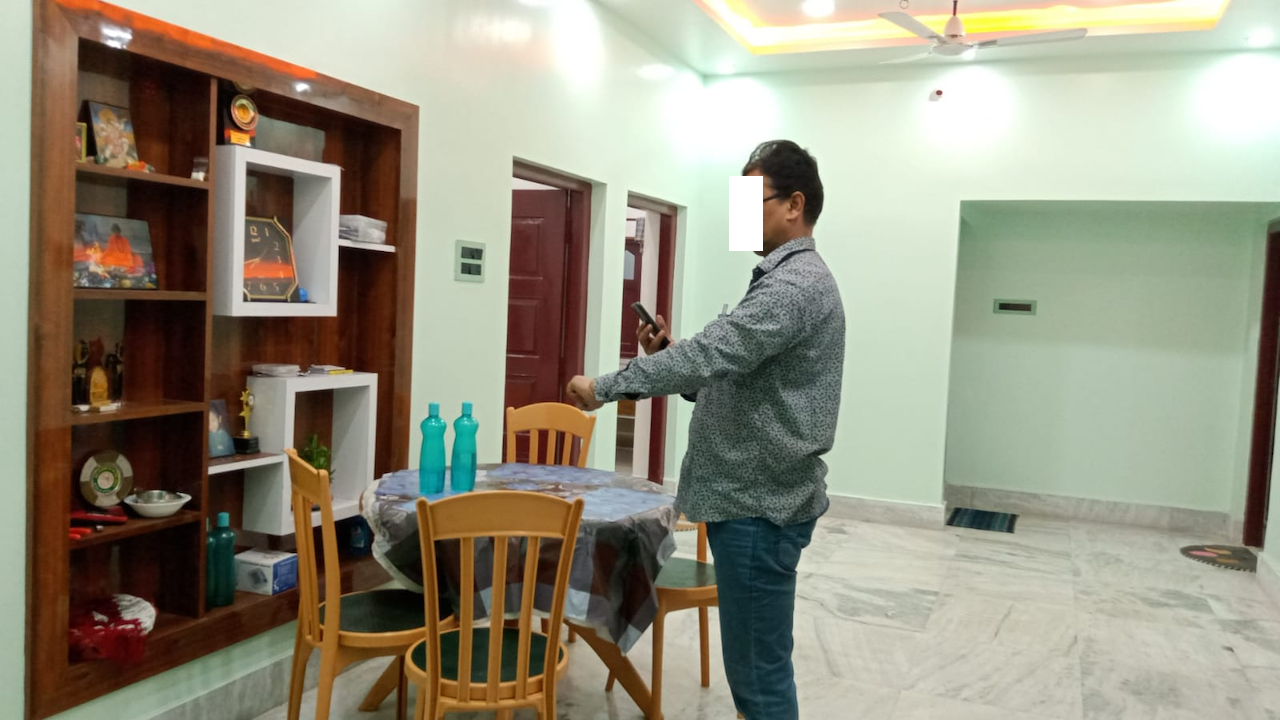}
		}\
            \subfloat[Diner\label{fig:diner}]{
			\includegraphics[width=0.45\columnwidth,keepaspectratio]{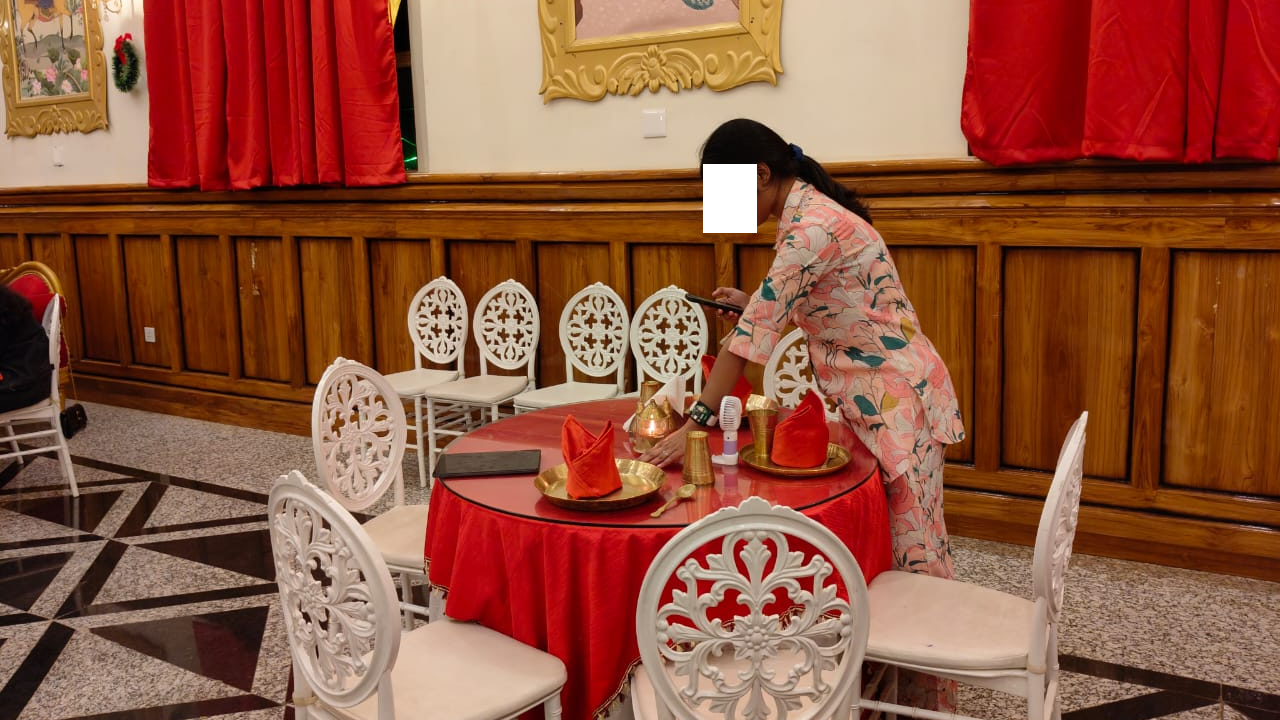}
		}
            \subfloat[Lab\label{fig:lab}]{
			\includegraphics[width=0.45\columnwidth,keepaspectratio]{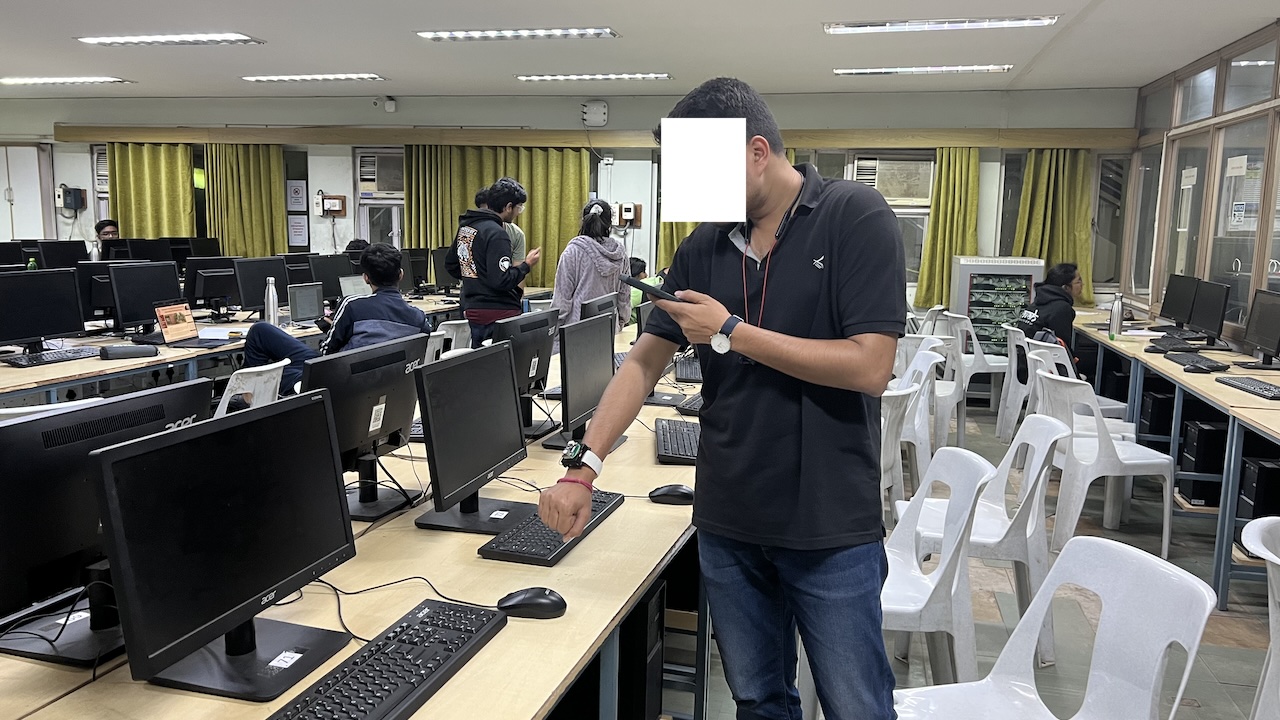}
		}
	\end{center}
	\caption{In-the-wild experiment conditions - (a) Working office, (b) Home with multiple rooms, (c) Diner, (d) Lab environment. The participant must first identify the dynamic pollution sources to effectively ventilate CO\textsubscript{2}.}
    \Description{Pictures of real-world experimental settings in four environments: an office, a home with multiple rooms, a diner, and a laboratory. These represent diverse conditions for in-the-wild testing of the AR system.}
	\label{fig:in_the_wild}
\end{figure}

\subsection{Study Conditions and Setup}
\label{sec:study_cond}
We have evaluated the system in both semi-controlled and in-the-wild settings. The semi-controlled experiments were conducted in two office rooms. In-the-wild experiments are conducted in office, household, diner, and lab environments. Next, we discuss the semi-controlled and in-the-wild setting in detail.

\subsubsection{Semi-controlled User Experiments}
\label{sec:semi-controlled}
We have taken two scenarios for these experiments: (i) a large office room (R1) with multiple windows, fans, and window ventilators, and (ii) a relatively small office room (R2) with only one window, fan, and window ventilator. The large room is $5$m $\times$ $8$m ($40$ m$^2$), and the small room is $3$m $\times$ $5$m ($15$ m$^2$). For the semi-controlled experiments, we ensure the formation of CO\textsubscript{2} bubbles in specific areas of the room by placing CO\textsubscript{2} sources (i.e., candles, heated baking soda, and indoor gatherings) in designated parts of the room. The participants act independently to figure out the bubbles and execute their ventilation strategies to reduce CO\textsubscript{2} concentration.

\subsubsection{In-the-wild User Experiments}
\label{sec:in-the-wild}
We have conducted uncontrolled in-the-wild user experiments in a working office, household, diner, and lab environments, as shown in the \figurename~\ref{fig:in_the_wild}. The pollution sources in these environments are dynamic and depend on the other occupants and their activities (e.g., cooking, burning, gathering). For instance, in a diner, CO\textsubscript{2} bubbles can increase when the diner staff decides to put up candles at each table or serve on hot plates (like sizzlers). Moreover, CO\textsubscript{2} can accumulate in the corner of the office or the lab due to gathering. Therefore, the participant must find instances where CO\textsubscript{2} bubbles are formed with the \ourmethod{} wrist-wearable and the AR app. Subsequently, they can employ effective ventilation strategies to mitigate the accumulated CO\textsubscript{2} bubbles from the space.

\subsection{Participants}
We conducted an a priori power analysis to determine the minimum sample size required for our within-group mixed/augmented reality study. Following the guidelines for AR/MR experimentation~\cite{ortloff2025small}, we adopt recommended effect size thresholds of $0.40$ (small), $0.81$ (medium), and $1.55$ (large) for Cohen’s $d$, and $0.17$ (small), $0.33$ (medium), and $0.54$ (large) for Cramer’s $V$. We power our analyses such that the intervention is likely to elicit medium to large effects. For the Welch’s t-tests, detecting a medium-to-large effect (Cohen’s $d > 0.81$) at $80\%$ statistical power and significance ($\alpha$) $0.05$ requires a minimum of $15$ participants. Similarly, for categorical comparisons using $\chi^2$-tests, detecting large (Cramer’s $V > 0.54$) to medium (Cramer’s $V>0.33$) effects requires at least $17$ to $45$ participants, respectively. From~\cite{ortloff2025small}, which summarizes proceedings of CHI from 2019 to 2023 for effect sizes, within-group studies typically include $28.5$ median participants (AR/MR-specific range: $7$–$40$, median $20$). Based on this, we have recruited $35$ participants through a call for volunteers during the offline prestudy survey.

Most of the participants were undergraduate and graduate students. Therefore, most of them are accustomed to playing smartphone games and are used to wearing smartwatches. Their age ranges from $20$ to $48$ years ($\mu = 25.11$ years, $\sigma = 6.23$ years). 30 (85.7\%) of the participants are identified as male, and 5 (14.3\%) as female. Most of the participants already have fair experience with smartphone games. Two participants play smartphone games every day. Seven participants play weekly. Five participants play monthly at least once. Moreover, $12$ participants reported that they play smartphone games rarely. However, $9$ participants do not play smartphone games. The participants have limited or no prior experience with smartphone AR games. 

In the semi-controlled user experiments, among $35$ participants, 31 participated in session S1, and 34 participated in session S2 (i.e., $30$ participated in both S1 and S2, achieving 80\% statistical power for medium-to-large effects). Moreover, the in-the-wild user experiments were conducted opportunistically in real-world indoor setups (i.e., dinner, research lab, office, and households). Due to the sporadic availability of these spaces and overlap with participants' availability, $20$ participants who participated in both S1 and S2 (i.e., among $30$) took part in these sessions. Finally, these participants who took part in all the AR sessions, baseline our approach with a generic 2D pollution heatmap visualization, ensuring sufficient power to detect medium-to-large effects for analyzing user experience and perception.

\begin{figure*}
    \centering
    \includegraphics[width=0.95\linewidth]{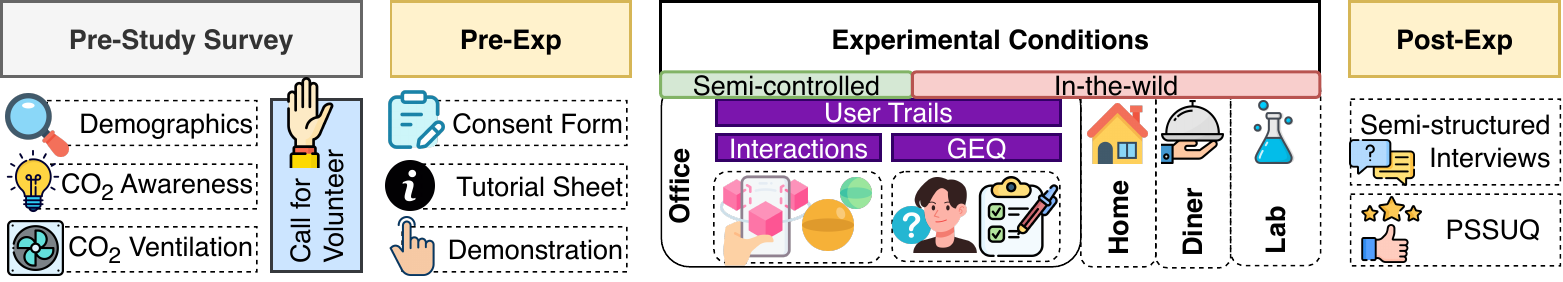}
    \caption{Overall study procedure - We estimated user awareness and recruited participants with a pre-study survey. We took consent from the participants, and after each user trial, we took an experience survey. Lastly, the participants took part in semi-structured interviews and usability surveys.}
    \Description{An overview of the three-phase study procedure. Phases include recruiting participants, pre-study surveys and consent, experimental sessions with the AR app, and post-experiment surveys with semi-structured interviews.}
    \label{fig:system}
\end{figure*}

\subsection{Study Procedure}
\label{sec:study}
The study contained three phases after recruiting the volunteers: (1) Pre-experiment activities (i.e., informed consents, explanation of the experiments, demonstration of the AR app, etc.), (2) User experiments (i.e., live interaction and game experience survey), and (3) Post-experiment activities (i.e., semi-structured interviews and system usability survey). \figurename~\ref{fig:system} depicts the overall study procedure. All survey questionnaires are included in the Appendix~\ref{sec:appendix}.

\subsubsection{Pre-experiment Activities}
First, an information sheet is handed out to the participants that describes the study overview, the objective of the participant during the study, how to use the smartphone AR gaming app, and the \ourmethod{} wrist-wearable to measure CO\textsubscript{2} concentration at any location of an indoor space. The information sheet also mentions the deployed sensors and the modalities being collected from the participants. Thereafter, participants signed a consent form to collect personal and experimental data during this study. Next, a research associate provided a brief tutorial to the new participants in a session by demonstrating the AR app (i.e., how to place the AR bubbles, how participants must align their hand to co-locate the \ourmethod{} wrist-wearable and the AR bubble to observe the changes in bubble color and size with the accumulation of pollutants over time).

\subsubsection{User Experiments}
A research associate helped the participants to put on \ourmethod{} wrist-wearable, an Empatica E4 watch, and a body camera to capture the participant's personalized CO\textsubscript{2} exposure, physiological data, and a first-person view of their actions. In a semi-controlled session, the research associate ensures that the CO\textsubscript{2} sources are active in their designated locations. Whereas, CO\textsubscript{2} sources are dynamic for in-the-wild sessions and depend on occupants' indoor activities as discussed in section~\ref{sec:study_cond}. We designed these sessions as gaming experiences for the participants, where they explore the indoor space and identify areas with higher CO\textsubscript{2} concentration by placing representative AR bubbles using the app. Further, the participants utilized the available tools (i.e., hand fans, battery-operated fans, etc.) and ventilation equipment (i.e., ceiling fan, pedestal fan, window ventilator, etc.) to direct airflow towards the AR bubbles, reducing CO\textsubscript{2} concentration in the space as discussed in section~\ref{sec:interact}. We kept $800$ ppm indoor CO\textsubscript{2} concentration as the sessions' stopping criterion (softbound). However, the participants can continue and further reduce the pollutants. A particular session can last between $ 20$ and $40$ minutes, including the pre-experiment activities. After each session, the participant completed the \textit{Game Experience Questionnaire} (GEQ)~\cite{ijsselsteijn2013game}, included in Appendix~\ref{appendix:post_study}, to assess their experiences and interactions with ventilating CO\textsubscript{2} using AR bubbles. We have used the (i) in-game GEQ and (ii) the post-game GEQ Questionnaires. Note that different user sessions are conducted on separate days.

\subsubsection{Post-experiment Activities}
After the user experiments, the participants were subjected to \textit{Post Study System Usability Questionnaire} (PSSUQ)~\cite{sauro2016quantifying} on the AR app. PSSUQ determines user-perceived system satisfaction with $16$ questions on a 7-point Likert scale (included in Appendix~\ref{appendix:pssuq}), where a lower score indicates better usability. The participants also provided feedback on: (i) whether their view on air pollutants improved compared to the prestudy survey, and (ii) features they would like to see in future versions of the AR app. Next, we organized \textit{Semi-structured Interviews} as focus group discussions among three participant groups and one research associate to understand their experience with different aspects of the system. The research associate moderated a discussion on how in-game interactions affected participants' perceptions of air pollutants and their gaming experience, along with suggestions to improve the current platform (discussion topics are included in Appendix~\ref{appendix:interview}). Participants can speak up in any order about the currently discussed topic and share their views without a time limit. We recorded the transcripts of the participants' opinions on their overall experience with the AR app and suggestions for improving the system. We next discuss our observations from these user experiments, surveys, and interviews.

\section{Study Results}
\label{sec:res}
This section analyzes how the participants perceived the visualization, gameplay, and overall experience during the semi-controlled and in-the-wild user experiment sessions.

\begin{figure*}
	\captionsetup[subfigure]{}
	\begin{center}
             \subfloat[Reduction in CO\textsubscript{2} concentration for each participant in S1\label{fig:min_max_co2_s1}]{
			\includegraphics[width=0.475\linewidth,keepaspectratio]{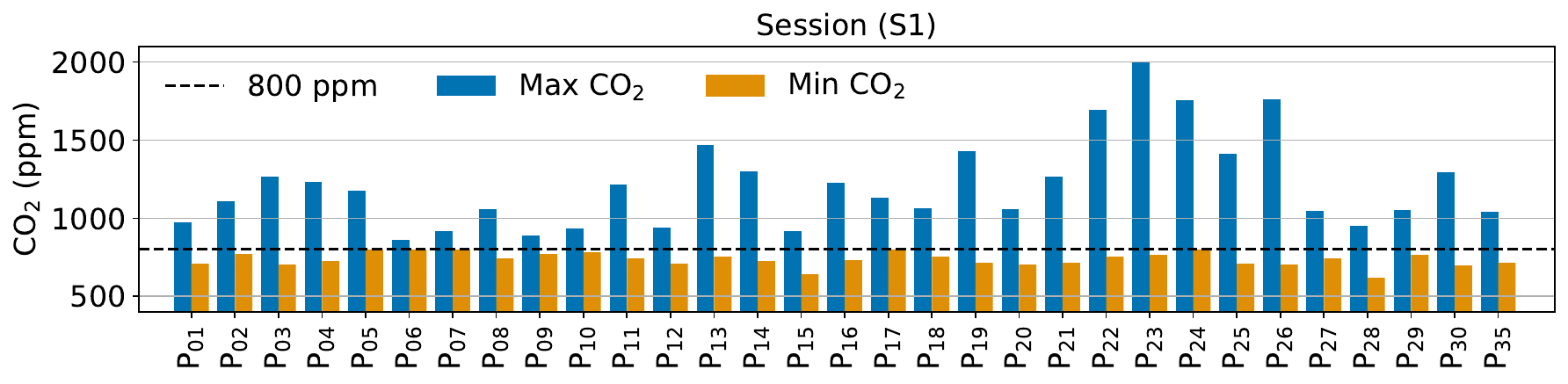}
		}
            \subfloat[Reduction in CO\textsubscript{2} concentration for each participant in S2\label{fig:min_max_co2_s2}]{
			\includegraphics[width=0.475\linewidth,keepaspectratio]{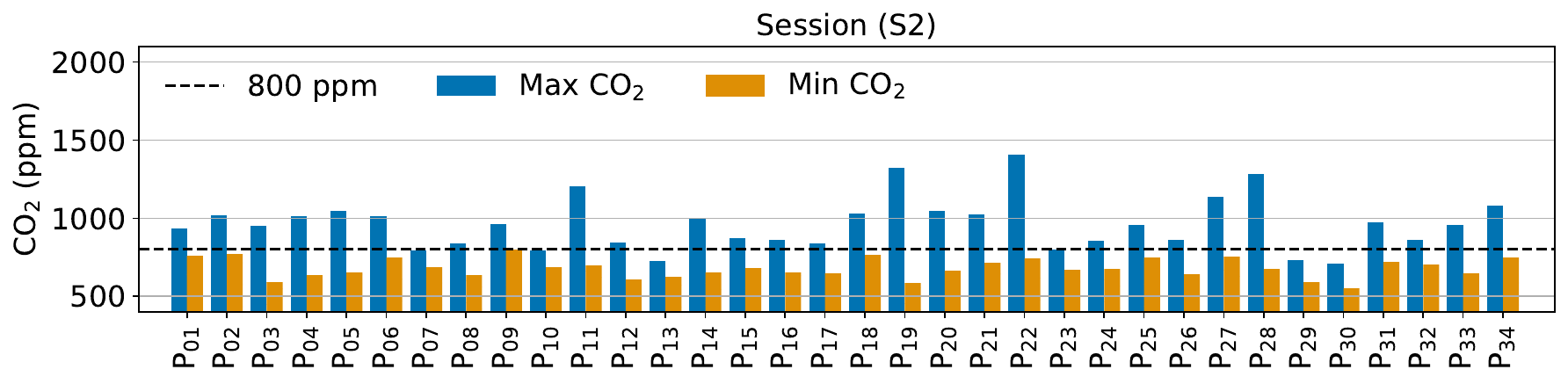}
		}\
            \subfloat[Average CO\textsubscript{2} reduction of in all AR settings\label{fig:avg_reduction_mr}]{
			\includegraphics[width=0.5\linewidth,keepaspectratio]{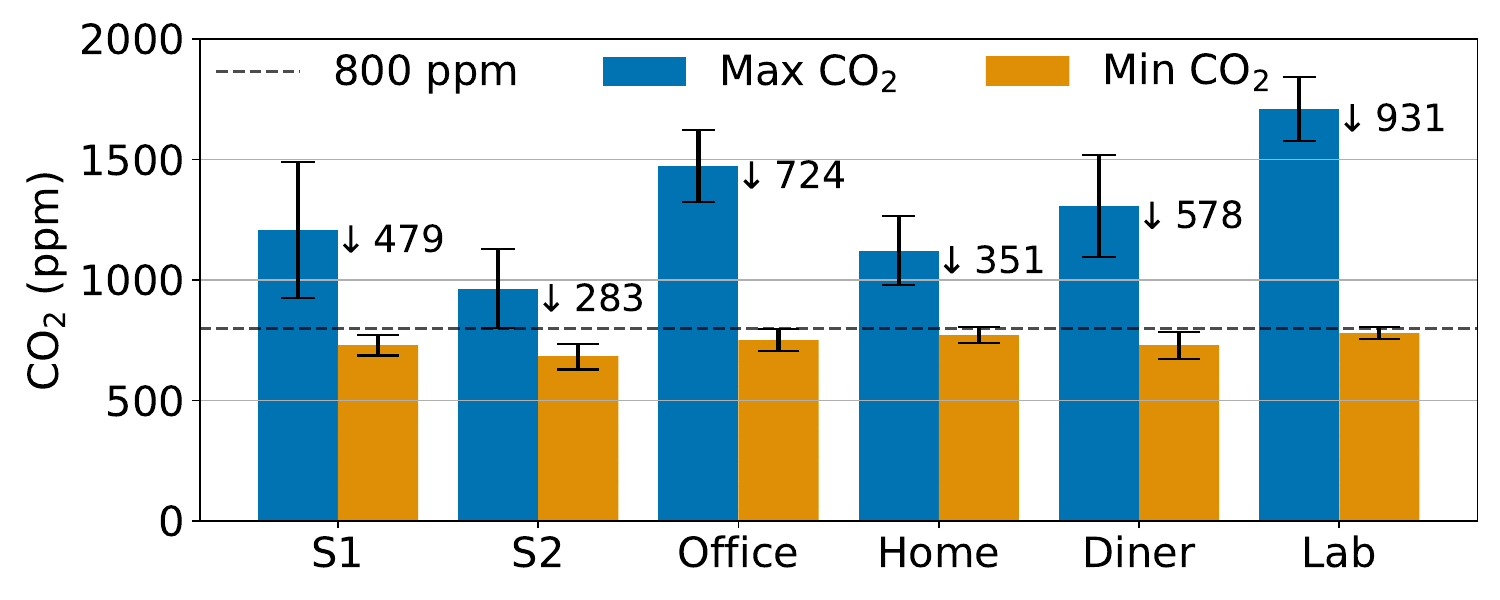}
		}
            \subfloat[Average CO\textsubscript{2} reduction in AR vs Heatmap\label{fig:avg_reduction_mrv2d}]{
			\includegraphics[width=0.35\linewidth,keepaspectratio]{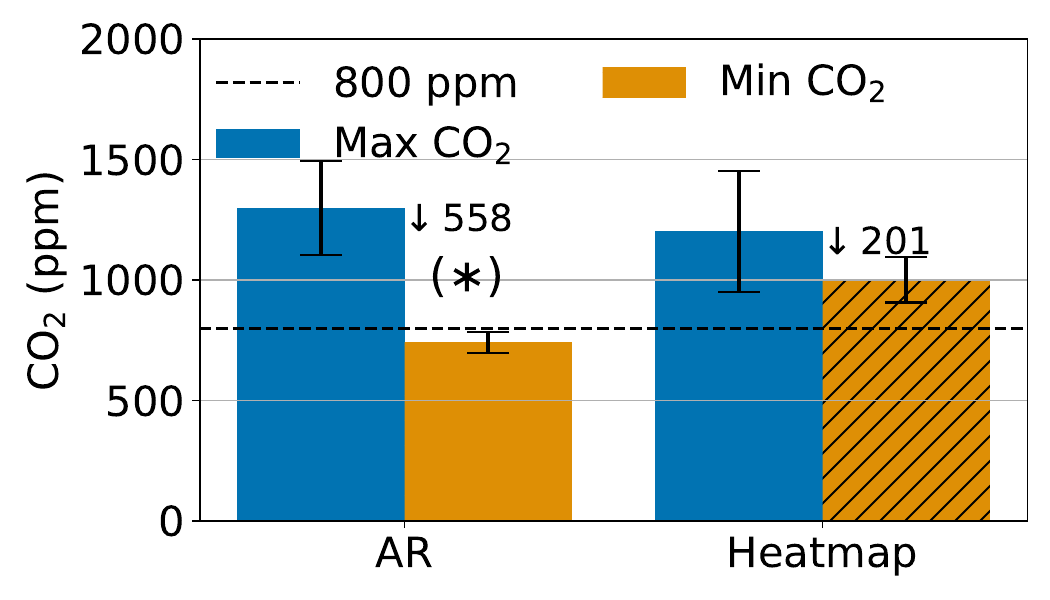}
		}
	\end{center}
	\caption{Effectiveness of the \ourmethod{} wearable and the AR app - (a,b) maximum CO\textsubscript{2} readings during the session S1, S2 and CO\textsubscript{2} reading at the end of the session after using the AR app to visualize and ventilate the pollutants, (c) average CO\textsubscript{2} reduction in each augmented reality setting, (d) average CO\textsubscript{2} reduction in augmented reality vs baseline 2D heatmap visualization. ($\ast$) indicates statistically significant CO\textsubscript{2} reduction with AR app.}
    \Description{Four bar charts summarizing the effectiveness of the AR application across sessions and environments. Charts (a)–(c) show reductions in CO2 concentration during experiments in semi-controlled and in-the-wild conditions. Chart (d) compares AR with a 2D heatmap, showing that AR enabled significant CO2 reductions, while the heatmap led to marginal improvements in air quality.}
	\label{fig:task_completion}
\end{figure*}

\subsection{Effective CO\textsubscript{2} Reduction}

\subsubsection{Effectiveness of the Augmented Reality Application}
In session S1, the average starting CO\textsubscript{2} concentration was $1207$ ppm, and the ending concentration was $728$ ppm. We observe an average reduction of $479$ (\textpm $281$) ppm. Individual reduction for each participant is shown in \figurename~\ref{fig:min_max_co2_s1}. In S1, most of the participants were newly introduced to the AR app. They took around $2.14$ (\textpm $1.6$) minutes per $100$ ppm of CO\textsubscript{2} reduction. The maximum and minimum time participants took to ventilate $100$ ppm CO\textsubscript{2} in S1 were $6$ minutes and $0.34$ minutes, respectively. In session S2, the average starting CO\textsubscript{2} concentration was $963$ ppm, and the ending concentration was $680$ ppm. We observe, on average, $283$ (\textpm $155$) ppm reduction during the session as shown in \figurename~\ref{fig:min_max_co2_s2}. In S2, $30$ participants were familiar with the AR app from S1, so we observed faster CO\textsubscript{2} ventilation. Four newly introduced participants were also able to reduce CO\textsubscript{2} levels below the $800$ ppm target. Participants took around $1.48$ (\textpm $1.11$) minutes (i.e., approx. $40$ seconds less than S1) to ventilate $100$ ppm of CO\textsubscript{2}. The participants took a maximum of $5.7$ minutes and a minimum of $0.3$ minutes to ventilate $100$ ppm CO\textsubscript{2} in S2. Similarly, during the in-the-wild experiments, we observe an average CO\textsubscript {2} reduction of $724$ ppm, $351$ ppm, $578$ ppm, and $931$ ppm in Office, Home, Diner, and Lab environments, respectively. \figurename~\ref{fig:avg_reduction_mr} shows the average starting and ending CO\textsubscript{2} concentration from the user experiments in different indoor setups. We observe a significant reduction (i.e., using Welch’s t-test, t-statistic $6.54$, p$<.01$, medium-to-large effect with Cohen's $d=-1.36$) of $558$ ppm CO\textsubscript{2} on average with the AR app in indoor environments as shown in \figurename~\ref{fig:avg_reduction_mrv2d}. Thus, the AR app and \ourmethod{} wrist-wearable have effectively represented CO\textsubscript{2} bubbles for participants to act upon and ventilate the pollutant from the indoor space.

\subsubsection{Effectiveness of 2D Spatial Heatmap}
To compare the effectiveness of the AR app, we tested with a 2D spatial heatmap visualization of CO\textsubscript{2} like \figurename~\ref{fig:co2_manage} in Section~\ref{sec:pilot_lab}. We deployed six static CO\textsubscript{2} sensors to generate a real-time pollution map of the office room (R1). Further, we conducted user experiments to understand how participants perceive the 2D pollution map and interact with the ventilation tools and equipment available near them. The 2D heatmap user experiments were conducted similarly to the semi-controlled AR sessions, and a research associate demonstrated the user interface to the participants before the experiments. We observed that participants faced difficulties planning targeted ventilation strategies with the heatmap visualization, resulting in longer session durations and inadequate ventilation. With the \ourmethod{} wearable and the AR app, we observe a significant CO\textsubscript{2} ventilation (i.e., $558$ ppm CO\textsubscript{2} on average); however, with the heatmap, participants can only ventilate up to $201$ ppm CO\textsubscript{2} on average with no statistical difference between the starting and ending CO\textsubscript{2} concentration, as shown in \figurename~\ref{fig:avg_reduction_mrv2d}. Most participants could not achieve the $800$ ppm target CO\textsubscript{2} level with the heatmap, even with longer session duration.

\subsection{Impact on User's Awareness}
\label{sec:awareness}
\begin{figure*}
	\captionsetup[subfigure]{}
	\begin{center}
            \subfloat[Improved my understanding\label{fig:understanding}]{
			\includegraphics[width=0.31\linewidth,keepaspectratio]{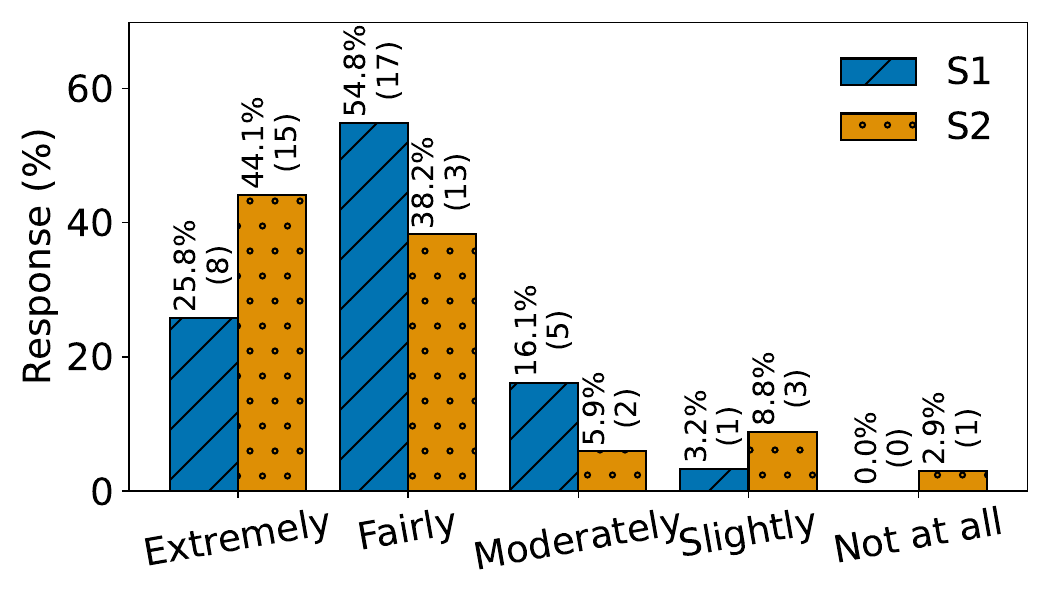}
		}
            \subfloat[Like to use in my home\label{fig:useinhome}]{
			\includegraphics[width=0.30\linewidth,keepaspectratio]{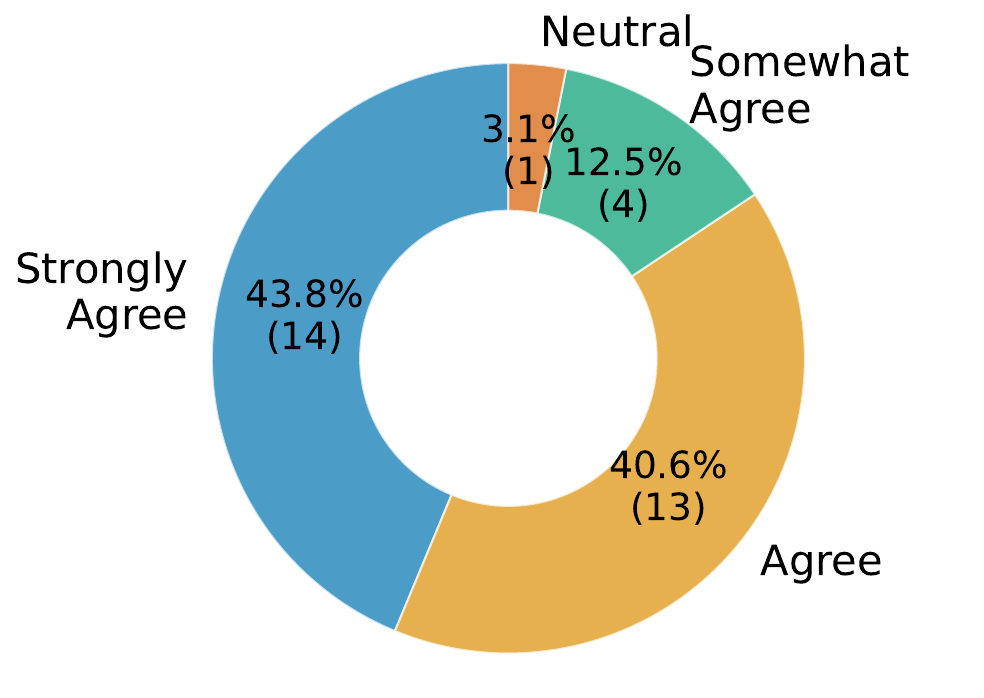}
		}
            \subfloat[Would recommend to a friend\label{fig:recommendtofriend}]{
			\includegraphics[width=0.30\linewidth,keepaspectratio]{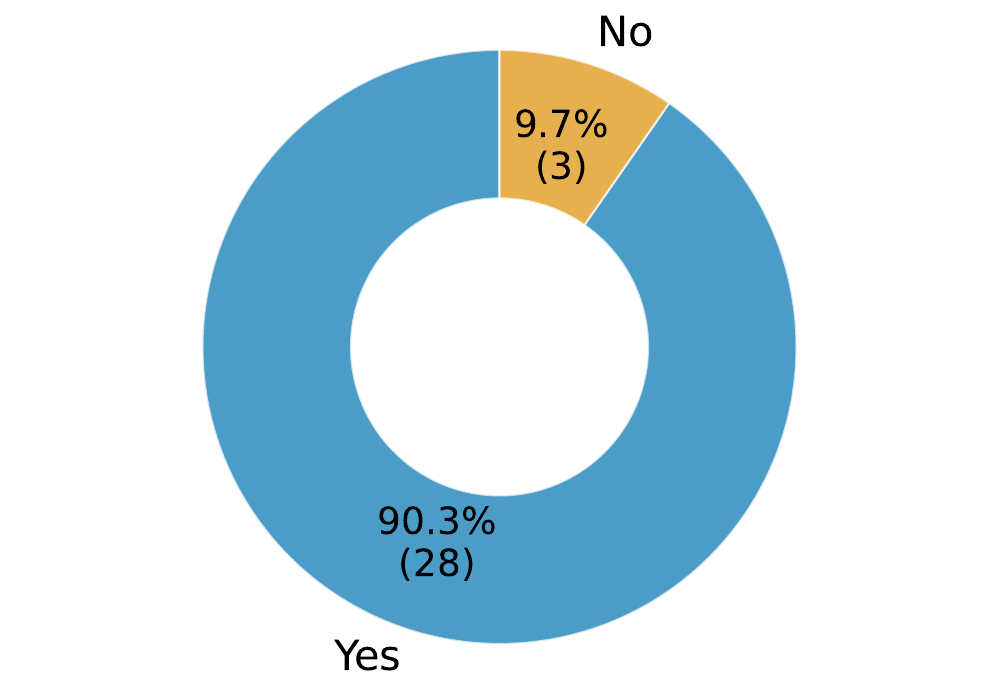}
		}
	\end{center}
	\caption{Understanding and awareness responses - (a) the AR app and the \ourmethod{} improved the understanding of the participants about indoor pollution, (b) the participants would like to use this app in their house to reduce pollution in kitchen and living room, (c) he participants would recommend this app to their friends and family to make them more aware about indoor pollution.}
    \Description{One grouped bar chart and two pie charts showing that most participants reported: (a) improved awareness of indoor pollutants after playing the AR game in S1 and S2, (b) wanting to use the system in their daily life, and (c) excitement of sharing the experience with their friends.}
	\label{fig:awareness}
\end{figure*}

As shown in \figurename~\ref{fig:understanding}, most participants reported that the smartphone AR app improved their overall understanding and awareness about indoor pollutants over the semi-controlled sessions. In S1, $8$ ($25.8\%$) \textit{extremely agree}, $17$ ($54.8\%$) \textit{fairly agree}, and in S2, $15$ ($44.1\%$) \textit{extremely agree}, $13$ ($38.2\%$) \textit{fairly agree} on the same. We observe a large association in the understanding of indoor pollutants ($\chi^2=41.64$,  p$<.01$, Cramer's $V=0.68$) from session S1 to S2 among the $30$ participants who attended both the sessions. An interesting participant comment related to awareness (AC) on how the app improves their understanding of indoor pollutants is as follows.

\noindent\textit{AC\#1: ``This fact, I didn't know at all, indoor pollution is a thing that we should be discussing. We always talk about outdoor pollution, but we never talk about indoor pollution. We might think, okay, candles and cooking, how much pollution can that be, but this made us realize that, just with two or three candles burning, this is the amount of ppm that you can get.''}

Moreover, $54.5\%$ \textit{strongly agree}, $30.3\%$ \textit{agree}, and $9\%$ \textit{somewhat agree} that this game makes the participants more aware of the pollution sources (like indoor gatherings, candles, etc) indoors. Among the participants, $14$ ($43.8\%$) \textit{strongly agree}, $13$ ($40.6\%$) \textit{agree} and $4$ ($12.5\%$) \textit{somewhat agree} to use this AR app in their home to reduce pollutants in places such as kitchen during cooking, and living room during a family get-together as shown in \figurename~\ref{fig:useinhome}. Notably, $28$ ($90.3\%$) participants wanted to recommend the app to their family members and friends (see \figurename~\ref{fig:recommendtofriend}) to make them more aware of pollution in their homes. Some participant comments related to indoor pollution sources are as follows:

\noindent\textit{AC\#2: ``We usually don't think about this. The small things, such as indoor group meetings, generate that high concentration of CO\textsubscript{2}. I guess the application has placed a sense and the fact that we usually feel tired and sleepy because of pollutants.''}

\noindent\textit{AC\#3: ``It increased really quickly, I had not heard that indoor pollution has a higher carbon dioxide level than outdoor pollution. It helps in reducing the carbon dioxide level, and also it makes me aware of how to deal with trapped pollutants.''}

\subsection{Quantitative User Feedback}
Here, we analyze the game enjoyment of the participants in terms of immersiveness and how competent and skillful they feel during the sessions, their interest and success in ventilating CO\textsubscript{2}, and the degree of positive experiences and associated physical (i.e., tiredness) and physiological overhead (i.e., tension, challenge, negative experiences, etc.).

\subsubsection{Competence}
\begin{figure*}
	\captionsetup[subfigure]{}
	\begin{center}
             \subfloat[I was interested in the game\label{fig:interest}]{
			\includegraphics[width=0.31\linewidth,keepaspectratio]{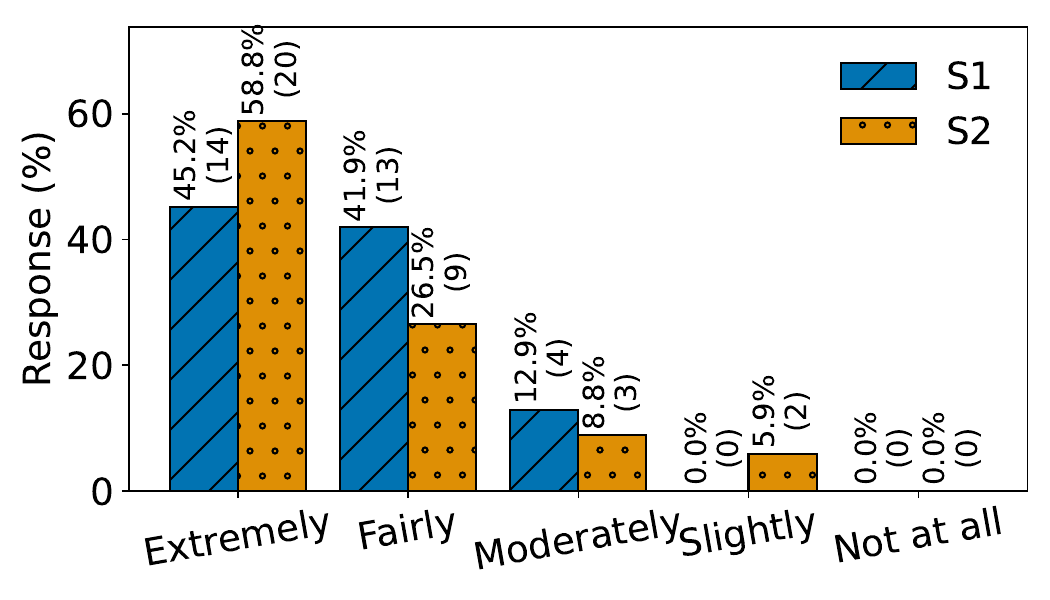}
		}
            \subfloat[I felt successful\label{fig:success}]{
			\includegraphics[width=0.31\linewidth,keepaspectratio]{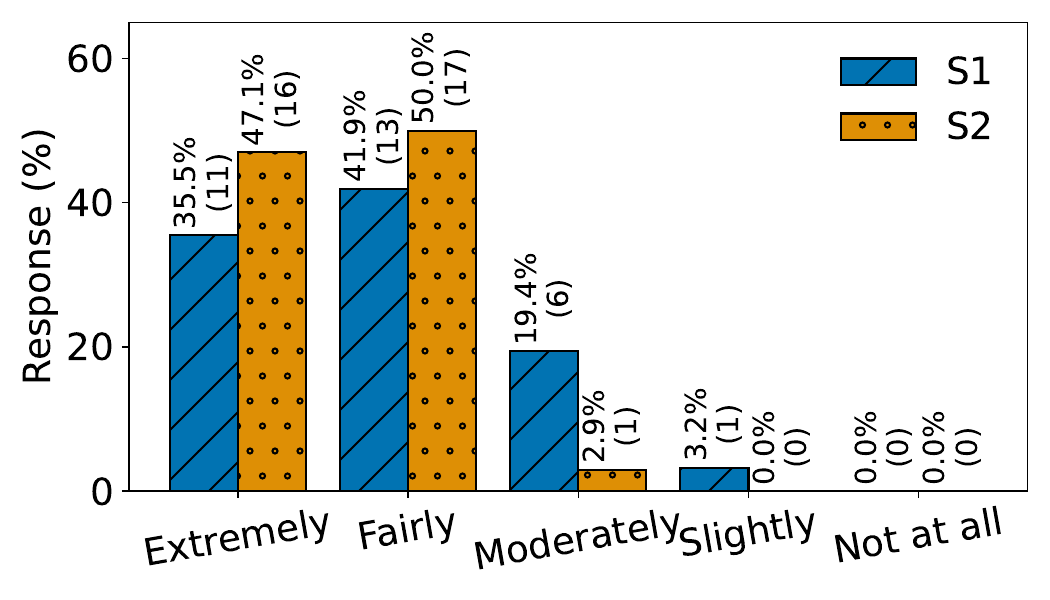}
		}
            \subfloat[Competence and immersion\label{fig:com_imm_mr_hm}]{
			\includegraphics[width=0.325\linewidth,keepaspectratio]{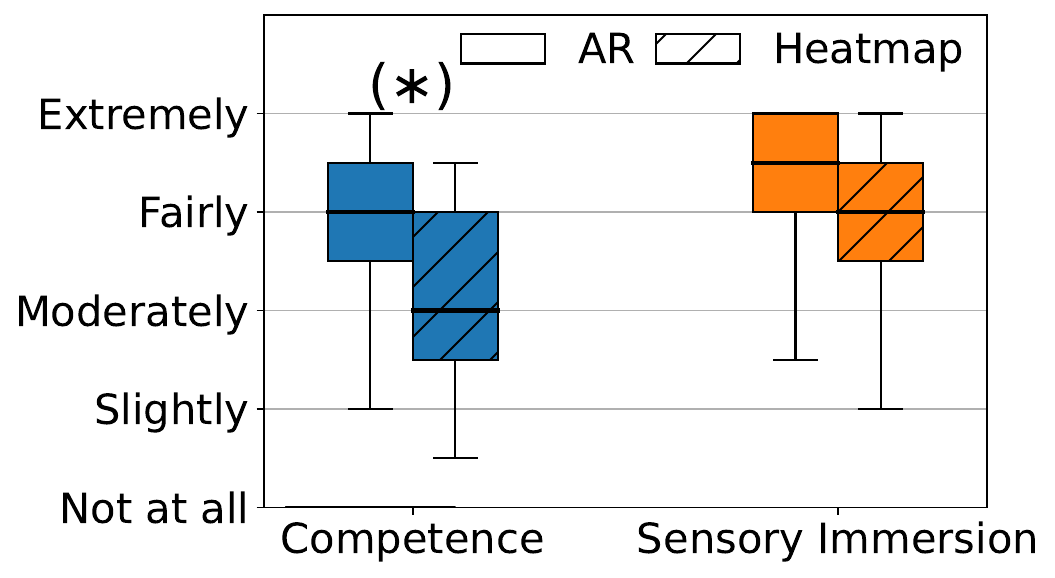}
		}
	\end{center}
	\caption{Participant's perception - (a) interest in the game story, (b) feeling of achievement and success after playing the game. In (c), competence and immersion in the AR app vs the baseline 2D spatial heatmap visualization. ($\ast$) indicates a statistically significant difference in competence among $20$ participants who attend all sessions.}
    \Description{Two grouped bar charts present participants' (a) interest, (b) perceived success after playing the AR game. Most report a stronger sense of success with AR bubbles than with the heatmap visualization. (c) Box plots that compare competence and immersion levels reported by participants after using the AR app and the 2D spatial heatmap visualization. Results indicate a statistically significant difference in competence due to the visualization technique for those attending all the sessions.}
	\label{fig:game_enjoyment_answers}
\end{figure*}

Competence represents how capable and skilled the player feels after completing the game. We found that in S1, most participants ($18$, $58\%$) showed a fair to an extreme level of competence. Whereas $10$ ($32.2\%$) showed moderate to fair and only three ($9.6\%$) showed slight to moderate competence levels during the gameplay. Similarly, in S2, $23$ ($67.6\%$) show fair to extreme, $9$ ($26.4\%$) show moderate to fair, and only two ($5.8\%$) participants show slight to moderate competence. Among the $30$ participants (P\textsubscript{01} to P\textsubscript{30}) who participated in both the semi-controlled sessions, $10$ ($33.3\%$) have improved competence from S1 to S2. While $18$ ($60\%$) participants show no significant change in their competence across the sessions, and only two ($6.6\%$) experience a slight reduction in their competence score. Therefore, with more practice sessions, one can improve their competence with the AR app to effectively reduce CO\textsubscript{2}. The participants show fair competence across the semi-controlled and in-the-wild experiments, as shown in \figurename~\ref{fig:com_imm_mr_hm} and \figurename~\ref{fig:pros_cons}. In comparison, participants show moderate competence with the baseline spatial 2D heatmap visualization. \figurename~\ref{fig:com_imm_mr_hm} indicates a statistically significant (using Welch’s t-test, t-statistic $2.64$, p$<.05$, medium-to-large effect with Cohen's $d=1.27$) difference in competence due to visualization technique among the $20$ participants who attended all AR and baseline 2D heatmap sessions.

\subsubsection{Sensory and Imaginative Immersion}
Immersion represents the degree of absorption in the game world. $26$ ($83.8\%$) participants in S1 and $28$ ($82.3\%$) participants in S2 experiences \textit{fair} to \textit{extreme} immersion. While immersion remains consistent across sessions for most of the participants, eight ($26.6\%$) participants experienced improved immersion levels from \textit{slight} to \textit{fair} or \textit{fair} to \textit{extreme}. Thus, for some participants, the game becomes more enjoyable with subsequent sessions. Overall, the participants show fair to extreme median immersion across the semi-controlled and in-the-wild experiments, as shown in \figurename~\ref{fig:com_imm_mr_hm}. In comparison, participants show fair median immersion with the baseline 2D heatmap, with no statistically significant difference due to the visualization technique.

\subsubsection{Participant Interest and Immediate Rewards}
During the sessions, participants must locate the CO\textsubscript{2} bubbles in an indoor space and ventilate them, thereby reducing the CO\textsubscript{2} concentration. As shown in \figurename~\ref{fig:interest}, $27$ ($87.1\%$) and $29$ ($85.3\%$) participants show \textit{extreme} and \textit{fair} interest for the two sessions, respectively. While four ($12.9\%$) participants show \textit{moderate} interest during S1. The participants who had already participated in S1 were more interested in S2. $20$ ($58.8\%$) were extremely, and $9$ ($26.5\%$) were \textit{fairly} interested in the game's story. Among the participants (P\textsubscript{31} to P\textsubscript{34}) who only participated in S2, two ($5.9\%$) show slight interest.

Almost all the participants in S2 felt successful in reducing the pollutants. As shown in \figurename~\ref{fig:success}, $16$ ($47\%$) felt \textit{extremely}, $17$ ($50\%$) felt \textit{fairly}, and only one ($2.9\%$) felt \textit{moderately} successful. However, in S1, $11$ ($35.5\%$) and $13$ ($41.9\%$) felt \textit{extreme} and \textit{fair} success, respectively, and $6$ ($19.4\%$) felt \textit{moderate} success. As mentioned earlier, a research associate demonstrated the user interface and functionality of the AR app to the participants in S1, as they were newly introduced to the AR app and the wearable device. In subsequent sessions, participants could use the AR app and reduce CO\textsubscript{2} bubbles without any tutorial, thereby improving their sense of achievement and success. Whereas, with the baseline 2D heatmap, only two ($10\%$) participants \textit{extremely}, $12$ ($60\%$) \textit{fairly}, and $6$ ($30\%$) \textit{slightly} reduce the CO\textsubscript{2} concentration within the session. The primary reason is a missing sense of urgency and actionable local pollution context. Thus, participants faced difficulties in planning and acting on one strategy, resulting in longer session durations and inadequate ventilation.

\begin{figure}
	\captionsetup[subfigure]{}
	\begin{center}
            \subfloat[Positive and negative metrics with AR vs heatmap\label{fig:reported_pros_cons}]{
			\includegraphics[width=0.8\linewidth,keepaspectratio]{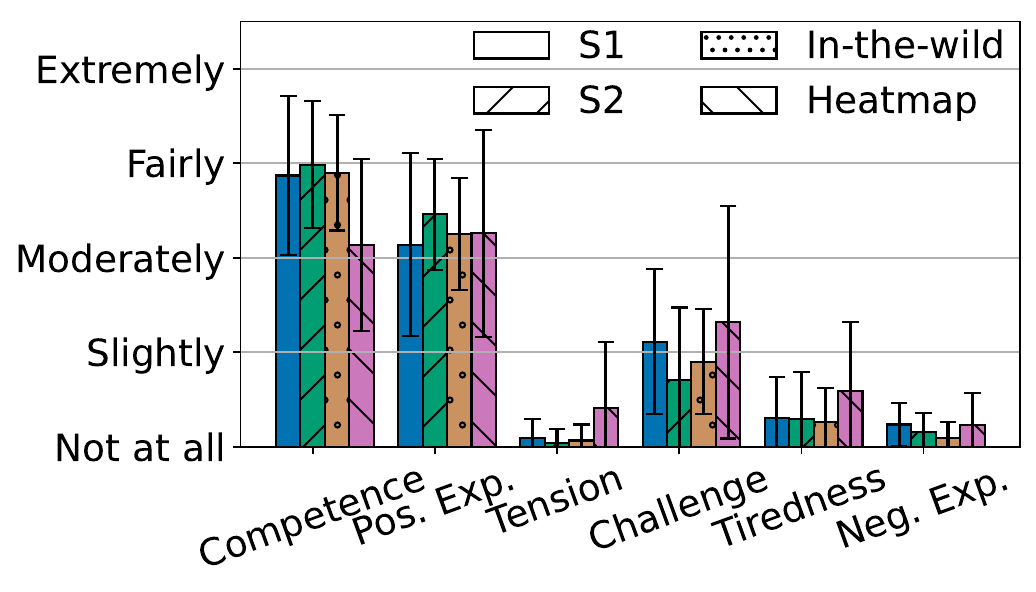}
		}\
            \subfloat[AR positive experiences with gaming habits\label{fig:game_freq_pos}]{
			\includegraphics[width=0.8\linewidth,keepaspectratio]{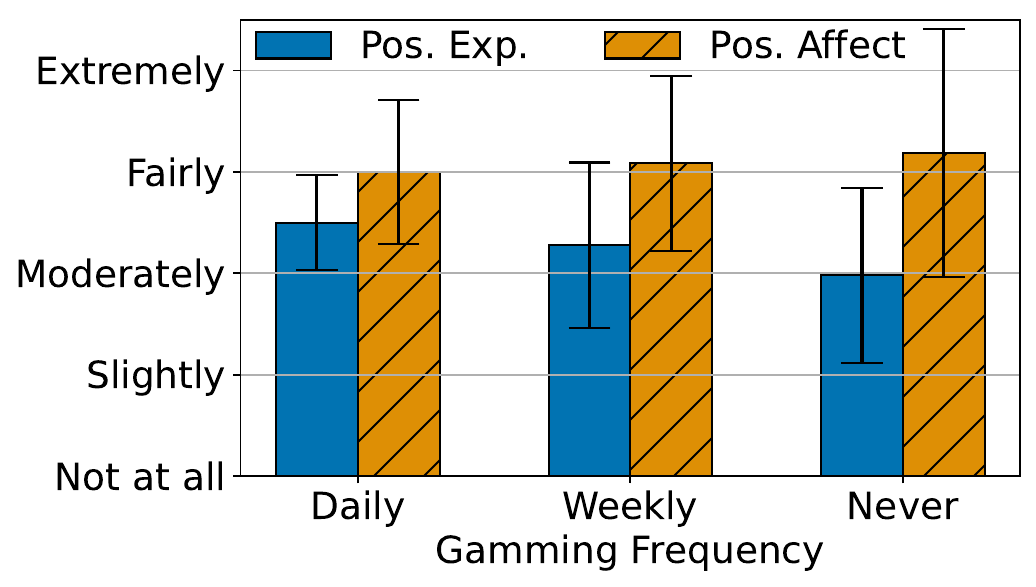}
		}
	\end{center}
	\caption{Game experience - (a) positive and negative metrics with the AR app vs the baseline heatmap, (b) AR positive experience with gaming habits of the participants.}
    \Description{Two bar charts for (a) positive and negative metrics with the AR application vs the baseline heatmap, (b) augmented reality positive experience with gaming habits of the participants.}
	\label{fig:pros_cons}
\end{figure}

\subsubsection{Positive and Negative Experience}
We observe that across the sessions S1, S2, and in-the-wild, the average competence with the game and positive experience of the participants remain fairly high with the AR app, as shown in \figurename~\ref{fig:pros_cons}. Whereas undesired experiences, such as in-game tension, tiredness, and negative experience after the gameplay session, were negligible. Moreover, the gameplay challenge reduces from S1 to S2, indicating an easy learning curve to use the AR app and the \ourmethod{} wearable for the participants. Thus, in S2, participants faced fewer challenges in finding the CO\textsubscript{2} bubbles. During the in-the-wild experiments, the gameplay challenge increases marginally due to dynamic pollution sources. However, the participants could find, ventilate, and reduce the CO\textsubscript{2} bubbles on their own (i.e., $45.4\%$ strongly agree, $33.3\%$ agree, and $6\%$ somewhat agree). Whereas, with the baseline 2D heatmap visualization, the gameplay challenges and tiredness increase, resulting in lower in-game competence and post-game tiredness.

Moreover, we also observe that participants gain consistent in-game short-term positive affect and long-term positive experience, irrespective of their gaming frequency. This indicates an easy-to-learn AR gaming experience for new users with little to no prior gaming expertise. \figurename~\ref{fig:game_freq_pos} shows that the participants who do not play mobile games gain a statistically similar long-term positive affect and long-term positive experience during the AR gameplay session as those who play games on a daily basis.

\begin{table*}
\centering
\scriptsize
\caption{Chi-square ($\chi^{2}$) statistic indicates strong association among GEQ metrics and participants' awareness, interest, and success in the AR app compared to baseline 2D heatmap visualization. Here, $\ast$ means p$<.05$, $\ast\ast$ means p$<.01$, and reports Cramer's $V$ with small ($0.17$), medium ($0.33$), and large ($0.54$) effect size thresholds for within-group AR studies~\cite{ortloff2025small}.}
\Description{This table presents chi-square statistics and Cramer's V effect size showing the association between game experience metrics (competence, immersion, challenge, positive affect, positive experience, tiredness) and participant outcomes (awareness, interest, and success) for both the AR application and the baseline 2D heatmap.}
\label{tab:chi2_geq}
\begin{tabular}{|l|cccc|cc|}
\hline
\multirow{2}{*}{\textbf{AR App}}  & \multicolumn{4}{c|}{\textbf{In-Game GEQ}}                                                              & \multicolumn{2}{c|}{\textbf{Post-Game GEQ}}      \\ 
\cline{2-7}
     & \textbf{Competence}       & \textbf{Immersion}        & \textbf{Challenge} & \textbf{Pos. Affect}      & \textbf{Pos. Exp}         & \textbf{Tiredness}  \\ 
\hline
Awareness           & 21.42$^\ast$ (0.33)              & \textbf{21.70}$^{\ast\ast} \textbf{(0.41)}$          & 7.30               & 17.19                     & 19.66                     & \textbf{17.26}$^\ast$ \textbf{(0.36)}       \\
Interest            & 20.88$^\ast$ (0.33)              & \textbf{63.17}$^{\ast\ast} \textbf{(0.7)}$ & 15.39              & 21.39$^\ast$ (0.33)             & 16.48                     & 11.84               \\
Success             & \textbf{85.13}$^{\ast\ast}$ \textbf{(0.66)} & 17.01$^{\ast\ast}$ (0.36)          & 18.11$^\ast$ (0.3)      & \textbf{82.33}$^{\ast\ast}$ \textbf{(0.65)} & \textbf{32.26}$^{\ast\ast}$ \textbf{(0.41)} & 3.05                \\ 
\hline
\multirow{2}{*}{\textbf{2D Heatmap}}   &                           &                           &                    &                           &                           &                     \\
 & \textbf{Competence}       & \textbf{Immersion}        & \textbf{Challenge} & \textbf{Pos. Affect}      & \textbf{Pos. Exp}         & \textbf{Tiredness}  \\ 
\hline
Awareness           & 12.83                     & 10.77                     & 20.62              & 9.17                      & 16.50                     & 4.81                \\
Interest            & \textbf{23.76}$^\ast$ \textbf{(0.85)}     & \textbf{24.20}$^{\ast\ast}$ \textbf{(0.86)} & 16.78              & 13.57                     & 20.49                     & 12.10               \\
Success             & 8.38                      & 7.25                      & 12.22              & 4.19                      & 11.17                     & 6.81                \\
\hline
\end{tabular}
\end{table*}

\subsection{Gameplay Experience and User Perceptions}
\tablename~\ref{tab:chi2_geq} compares the associations between different game experience metrics and the participants' awareness, interest, and success, using chi-square tests for both the AR app and the 2D heatmap visualization. In the AR app, competence and immersion are significantly related to \textit{awareness}, where immersion ($\chi^2=21.70$, p$<.01$) shows medium-to-large and competence ($\chi^2=21.42$, p$<.05$) shows medium association. Moreover, tiredness ($\chi^2=17.26$, p$<.05$) also shows medium-to-large association to awareness. Therefore, as the participants explore indoor space, they become more aware of the pollution dynamics. \textit{Interest} in the AR app shows large association with immersion ($\chi^2=63.17$, p$<.01$) along with medium association with competence ($\chi^2=20.88$, p$<.05$) and postitve affect ($\chi^2=21.39$, p$<.05$) while proactively ventilating CO\textsubscript{2} bubbles. \textit{Success} shows the large association with competence ($\chi^2=85.13$, p$<.01$) and positive affect ($\chi^2=82.33$, p$<.01$), which leads to medium-to-large effect on long-term positive experiences ($\chi^2=32.26$, p$<.01$), highlighting the effectiveness of the AR app in reducing CO\textsubscript{2}.

In contrast, the 2D heatmap exhibits weaker associations. \textit{Interest} is largely connected to both competence ($\chi^2=23.76$, p$<.05$) and immersion ($\chi^2=24.20$, p$<.01$). However, success displays insignificant associations across all game experience metrics, indicating that the 2D heatmap is less effective in enhancing users' feelings of achievement. These findings suggest that the AR app enhances game experience metrics like \textit{immersion} and \textit{competence}, nudging users' \textit{awareness} and \textit{interest} towards concrete actions, such as engaging with ventilation systems, windows, or directing airflow (validating \textbf{H2}, from awareness to actionability), to \textit{successfully} reduce indoor CO\textsubscript{2} bubbles.

\subsection{Post-Study System Usability of \ourmethod{}}
\begin{figure}
	\captionsetup[subfigure]{}
	\begin{center}
        \subfloat[Usability Questions\label{fig:pssuq_questions}]{
			\includegraphics[width=0.80\linewidth,keepaspectratio]{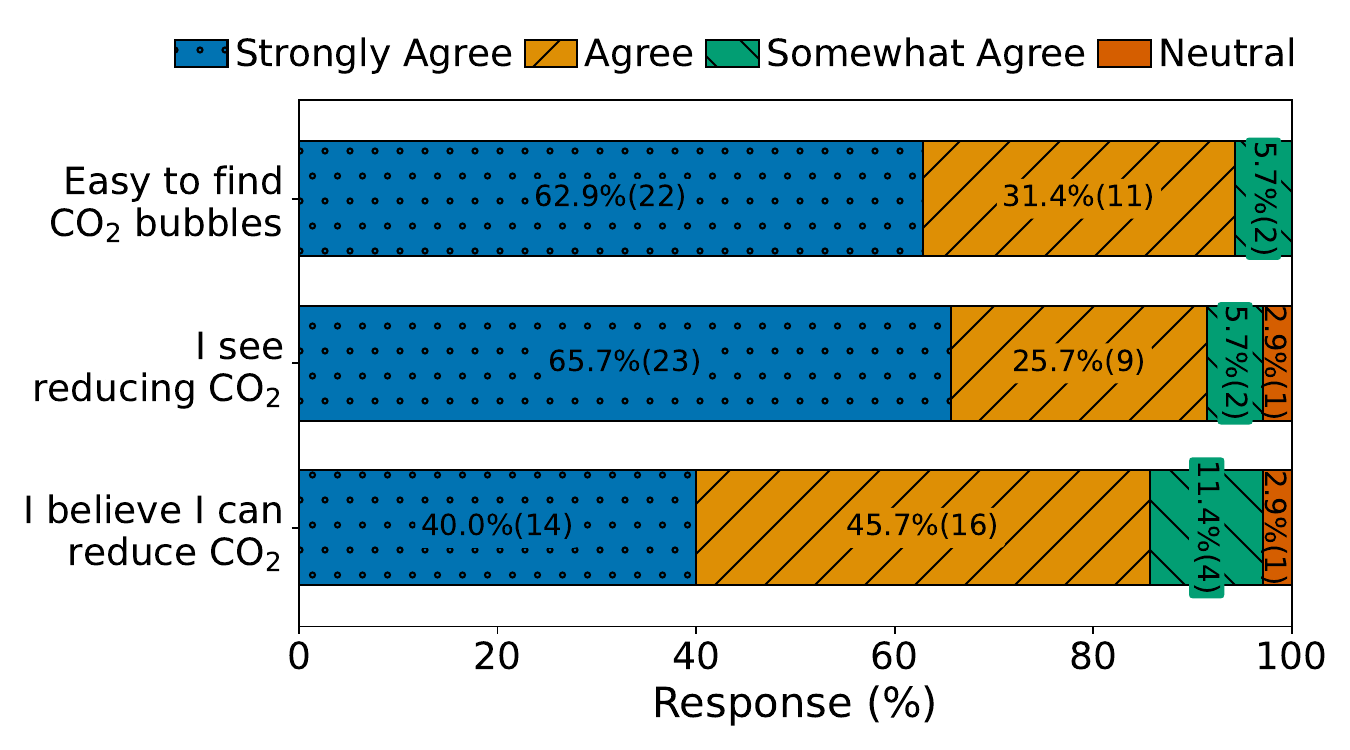}
		}\
            \subfloat[Usability Score\label{fig:pssuq_scores}]{
			\includegraphics[width=0.7\linewidth,keepaspectratio]{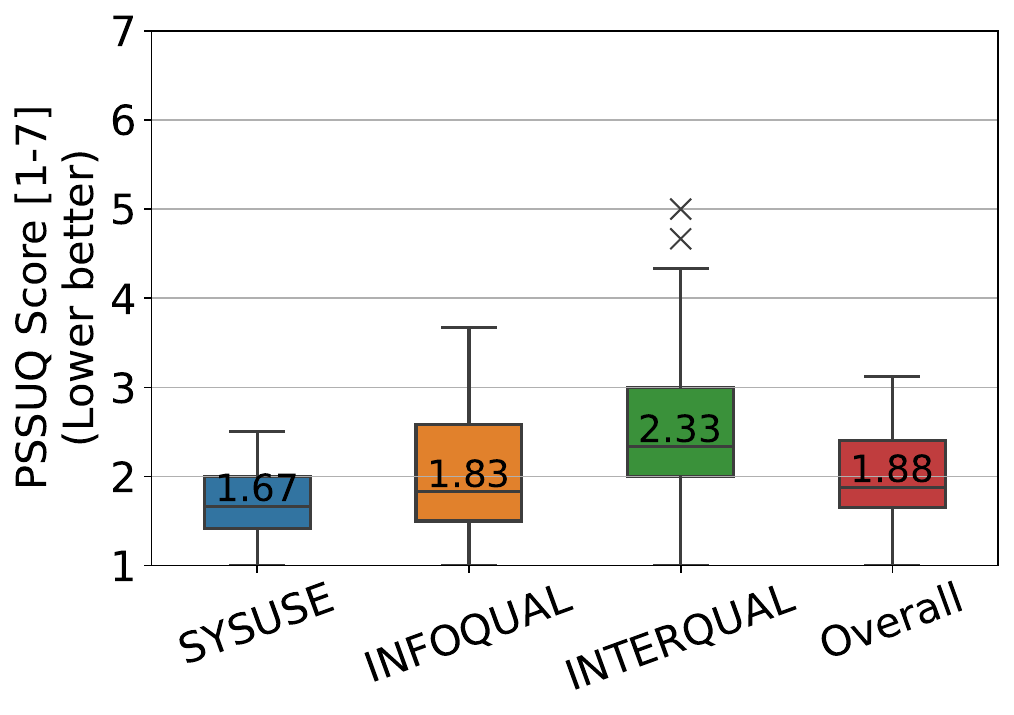}
		}
	\end{center}
	\caption{Post study system usability: (a) response to usability questions (b) Computed usability scores for semi-controlled and in-the-wild user experiments.}
    \Description{Post study system usability: (a) horizontally stacked bar chart shows participant confidence in three usability questions: (i) Easy to find CO2 bubbles, (ii) I see reducing CO2, (iii) I believe I can reduce CO2, and (b) box plots display usability scores for the system based on post-study system usability surveys.} 
	\label{fig:usibility}
\end{figure}

Here, we analyze the perceived satisfaction level of the participants and whether they were confident in ventilating accumulated CO\textsubscript{2} bubbles using the \ourmethod{} wrist-wearable and the AR app. As shown in \figurename~\ref{fig:pssuq_questions}, among all the participants across the sessions, $22$ ($62.9\%$) participants strongly agreed, $11$ ($31.4\%$) participants agreed, and the remaining ($5.7\%$) somewhat agreed that it was easy to find the CO\textsubscript{2} bubbles. They can clearly observe the gradual changes in CO\textsubscript{2} concentration. $23$ ($65.7\%$) strongly agree, $9$ ($25.7\%$) agree, and two ($5.7\%$) somewhat agree that they can see the reduction in CO\textsubscript{2} readings after they start the window ventilator, or open the window, etc. The participants were confident, $14$ ($40\%$) strongly believed, $16$ ($45.7\%$) believed, and $4$ ($11.4\%$) somewhat believed that they could reduce CO\textsubscript{2} on their own.

In \tablename~\ref{tab:pssuq_chi2}, we present the relationships between PSSUQ questions and user perception, like competence, immersion, awareness, and effective CO\textsubscript{2} reduction when using the AR app. For \textit{competence}, the most substantial associations are observed with comfort using the AR app (Q4, $\chi^2=22.50$, p$<.05$) and belief in reducing pollutants (Q6, $\chi^2=29.42$, p$<.05$). Notably, the satisfaction with the app's overall experience (Q16) also shows a large association ($\chi^2=27.66$, p$<.01$). These results imply that competence is significantly linked to ease of use and perceived effectiveness in CO\textsubscript{2} ventilation. \textit{Immersion} shows a significant association specifically with the simplicity of using the AR app (Q2, $\chi^2=18.42$, p$<.05$), highlighting that an easy-to-use interface enhances user immersion in the experience. \textit{Awareness} exhibits a strong link with the ease of finding CO\textsubscript{2} bubbles (Q10, $\chi^2=19.45$, p$<.01$). For CO\textsubscript{2} \textit{reduction}, significant associations are found with ease of learning (Q5, $\chi^2=6.34$, p$<.05$) and users' confidence with the AR app (Q6, $\chi^2=11.61$, p$<.01$). Additionally, error handling (Q7, $\chi^2=12.94$, p$<.05$) largely contributes to the perceived effectiveness in reducing CO\textsubscript{2}.

Finally, \figurename~\ref{fig:pssuq_scores} shows box plots of the PSSUQ (Post-Study System Usability Questionnaire) scores reported by the participants; note that a lower PSSUQ score means better. Median system usability (SYSUSE) is $1.67$, indicating the practicality of the AR visualization. Further, the information, such as the information sheet (i.e., shared during first-time participation) and AR app demonstration, benefited the participants. The information quality (INFOQUAL) is $1.83$. The interface quality (INTERQUAL) is $2.33$, indicating a reasonably good and responsive application interface. Overall, the wearable and the AR app achieve a median PSSUQ score of $1.88$, which is perceived as usable and compelling to the participants. In a nutshell, the AR app's design elements, such as ease of use, clear visual representation, and effective error handling, strongly influence user engagement, perceived competence, environmental awareness, and impact. However, there is room for further improvement as discussed in the following section. 

\begin{table*}
\centering
\scriptsize
\caption{Chi-square ($\chi^2$) statistic indicates strong association among system usability metrics and participant engagement factors. Here, $\ast$ means p$<.05$, $\ast\ast$ means p$<.01$, and reports Cramer's $V$ with small ($0.17$), medium ($0.33$), and large ($0.54$) effect size thresholds for within-group AR studies~\cite{ortloff2025small}. PSSUQ questionnaire is included in Appendix~\ref{appendix:pssuq}.}
\Description{This table presents chi-square statistics and Cramer's V effect size showing the association between usability metrics (system usefulness, information quality, interface quality) and user perceptions (competence, immersion, awareness, effective CO2 reduction). Each usability aspect is represented with detailed question-specific results (Q1–Q16) to identify which features of the AR app most influence successful user engagement and confidence.}
\label{tab:pssuq_chi2}
\setlength{\tabcolsep}{3pt}
\begin{tabular}{|p{1.7cm}|cccccc|cccccc|cccc|}
\hline
\multirow{2}{*}{\textbf{Factors}}          & \multicolumn{6}{c|}{\textbf{SYSUSE}}                                              & \multicolumn{6}{c|}{\textbf{INFOQUAL}}                                               & \multicolumn{4}{c|}{\textbf{INTERQUAL }}                    \\ 
\cline{2-17}
& \textbf{Q1} & \textbf{Q2} & \textbf{Q3} & \textbf{Q4} & \textbf{Q5} & \textbf{Q6} & \textbf{Q7} & \textbf{Q8} & \textbf{Q9} & \textbf{Q10} & \textbf{Q11} & \textbf{Q12} & \textbf{Q13} & \textbf{Q14} & \textbf{Q15} & \textbf{Q16}  \\ 
\hline
Competence                                                                  & 10.45       & 10.85       & 18.32       & \begin{tabular}{@{}c@{}} 22.50$^{\ast}$ \\ (0.6) \end{tabular}     & 16.60       & \begin{tabular}{@{}c@{}} \textbf{29.42}$^{\ast}$ \\ \textbf{(0.56)}\end{tabular}      & 20.07       & 34.32       & 17.98       & 16.06        & 23.05        & 31.00        & \begin{tabular}{@{}c@{}} \textbf{42.50}$^{\ast}$ \\ \textbf{(0.59)} \end{tabular}      & 43.44        & 21.90        & \begin{tabular}{@{}c@{}} 27.66$^{\ast\ast}$ \\ (0.67) \end{tabular}       \\
Immersion                                                                   & 8.63        & \begin{tabular}{@{}c@{}} \textbf{18.42}$^{\ast}$ \\ \textbf{(0.55)}\end{tabular} & 12.42       & 14.39       & 3.99        & 8.80        & 28.10       & 28.16       & 28.80       & 9.12         & 22.65        & 21.87        & 20.59        & 18.08        & 19.75        & 11.91         \\
Awareness                                                                   & 2.24        & 5.24        & 4.93        & 5.52        & 5.62        & 9.80        & 11.61       & 17.40       & 8.51        & \begin{tabular}{@{}c@{}} \textbf{19.45}$^{\ast\ast}$ \\ \textbf{(0.56)}\end{tabular}      & 8.61         & 11.74        & 9.50         & 7.08         & 4.15         & 2.97          \\
CO\textsubscript{2}$\downarrow$($>$100ppm/m) & 0.89        & 0.12        & 2.22        & 0.67        & \begin{tabular}{@{}c@{}}6.34$^{\ast}$ \\ (0.45)\end{tabular}       & \begin{tabular}{@{}c@{}}\textbf{11.61}$^{\ast\ast}$ \\ \textbf{(0.61)}\end{tabular}     & \begin{tabular}{@{}c@{}} \textbf{12.94}$^{\ast}$ \\ \textbf{(0.65)}\end{tabular}      & 10.39       & 2.47        & 0.96         & 6.52         & 6.91         & 3.00         & 4.17         & 1.37         & 0.42          \\
\hline
\end{tabular}
\end{table*}

\subsection{Qualitative User Feedback}
\label{sec:part_comments}

As mentioned earlier in section~\ref{sec:study}, we moderated three-member focus group discussions over the following topics: (i) their overall experience with the AR game, (ii) how the in-game experience impacted their perception of air pollution, and (iii) the suggestions to improve the platform. See Appendix~\ref{appendix:interview} for topic details. \tablename~\ref{tab:sentiment} presents the summary of sentiments over different aspects of the study. We next discuss the core observations from the discussions.

\subsubsection{Experience with the AR Application} 
The participants were overall satisfied with the AR app. Notably, using an AR app and playing an AR game was a first-time experience for many of our participants, and the quantitative feedback says that they enjoyed the application. During the focus group discussions, almost $75\%$ of participant comments were positive (i.e., keywords such as good, smooth, interesting, and great) as shown in \tablename~\ref{tab:sentiment}.

\noindent\textit{EX\#1: ``The best part was that it was very interactive and creative. It's like a camera we have for the surrounding CO\textsubscript{2} level.''}

Notably, the participants were excited about observing a visual representation of their personal pollution exposure through the AR bubbles, and they were enthusiastic about exploring the CO\textsubscript{2} concentration at various corners of the rooms. In section~\ref{sec:awareness}, more than $80\%$ of participants agreed on improved overall understanding of indoor pollutants. During the focus group discussions, almost $85\%$ of participants shared strong positive remarks about increased awareness. Participants were surprised by how quickly indoor CO\textsubscript{2} can reach unhealthy levels. Moreover, participants found the AR app engaging, and almost $40\%$ of them wanted a longer gameplay session.

\noindent\textit{EX\#2: ``Of course, because I haven't seen any numerical or quantitative value of pollution before that. So, I can see those pollutants, not with my naked eyes, but with the AR app. That was very cool.''}

\begin{table*}[]
\centering
\scriptsize
\caption{Summary of sentiments in the semi-structured interviews for different aspects of the study.}
\Description{This summary table presents feedback sentiment from post-study interviews, categorizing participant comments into different aspects such as overall experience, awareness/learning, AR visualization, responsiveness, interface usability, and game duration. For each aspect, the proportion of positive and negative/mixed sentiment is shown, accompanied by concise feedback summaries.}
\label{tab:sentiment}
\begin{tabular}{|l|c|p{7cm}|}
\hline
\textbf{Aspect} & \textbf{Sentiment ($+$\%)} & \textbf{Feedback Summary} \\
\hline
Overall Experience & Positive (75\%) & Majority of comments were positive about the experience (``good," ``smooth," ``interesting", ``great"), with some negatives about sensing delay or learning curve. \\ \hline
Awareness/Learning & Positive (85\%) & Strong positive remarks about increased awareness, surprise at indoor CO\textsubscript{2} levels. \\ \hline
AR Visualization & Positive (70\%) & AR bubbles are intuitive and color/size changes are helpful, but some concerns about clutter and bubble size filling the smartphone screen. \\ \hline
Responsiveness & Mixed/Negative (40\%) & Mixed sentiment with negative feedback about sensing delay of approx. 30 sec to adjust to CO\textsubscript{2} concentration at a new location, desire for real-time change. \\ \hline
Interface Usability & Mixed/Negative (45\%) & Positive about interface clarity, Non-intuitive for new participants, requests for tutorials and start-up instructions to reduce the learning curve. \\ \hline
Game Duration & Mixed (60\%) & Some participants wanted longer gameplay sessions, but overall found it engaging. \\
\hline
\end{tabular}
\end{table*}

\subsubsection{Bubble Visualization and User Interface of the Application}
\label{sec:bubble_viz_interview}
Participants have focused on the dynamic size and color of the AR bubbles for real-time monitoring of CO\textsubscript{2}, which has turned out to be an exciting feature for them. They felt that the AR bubbles helped them interact more effectively with the CO\textsubscript{2} bubbles around them, and they were motivated to trigger directed airflow to reduce the CO\textsubscript{2} bubbles in the AR app. Almost $70\%$ of participant comments about the AR visualization were positive during the focus group discussions, as shown in \tablename~\ref{tab:sentiment}. We observed a mixed response (i.e., $45\%$ positive) for the user interface. While participants were positive about the interface clarity, they also raised concerns about the app's non-intuitive nature at the beginning and suggested the inclusion of startup tutorials.

\noindent\textit{BV\#1: ``In responsiveness and clarity, it is very good. In addition to the clarity, whenever I place a bubble, the size of the bubble increases, and then it shows how much PPM. It was also increasing. So, it's clear that you're searching for CO\textsubscript{2}.''}

\noindent\textit{BV\#2: ``The bubbles were decreasing when we were opening the window or AC or switching on the fan. Well made.''}

Moreover, $60\%$ of participants also felt that a higher responsiveness of the AR bubbles would have been better for triggering quicker actions. Notably, when the participant moves away from an AR bubble and comes back later on to check, the \ourmethod{} wrist-wearable takes around $30$ seconds to adjust to the CO\textsubscript{2} concentration at that location and update the bubble size and color in the app.

\subsubsection{Suggestions and Feedback from the Participants}
We received constructive feedback from the participants during the focus group discussions. First, a few participants felt that the first-time tutorial instructions could be directly embedded in the app, covering some basic ideas about pollutants and their equivalent representations in AR bubbles. Second, rather than creating multiple AR bubbles at the same location with multiple taps and subsequently cluttering the app interface, the participants suggested merging such repeated bubbles to represent the CO\textsubscript{2} concentration around that location. Finally, a few participants indicated having a lifetime of AR bubbles and representing it using opaqueness, which adds another dimension to the bubbles' trustfulness. For instance, a bubble placed long back might not represent the actual CO\textsubscript{2} concentration at that location, so over time, it may get opaque to indicate the trustfulness of the data.

\section{Discussion}
\label{sec:discuss}
In this section, we discuss some important takeaways based on our experience on the overall development of \ourmethod{}.

\subsection{Advantages over the Baseline 2D Heatmap Visualization}
Traditional 2D heatmaps visualize CO\textsubscript{2} concentration within the spatial dimension of the indoor space, which limits user engagement and poses a critical problem of sensor placement. CO\textsubscript{2} bubbles are formed at source heights and get trapped in different parts of the indoor space (see pilot experiment in section~\ref{sec:pilot_lab}). Hence, the optimal placement of static sensors is very challenging, and we must follow a dense deployment strategy, incurring high financial costs for accurate ambient monitoring. In contrast, the proposed AR-based approach uses low-cost \ourmethod{} wrist-wearable and provides clearer 3D awareness by visualizing CO\textsubscript{2} concentrations at varying heights, including bubbles that form near occupants' breathing zones. As users can physically move around to inspect real-time changes, they gain more dynamic insight than static heatmaps allow. Moreover, we observe that AR bubbles heightened the sense of presence and urgency (see section~\ref{sec:awareness}), which leads to quicker responses, such as opening windows or turning on fans. 

However, such wearable-based ambient sensing relies on the user to continuously move around the indoor space for the trustworthiness of the data. Thus, both low-cost wearable and static pollution sensors must function jointly as a trustworthy data source for long-term human-centered applications, covering larger indoor spaces than the sum of their components to enable continuous ambient sensing and immersive AR visualization~\cite{vatavu2022ambient}. We left this for future work.

\subsection{Trade-off Between Ventilation, Thermal Comfort, and Energy}
While \ourmethod{}’s AR visualization encourages users to ventilate when CO\textsubscript{2} levels rise, in practice, such decisions are in direct conflict with thermal comfort and energy consumption. Studies~\cite{karmakar2024exploring,zhong2020hilo} have consistently found that occupants avoid opening windows in cold climates because doing so introduces thermal discomfort, even when air quality is poor. This is a recurring theme in real-world settings, where occupants explicitly prioritize warmth over ventilation in winter. In shared environments like classrooms~\cite{gao2020n} and offices~\cite{zhong2021complexity,SHEIKHKHAN2021111363}, these tensions are amplified as one person’s decision to ventilate can negatively impact others’ comfort, leading to disagreements and complaints about cold air or drafts. Reluctance to ventilation is further linked to increased heating demand to maintain comfortable indoor temperatures, significantly raising energy costs~\cite{hussain2023indoor}. Acknowledging this difficult trade-off between ventilation, thermal comfort, and energy costs clarifies why visualizing pollutants alone may not always lead to ventilation actions by occupants in real-world environments. In future versions of \ourmethod{}, we plan to integrate temperature data to offer more context‑aware recommendations, helping users navigate these competing priorities rather than treating ventilation as the only response to indoor pollutants.

\subsection{Practical Limitations of the Current Setup}
We observed that participants appreciated the concept of using AR visualization to provide real-time insights into air pollutants and using game-based prevention to reduce pollution exposure. At the same time, they have also complained about the system's responsiveness (approx. $30$ sec waiting time for getting the updated CO\textsubscript{2} values), the excessive density of the spheres (ability to place multiple spheres at a close location), and supporting on-app tutorials for new users. While some of them can be fixed easily from the software side, improving the system's responsiveness is more of a hardware issue that needs careful consideration of sensor price and accuracy trade-offs, which is left for future work.

\subsection{Extending the \ourmethod{} Platform}
The \ourmethod{} platform can be extended beyond CO\textsubscript{2} to support other indoor pollutants, such as VOCs and PM\textsubscript{2.5}. Doing so would require integrating additional sensors into the wearable, developing visualization strategies that capture each pollutant’s unique dispersion patterns (e.g., the wave-like spread of VOCs~\cite{hekiem2021advanced}), and designing pollutant-specific interaction models (e.g., ceiling fans can disperse gases but may pull PM\textsubscript{2.5}~\cite{liu2021residential,karmakar2024exploring}). The platform can also be expanded into a multiplayer system, where readings from multiple wearables are combined to create richer, shared views of exposure. Such collaborative interactions, like one participant directing airflow with a pedestal fan while another operates window ventilation, can improve indoor air quality while fostering cooperation, social engagement, and environmental awareness~\cite{mittmann2022lina}.

\section{Conclusion}
Illustrating indoor CO\textsubscript{2} levels within physical spaces is instrumental in raising awareness of associated health risks. We introduced an augmented reality (AR) experience that combines real-time CO\textsubscript{2} monitoring through a wrist-worn \textit{CoWear} sensor with immersive, game-like interactions, with the goal of encouraging healthier indoor habits. From a study involving $35$ participants, our findings revealed an increased engagement and a better understanding of indoor pollutants. Moreover, with a median usability score of 1.88, the system underscores both its user-friendliness and its potential to be applied in real-world settings for promoting improved indoor air quality.
\section*{Ethical Considerations}
\label{ethics}
The institute's ethical review committee has approved this study (Order No: IIT/SRIC/DEAN/2023, dated July 31, 2023). Moreover, we have made significant efforts to anonymize the participants to preserve privacy while providing the necessary details on the study methodology. All participants signed forms consenting to the use of collected pollutant measurements and video, audio, and physiological measurements for non-commercial research purposes.
\begin{acks}
The authors acknowledge the use of OpenAI's ChatGPT\footnote{\url{https://chatgpt.com/}} to assist with grammar correction and stylistic refinement. No sections of the manuscript were wholly generated by AI tools. All content was critically reviewed and verified by the authors to ensure that it accurately represents the research findings and scholarly intent. The authors would like to thank the anonymous reviewers for the constructive comments, which have helped to improve the overall presentation of the paper. The research of the first author is supported by the Prime Minister's Research Fellowship (PMRF) in India through grant number IIT/Acad/PMRF/SPRING/2022-23, dated 24 March 2023. The work is also supported by AI4ICPS IIT Kharagpur Technology Innovation Hub, under award number TRP3RD3223323, dated 22 February 2024.
\end{acks}

\bibliographystyle{ACM-Reference-Format}
\bibliography{reference}

@String{Computing = "Computing" }

@String{Computer = "{IEEE} Computer" }

@String{Springer = "Springer-Verlag" }

@inproceedings{ortloff2025small,
  title={Small, Medium, Large? A Meta-Study of Effect Sizes at CHI to Aid Interpretation of Effect Sizes and Power Calculation},
  author={Ortloff, Anna-Marie and Martius, Florin and Meier, Mischa and Raimbault, Theo and Geierhaas, Lisa and Smith, Matthew},
  booktitle={Proceedings of the 2025 CHI Conference on Human Factors in Computing Systems},
  pages={1--28},
  year={2025}
}

@inproceedings{hussain2023indoor,
  title={The indoor Air quality trilemma: Improving Air quality, using less energy, and meeting stakeholder requirements},
  author={Hussain, Iman and Friday, Adrian and Booker, Douglas},
  booktitle={Extended Abstracts of the 2023 CHI Conference on Human Factors in Computing Systems},
  pages={1--6},
  year={2023}
}

@article{gao2020n,
  title={n-gage: Predicting in-class emotional, behavioural and cognitive engagement in the wild},
  author={Gao, Nan and Shao, Wei and Rahaman, Mohammad Saiedur and Salim, Flora D},
  journal={Proceedings of the ACM on Interactive, Mobile, Wearable and Ubiquitous Technologies},
  volume={4},
  number={3},
  pages={1--26},
  year={2020},
  publisher={ACM New York, NY, USA}
}

@article{SHEIKHKHAN2021111363,
title = {Application of an occupant voting system for continuous occupant feedback on thermal and indoor air quality – Case studies in office spaces},
journal = {Energy and Buildings},
volume = {251},
pages = {111363},
year = {2021},
issn = {0378-7788},
doi = {https://doi.org/10.1016/j.enbuild.2021.111363},
url = {https://www.sciencedirect.com/science/article/pii/S0378778821006472},
author = {Donya {Sheikh Khan} and Jakub Kolarik and Peter Weitzmann}
}

@misc{epa_indoor,
	author = {EPA},
	title = {{The Inside Story: A Guide to Indoor Air Quality}},
	howpublished = "\url{https://www.epa.gov/indoor-air-quality-iaq/inside-story-guide-indoor-air-quality}",
	year = {2024}, 
}

@article{patel2018physiology,
  title={{Physiology, carbon dioxide retention}},
  author={Patel, Shivani and Miao, Julia H and Yetiskul, Ekrem and Anokhin, Anya and Majmundar, Sapan H},
  year={2018}
}

@inproceedings{vatavu2022ambient,
  title={Are ambient intelligence and augmented reality two sides of the same coin? Implications for human-computer interaction},
  author={Vatavu, Radu-Daniel},
  booktitle={CHI Conference on Human Factors in Computing Systems Extended Abstracts},
  pages={1--8},
  year={2022}
}

@misc{SENSIRION_co2_sensor,
	author = {SENSIRION},
	title = {{SCD41, Improved CO2 accuracy with extended measurement range and single-shot mode}},
	howpublished = "\url{https://sensirion.com/products/catalog/SCD41}",
	year = {2022}, 
}

@misc{whoindoor,
	author = {WHO},
	title = {{Household air pollution}},
	howpublished = "\url{https://www.who.int/news-room/fact-sheets/detail/household-air-pollution-and-health}",
	year = {2022}, 
}

@book{sauro2016quantifying,
  title={Quantifying the user experience: Practical statistics for user research},
  author={Sauro, Jeff and Lewis, James R},
  year={2016},
  publisher={Morgan Kaufmann}
}

@article{ijsselsteijn2013game,
  title={The game experience questionnaire},
  author={IJsselsteijn, Wijnand A and De Kort, Yvonne AW and Poels, Karolien},
  year={2013},
  publisher={Technische Universiteit Eindhoven}
}

@article{maag2018w,
  title={W-air: Enabling personal air pollution monitoring on wearables},
  author={Maag, Balz and Zhou, Zimu and Thiele, Lothar},
  journal={Proceedings of the ACM on Interactive, Mobile, Wearable and Ubiquitous Technologies},
  volume={2},
  number={1},
  pages={1--25},
  year={2018},
  publisher={ACM New York, NY, USA}
}

@article{moore2018managing,
  title={Managing in-home environments through sensing, annotating, and visualizing air quality data},
  author={Moore, Jimmy and Goffin, Pascal and Meyer, Miriah and Lundrigan, Philip and Patwari, Neal and Sward, Katherine and Wiese, Jason},
  journal={Proceedings of the ACM on interactive, mobile, wearable and ubiquitous technologies},
  volume={2},
  number={3},
  pages={1--28},
  year={2018},
  publisher={ACM New York, NY, USA}
}

@inproceedings{jiang2011maqs,
  title={MAQS: a personalized mobile sensing system for indoor air quality monitoring},
  author={Jiang, Yifei and Li, Kun and Tian, Lei and Piedrahita, Ricardo and Yun, Xiang and Mansata, Omkar and Lv, Qin and Dick, Robert P and Hannigan, Michael and Shang, Li},
  booktitle={Proceedings of the 13th international conference on Ubiquitous computing},
  pages={271--280},
  year={2011}
}

@inproceedings{fang2016airsense,
  title={AirSense: an intelligent home-based sensing system for indoor air quality analytics},
  author={Fang, Biyi and Xu, Qiumin and Park, Taiwoo and Zhang, Mi},
  booktitle={Proceedings of the 2016 ACM International joint conference on pervasive and ubiquitous computing},
  pages={109--119},
  year={2016}
}

@inproceedings{kim2020awareness,
  title={Awareness, understanding, and action: a conceptual framework of user experiences and expectations about indoor air quality visualizations},
  author={Kim, Sunyoung and Li, Muyang},
  booktitle={Proceedings of the 2020 CHI Conference on Human Factors in Computing Systems},
  pages={1--12},
  year={2020}
}

@inproceedings{margariti2024evaluating,
  title={Evaluating ActuAir: Building Occupants' Experiences of a Shape-Changing Air Quality Display},
  author={Margariti, Eleni and Vlachokyriakos, Vasilis and Durrant, Abigail C and Kirk, David},
  booktitle={Proceedings of the CHI Conference on Human Factors in Computing Systems},
  pages={1--21},
  year={2024}
}

@inproceedings{chae2022virtual,
  title={Virtual air conditioner’s airflow simulation and visualization in ar},
  author={Chae, Joohwan and Kim, Donghan and Jeong, Wooseok and Jo, Eunchan and Jeong, Won-Ki and Choi, JunYoung and Kim, Seung-wook and Kim, Myoung Gon and Lee, Jae-Won and Lee, Hyechan and others},
  booktitle={Proceedings of the 28th ACM Symposium on Virtual Reality Software and Technology},
  pages={1--11},
  year={2022}
}

@inproceedings{bauer2023improving,
  title={Improving Environmental Knowledge with a Serious Game: An experimental study},
  author={Bauer, Mathias and Wei{\ss}, Malte},
  booktitle={Extended Abstracts of the 2023 CHI Conference on Human Factors in Computing Systems},
  pages={1--10},
  year={2023}
}

@inproceedings{albar2024playful,
  title={A Playful Path to Sustainability: Synthesizing design strategies for children's environmental sustainability learning through gameful interventions},
  author={Albar, Raghad and Gauthier, Andrea and Vasalou, Asimina},
  booktitle={Proceedings of the 23rd Annual ACM Interaction Design and Children Conference},
  pages={201--217},
  year={2024}
}

@inproceedings{ishoj2021axo,
  title={AXO: a video game that encourages recycling to preteens},
  author={Ishoj-Paris, Yohanna and Gravel-Villeneuve, Ariane and Vermette-David, Francis and Dixon-Sequeira, Lissa and Dicaire, Maxime and Morin-Grandmont, Maxime and Jomphe, Naomi and Fournier, Nathaelle and Champeau-Fournier, Ophelie and Par{\'e}, Cassandra and others},
  booktitle={Extended Abstracts of the 2021 Annual Symposium on Computer-Human Interaction in Play},
  pages={350--355},
  year={2021}
}

@inproceedings{prophet2018cultivating,
  title={Cultivating environmental awareness: Modeling air quality data via augmented reality miniature trees},
  author={Prophet, Jane and Kow, Yong Ming and Hurry, Mark},
  booktitle={Augmented Cognition: Intelligent Technologies: 12th International Conference, AC 2018, Held as Part of HCI International 2018, Las Vegas, NV, USA, July 15-20, 2018, Proceedings, Part I},
  pages={406--424},
  year={2018},
  organization={Springer}
}

@inproceedings{lindrup2023carbon,
  title={Carbon Scales: Collective sense-making of carbon emissions from food production through physical data representation},
  author={Lindrup, Martin Valdemar Anker and Menon, Arjun Rajendran and Bi{\o}rn-Hansen, Aksel},
  booktitle={Proceedings of the 2023 ACM Designing Interactive Systems Conference},
  pages={1515--1530},
  year={2023}
}

@article{teles2020game,
  title={Game-like 3d visualisation of air quality data},
  author={Teles, Bruno and Mariano, Pedro and Santana, Pedro},
  journal={Multimodal Technologies and Interaction},
  volume={4},
  number={3},
  pages={54},
  year={2020},
  publisher={MDPI}
}

@article{li2023visual,
  title={Visual analysis of air pollution spatio-temporal patterns},
  author={Li, Jiayang and Bi, Chongke},
  journal={The Visual Computer},
  volume={39},
  number={8},
  pages={3715--3726},
  year={2023},
  publisher={Springer}
}

@article{relvas2024serious,
  title={A serious game for raising air pollution perception in children},
  author={Relvas, Tiago and Mariano, Pedro and Almeida, Susana Marta and Santana, Pedro},
  journal={Journal of Computers in Education},
  pages={1--31},
  year={2024},
  publisher={Springer}
}

@article{chen2019visualization,
  title={Visualization of real-time monitoring datagraphic of urban environmental quality},
  author={Chen, Pengyu},
  journal={Eurasip Journal on Image and Video Processing},
  volume={2019},
  number={1},
  pages={42},
  year={2019},
  publisher={Springer}
}

@inproceedings{jin2024exploring,
  title={Exploring the Opportunity of Augmented Reality (AR) in Supporting Older Adults to Explore and Learn Smartphone Applications},
  author={Jin, Xiaofu and Tong, Wai and Wei, Xiaoying and Wang, Xian and Kuang, Emily and Mo, Xiaoyu and Qu, Huamin and Fan, Mingming},
  booktitle={Proceedings of the CHI Conference on Human Factors in Computing Systems},
  pages={1--18},
  year={2024}
}

@article{feldpausch2013adventures,
  title={The Adventures of Carbon Bond: Using a melodramatic game to explain CCS as a mitigation strategy for climate change},
  author={Feldpausch-Parker, Andrea M and O'Byrne, Megan and Endres, Danielle and Peterson, Tarla R},
  journal={Greenhouse Gases: Science and Technology},
  volume={3},
  number={1},
  pages={21--29},
  year={2013},
  publisher={Wiley Online Library}
}

@article{ghahramani2019personal,
  title={Personal CO2 bubble: Context-dependent variations and wearable sensors usability},
  author={Ghahramani, Ali and Pantelic, Jovan and Vannucci, Matthew and Pistore, Lorenza and Liu, Shichao and Gilligan, Brian and Alyasin, Soheila and Arens, Edward and Kampshire, Kevin and Sternberg, Esther},
  journal={Journal of Building Engineering},
  volume={22},
  pages={295--304},
  year={2019},
  publisher={Elsevier}
}

@article{azuma2018effects,
  title={Effects of low-level inhalation exposure to carbon dioxide in indoor environments: A short review on human health and psychomotor performance},
  author={Azuma, Kenichi and Kagi, Naoki and Yanagi, U and Osawa, Haruki},
  journal={Environment international},
  volume={121},
  pages={51--56},
  year={2018},
  publisher={Elsevier}
}

@article{satish2012co2,
  title={Is CO2 an indoor pollutant? Direct effects of low-to-moderate CO2 concentrations on human decision-making performance},
  author={Satish, Usha and Mendell, Mark J and Shekhar, Krishnamurthy and Hotchi, Toshifumi and Sullivan, Douglas and Streufert, Siegfried and Fisk, William J},
  journal={Environmental health perspectives},
  volume={120},
  number={12},
  pages={1671--1677},
  year={2012},
  publisher={National Institute of Environmental Health Sciences}
}

@article{allen2016associations,
  title={Associations of cognitive function scores with carbon dioxide, ventilation, and volatile organic compound exposures in office workers: a controlled exposure study of green and conventional office environments},
  author={Allen, Joseph G and MacNaughton, Piers and Satish, Usha and Santanam, Suresh and Vallarino, Jose and Spengler, John D},
  journal={Environmental health perspectives},
  volume={124},
  number={6},
  pages={805--812},
  year={2016},
  publisher={National Institute of Environmental Health Sciences}
}

@article{karmakar2024exploring,
  title={Exploring Indoor Air Quality Dynamics in Developing Nations: A Perspective from India},
  author={Karmakar, Prasenjit and Pradhan, Swadhin and Chakraborty, Sandip},
  journal={ACM Journal on Computing and Sustainable Societies},
  year={2024},
  publisher={ACM New York, NY}
}

@article{gall2016real,
  title={Real-time monitoring of personal exposures to carbon dioxide},
  author={Gall, Elliott T and Cheung, Toby and Luhung, Irvan and Schiavon, Stefano and Nazaroff, William W},
  journal={Building and Environment},
  volume={104},
  pages={59--67},
  year={2016},
  publisher={Elsevier}
}

@article{strom2016effects,
  title={The effects of bedroom air quality on sleep and next-day performance},
  author={Str{\o}m-Tejsen, Peter and Zukowska, D and Wargocki, Pawel and Wyon, David Peter},
  journal={Indoor air},
  volume={26},
  number={5},
  pages={679--686},
  year={2016},
  publisher={Wiley Online Library}
}

@article{morales2021air,
  title={The air that we breathe: neutral and volatile {PFAS} in indoor air},
  author={Morales-McDevitt, Maya E and Becanova, Jitka and Blum, Arlene and Bruton, Thomas A and Vojta, Simon and Woodward, Melissa and Lohmann, Rainer},
  journal={Environmental science \& technology letters},
  volume={8},
  number={10},
  pages={897--902},
  year={2021},
  publisher={ACS Publications}
}

@article{ramalho2015association,
  title={Association of carbon dioxide with indoor air pollutants and exceedance of health guideline values},
  author={Ramalho, Olivier and Wyart, Guillaume and Mandin, Corinne and Blondeau, Patrice and Cabanes, Pierre-Andr{\'e} and Leclerc, Nathalie and Mullot, Jean-Ulrich and Boulanger, Guillaume and Redaelli, Matteo},
  journal={Building and Environment},
  volume={93},
  pages={115--124},
  year={2015},
  publisher={Elsevier}
}

@article{tsai2012office,
  title={Office workers’ sick building syndrome and indoor carbon dioxide concentrations},
  author={Tsai, Dai-Hua and Lin, Jia-Shiang and Chan, Chang-Chuan},
  journal={Journal of occupational and environmental hygiene},
  volume={9},
  number={5},
  pages={345--351},
  year={2012},
  publisher={Taylor \& Francis}
}

@inproceedings{zhong2021complexity,
  title={The Complexity of Indoor Air Quality Forecasting and the Simplicity of Interacting with It -- A Case Study of 1007 Office Meetings},
  author={Zhong, Sailin and Lalanne, Denis and Alavi, Hamed},
  booktitle={Proceedings of the 2021 CHI Conference on Human Factors in Computing Systems},
  pages={1--19},
  year={2021}
}

@inproceedings{snow2019performance,
  title={Performance by design: supporting decisions around indoor air quality in offices},
  author={Snow, Stephen and Oakley, Michael and Schraefel, MC},
  booktitle={Proceedings of the 2019 on Designing Interactive Systems Conference},
  pages={99--111},
  year={2019}
}

@article{enriquez2014quasi,
  title={The quasi-static growth of {CO2} bubbles},
  author={Enr{\'\i}quez, Oscar R and Sun, Chao and Lohse, Detlef and Prosperetti, Andrea and Van Der Meer, Devaraj},
  journal={Journal of fluid mechanics},
  volume={741},
  pages={R1},
  year={2014},
  publisher={Cambridge University Press}
}

@article{pantelic2020personal,
  title={Personal {CO2} cloud: laboratory measurements of metabolic {CO2} inhalation zone concentration and dispersion in a typical office desk setting},
  author={Pantelic, Jovan and Liu, Shichao and Pistore, Lorenza and Licina, Dusan and Vannucci, Matthew and Sadrizadeh, Sasan and Ghahramani, Ali and Gilligan, Brian and Sternberg, Esther and Kampschroer, Kevin and others},
  journal={Journal of exposure science \& environmental epidemiology},
  volume={30},
  number={2},
  pages={328--337},
  year={2020},
  publisher={Nature Publishing Group US New York}
}

@article{masic2020evaluation,
  title={Evaluation of optical particulate matter sensors under realistic conditions of strong and mild urban pollution},
  author={Masic, Adnan and Bibic, Dzevad and Pikula, Boran and Blazevic, Almir and Huremovic, Jasna and Zero, Sabina},
  journal={Atmospheric measurement techniques},
  volume={13},
  number={12},
  pages={6427--6443},
  year={2020},
  publisher={Copernicus GmbH}
}

@article{zheng2018field,
  title={Field evaluation of low-cost particulate matter sensors in high-and low-concentration environments},
  author={Zheng, Tongshu and Bergin, Michael H and Johnson, Karoline K and Tripathi, Sachchida N and Shirodkar, Shilpa and Landis, Matthew S and Sutaria, Ronak and Carlson, David E},
  journal={Atmospheric Measurement Techniques},
  volume={11},
  number={8},
  pages={4823--4846},
  year={2018}
}

@article{lee2022online,
  title={An online interactive dashboard to explore personal exposure to air pollution},
  author={Lee, Won Do and Schulte, Kayla and Schwanen, Tim},
  journal={Findings},
  volume={2022},
  year={2022},
  publisher={Findings Press}
}

@inproceedings{hsu2017community,
  title={Community-empowered air quality monitoring system},
  author={Hsu, Yen-Chia and Dille, Paul and Cross, Jennifer and Dias, Beatrice and Sargent, Randy and Nourbakhsh, Illah},
  booktitle={Proceedings of the 2017 CHI Conference on human factors in computing systems},
  pages={1607--1619},
  year={2017}
}

@inproceedings{schurholz2019myaqi,
  title={Myaqi: Context-aware outdoor air pollution monitoring system},
  author={Sch{\"u}rholz, Daniel and Nurgazy, Meruyert and Zaslavsky, Arkady and Jayaraman, Prem Prakash and Kubler, Sylvain and Mitra, Karan and Saguna, Saguna},
  booktitle={Proceedings of the 9th International Conference on the Internet of Things},
  pages={1--8},
  year={2019}
}

@inproceedings{mathews2021air,
  title={Air: An augmented reality application for visualizing air pollution},
  author={Mathews, Noble Saji and Chimalakonda, Sridhar and Jain, Suresh},
  booktitle={2021 IEEE Visualization Conference (VIS)},
  pages={146--150},
  year={2021},
  organization={IEEE}
}

@inproceedings{ceccarini2022escapecampus,
  title={{EscapeCampus}: exploiting a Game-based Learning tool to increase the sustainability knowledge of students},
  author={Ceccarini, Chiara and Prandi, Catia},
  booktitle={Proceedings of the 2022 ACM Conference on Information Technology for Social Good},
  pages={390--396},
  year={2022}
}

@inproceedings{zhong2020hilo,
  title={Hilo-wear: exploring wearable interaction with indoor air quality forecast},
  author={Zhong, Sailin and Alavi, Hamed S and Lalanne, Denis},
  booktitle={Extended Abstracts of the 2020 CHI Conference on Human Factors in Computing Systems},
  pages={1--8},
  year={2020}
}

@article{hekiem2021advanced,
  title={Advanced vapour sensing materials: Existing and latent to acoustic wave sensors for {VOC}s detection as the potential exhaled breath biomarkers for lung cancer},
  author={Hekiem, Nurul Liyana Lukman and Ralib, Aliza Aini Md and Ahmad, Farah B and Nordin, Anis Nurashikin and Ab Rahim, Rosminazuin and Za’bah, Nor Farahidah and others},
  journal={Sensors and Actuators A: Physical},
  volume={329},
  pages={112792},
  year={2021},
  publisher={Elsevier}
}

@article{liu2021residential,
  title={Residential building ventilation in situations with outdoor {PM2.5} pollution},
  author={Liu, Sumei and Song, Rui and Zhang, Tengfei Tim},
  journal={Building and Environment},
  volume={202},
  pages={108040},
  year={2021},
  publisher={Elsevier}
}

@inproceedings{cosio2023virtual,
  title={Virtual and augmented reality for environmental sustainability: A systematic review},
  author={Cosio, Laura D and Buruk, O{\u{g}}uz'Oz' and Fern{\'a}ndez Galeote, Daniel and Bosman, Isak De Villiers and Hamari, Juho},
  booktitle={Proceedings of the 2023 CHI conference on human factors in computing systems},
  pages={1--23},
  year={2023}
}

@article{cao2023mobile,
  title={Mobile augmented reality: User interfaces, frameworks, and intelligence},
  author={Cao, Jacky and Lam, Kit-Yung and Lee, Lik-Hang and Liu, Xiaoli and Hui, Pan and Su, Xiang},
  journal={ACM Computing Surveys},
  volume={55},
  number={9},
  pages={1--36},
  year={2023},
  publisher={ACM New York, NY}
}

@inproceedings{du2020depthlab,
  title={{DepthLab}: Real-time {3D} interaction with depth maps for mobile augmented reality},
  author={Du, Ruofei and Turner, Eric and Dzitsiuk, Maksym and Prasso, Luca and Duarte, Ivo and Dourgarian, Jason and Afonso, Joao and Pascoal, Jose and Gladstone, Josh and Cruces, Nuno and others},
  booktitle={Proceedings of the 33rd Annual ACM Symposium on User Interface Software and Technology},
  pages={829--843},
  year={2020}
}

@inproceedings{an2021arshoe,
  title={{ARShoe}: Real-time augmented reality shoe try-on system on smartphones},
  author={An, Shan and Che, Guangfu and Guo, Jinghao and Zhu, Haogang and Ye, Junjie and Zhou, Fangru and Zhu, Zhaoqi and Wei, Dong and Liu, Aishan and Zhang, Wei},
  booktitle={Proceedings of the 29th ACM International Conference on Multimedia},
  pages={1111--1119},
  year={2021}
}

@article{ahn2015supporting,
  title={Supporting healthy grocery shopping via mobile augmented reality},
  author={Ahn, Junho and Williamson, James and Gartrell, Mike and Han, Richard and Lv, Qin and Mishra, Shivakant},
  journal={ACM Transactions on Multimedia Computing, Communications, and Applications (TOMM)},
  volume={12},
  number={1s},
  pages={1--24},
  year={2015},
  publisher={ACM New York, NY, USA}
}

@article{clark2022articulate,
  title={{ARticulate}: one-shot interactions with intelligent assistants in unfamiliar smart spaces using augmented reality},
  author={Clark, Meghan and Newman, Mark W and Dutta, Prabal},
  journal={Proceedings of the ACM on Interactive, Mobile, Wearable and Ubiquitous Technologies},
  volume={6},
  number={1},
  pages={1--24},
  year={2022},
  publisher={ACM New York, NY, USA}
}

@article{mittmann2022lina,
  title={{LINA}-a social augmented reality game around mental health, supporting real-world connection and sense of belonging for early adolescents},
  author={Mittmann, Gloria and Barnard, Adam and Krammer, Ina and Martins, Diogo and Dias, Jo{\~a}o},
  journal={Proceedings of the ACM on Human-Computer Interaction},
  volume={6},
  number={CHI PLAY},
  pages={1--21},
  year={2022},
  publisher={ACM New York, NY, USA}
}

@inproceedings{prandi2019augmenting,
  title={Augmenting good behaviour: Mixing digital and reality to promote sustainability in a campus community},
  author={Prandi, Catia and Ceccarini, Chiara and Salomoni, Paola},
  booktitle={Proceedings of the 5th EAI International Conference on Smart Objects and Technologies for Social Good},
  pages={189--194},
  year={2019}
}

@inproceedings{silva2022understanding,
  title={Understanding {AR} activism: An interview study with creators of augmented reality experiences for social change},
  author={Silva, Rafael ML and Principe Cruz, Erica and Rosner, Daniela K and Kelly, Dayton and Monroy-Hern{\'a}ndez, Andr{\'e}s and Liu, Fannie},
  booktitle={Proceedings of the 2022 CHI Conference on Human Factors in Computing Systems},
  pages={1--15},
  year={2022}
}

@article{schaper2022addressing,
  title={Addressing waste separation with a persuasive augmented reality app},
  author={Schaper, Philipp and Riedmann, Anna and Oberd{\"o}rfer, Sebastian and Kr{\"a}he, Maileen and Lugrin, Birgit},
  journal={Proceedings of the ACM on Human-Computer Interaction},
  volume={6},
  number={MHCI},
  pages={1--16},
  year={2022},
  publisher={ACM New York, NY, USA}
}

@article{assor2024augmented,
  title={Augmented reality waste accumulation visualizations},
  author={Assor, Ambre and Prouzeau, Arnaud and Dragicevic, Pierre and Hachet, Martin},
  journal={ACM Journal on Computing and Sustainable Societies},
  volume={2},
  number={2},
  pages={1--29},
  year={2024},
  publisher={ACM New York, NY}
}

@inproceedings{sassmannshausen2021citizen,
  title={Citizen-centered design in urban planning: How augmented reality can be used in citizen participation processes},
  author={Sa{\ss}mannshausen, Sheree May and Radtke, J{\"o}rg and Bohn, Nino and Hussein, Hassan and Randall, Dave and Pipek, Volkmar},
  booktitle={Proceedings of the 2021 ACM Designing Interactive Systems Conference},
  pages={250--265},
  year={2021}
}

@inproceedings{alvarez2020store,
  title={In-store augmented reality-enabled product comparison and recommendation},
  author={{\'A}lvarez M{\'a}rquez, Jes{\'u}s Omar and Ziegler, J{\"u}rgen},
  booktitle={Proceedings of the 14th ACM Conference on Recommender Systems},
  pages={180--189},
  year={2020}
}

@inproceedings{roddiger2018armart,
  title={{ARMart}: {AR}-based shopping assistant to choose and find store items},
  author={R{\"o}ddiger, Tobias and Doerner, Dominik and Beigl, Michael},
  booktitle={Proceedings of the 2018 ACM international joint conference and 2018 international symposium on pervasive and ubiquitous computing and wearable computers},
  pages={440--443},
  year={2018}
}

@inproceedings{katsiokalis2023gonature,
  title={{GoNature AR}: Air Quality \& Noise Visualization Through a Multimodal and Interactive Augmented Reality Experience},
  author={Katsiokalis, Minas and Tsekeri, Elisavet and Lilli, Aikaterini and Gobakis, Konstantinos and Kolokotsa, Dionysia and Mania, Katerina},
  booktitle={Proceedings of the 2023 ACM International Conference on Interactive Media Experiences},
  pages={366--369},
  year={2023}
}

@inproceedings{krings2020development,
  title={Development framework for context-aware augmented reality applications},
  author={Krings, Sarah and Yigitbas, Enes and Jovanovikj, Ivan and Sauer, Stefan and Engels, Gregor},
  booktitle={Companion Proceedings of the 12th ACM SIGCHI Symposium on Engineering Interactive Computing Systems},
  pages={1--6},
  year={2020}
}

@article{hsu2020smell,
  title={Smell {Pittsburgh}: engaging community citizen science for air quality},
  author={Hsu, Yen-Chia and Cross, Jennifer and Dille, Paul and Tasota, Michael and Dias, Beatrice and Sargent, Randy and Huang, Ting-Hao and Nourbakhsh, Illah},
  journal={ACM Transactions on Interactive Intelligent Systems (TiiS)},
  volume={10},
  number={4},
  pages={1--49},
  year={2020},
  publisher={ACM New York, NY, USA}
}

@inproceedings{gupta2022human,
  title={The Human-Air Interface: Responding To Poor Air Quality Through Lived Experience and Digital Information},
  author={Gupta, Meghna and Eden, Grace},
  booktitle={Proceedings of the 2022 ACM Designing Interactive Systems Conference},
  pages={1085--1098},
  year={2022}
}

@inproceedings{liu2020making,
  title={Making air quality data meaningful: Coupling objective measurement with subjective experience through narration},
  author={Liu, Szu-Yu and Cranshaw, Justin and Roseway, Asta},
  booktitle={Proceedings of the 2020 ACM designing interactive systems conference},
  pages={1313--1326},
  year={2020}
}

@article{liu2018third,
  title={Third-eye: A mobilephone-enabled crowdsensing system for air quality monitoring},
  author={Liu, Liang and Liu, Wu and Zheng, Yu and Ma, Huadong and Zhang, Cheng},
  journal={Proceedings of the ACM on Interactive, Mobile, Wearable and Ubiquitous Technologies},
  volume={2},
  number={1},
  pages={1--26},
  year={2018},
  publisher={ACM New York, NY, USA}
}

@article{allen2015cognitive,
  title={Associations of cognitive function scores with carbon dioxide, ventilation, and volatile organic compound exposures in office workers: a controlled exposure study of green and conventional office environments},
  author={Allen, Joseph G and MacNaughton, Piers and Satish, Usha and Santanam, Suresh and Vallarino, Jose and Spengler, John D},
  journal={Environmental health perspectives},
  volume={124},
  number={6},
  pages={805--812},
  year={2016},
  publisher={National Institute of Environmental Health Sciences}
}

@article{bowen2020review,
  title={Indoor {CO2} concentrations and cognitive function: A critical review},
  author={Du, Bowen and Tandoc, Marlie C and Mack, Michael L and Siegel, Jeffrey A},
  journal={Indoor air},
  volume={30},
  number={6},
  pages={1067--1082},
  year={2020},
  publisher={Wiley Online Library}
}

@article{fan2023short,
  title={Short-term exposure to indoor carbon dioxide and cognitive task performance: A systematic review and meta-analysis},
  author={Fan, Yuejie and Cao, Xiaodong and Zhang, Jie and Lai, Dayi and Pang, Liping},
  journal={Building and Environment},
  volume={237},
  pages={110331},
  year={2023},
  publisher={Elsevier}
}

@article{csoma2022hypercapnia,
  title={Hypercapnia in {COPD}: causes, consequences, and therapy},
  author={Csoma, Bal{\'a}zs and Vulpi, Maria Rosaria and Dragonieri, Silvano and Bentley, Andrew and Felton, Timothy and L{\'a}z{\'a}r, Zs{\'o}fia and Bikov, Andras},
  journal={Journal of Clinical Medicine},
  volume={11},
  number={11},
  pages={3180},
  year={2022}
}

@article{wei2018aecopd,
  title={The features of {AECOPD} with carbon dioxide retention},
  author={Wei, Xia and Yu, Nan and Ding, Qi and Ren, Jingting and Mi, Jiuyun and Bai, Lu and Li, Jianying and Qi, Min and Guo, Youmin},
  journal={BMC Pulmonary Medicine},
  volume={18},
  number={1},
  pages={124},
  year={2018}
}

@article{permentier2017carbon,
  title={Carbon dioxide poisoning: a literature review of an often forgotten cause of intoxication in the emergency department},
  author={Permentier, Kris and Vercammen, Steven and Soetaert, Sylvia and Schellemans, Christian},
  journal={International journal of emergency medicine},
  volume={10},
  number={1},
  pages={14},
  year={2017},
  publisher={Springer}
}

\appendix
\section{Appendix}
\label{sec:appendix}

\subsection{Pre Study Survey Questionnaire}
\label{appendix:pre_study}

\subsubsection{Demographics}
\begin{enumerate}
    \item What is your age?
    \item What is your gender? \\
    $\circ$ Male $\circ$ Female $\circ$ Non-binary $\circ$ Prefer not to answer
    \item How would you describe your hometown? \\
    $\circ$ Rural $\circ$ Suburban $\circ$ Urban
    \item Do you live outside of your hometown for work? \\
    $\circ$ Yes $\circ$ No
    \item In which city are you currently residing? 
    \item What is the highest educational level you have completed? \\
    $\circ$ Primary $\circ$ Secondary $\circ$ Tertiary (College / University) $\circ$ Postgraduate (Master / PHD)
    \item Do you have any children? \\
    $\circ$ Yes $\circ$ No
    \item How many members are there in your family?
    \item Do you live with your family? \\
    $\circ$ Yes $\circ$ No
    \item What is your employment status? \\
    $\circ$ Employed $\circ$ Retired $\circ$ Housewife $\circ$ Student
    \item What is your household's monthly income (people living together as a family and sharing finances)?
    
    $\circ$ <250 USD $\circ$ 250 - 475 USD $\circ$ 475 -750 USD $\circ$ 750-950 USD $\circ$ >950 USD $\circ$ Prefer not to answer
    \item How would you describe your living place? \\
    $\circ$ One storey House $\circ$ Two storey House $\circ$ Flat $\circ$ 1BHK Apartment $\circ$ Hostel Room
    \item Do you / your family members have any respiratory disease/health condition that is caused by poor air quality? \\
    $\circ$ Yes $\circ$ No $\circ$ If yes, please specify the health condition.
    
\end{enumerate}

\subsubsection{Awareness on Indoor Pollution}
Responses are selected from options: $\circ$ TRUE $\circ$ FALSE $\circ$ N/A.

\begin{enumerate}
    \item[Q1] One third of the global population is affected by harmful household air pollutants
    \item[Q2] Approximately 11\% of lung cancer deaths in adults are attributable to exposure to carcinogens from household air pollution
    \item[Q3] Permissible carbon dioxide concentration in indoor spaces is 400 ppm for long-term
    \item[Q4] The UK government has pledged to implement new standards, guidelines, and regulations that will require all newly constructed homes from 2025 onward to generate 75-80\% fewer carbon emissions
    \item[Q5] Road accidents cause more deaths than respiratory diseases in your country
    \item[Q6] National Green Tribunal (NGT) recommended the government to mandate monitoring and reporting of Indoor Air Quality in all public buildings
    \item[Q7] Indoor Air Quality regulations are strictly followed in some states of your country
\end{enumerate}

\subsubsection{Understanding of Indoor Pollutants}
\begin{enumerate}
    \item Have you ever heard of sick building syndrome? \\
        $\circ$ Yes $\circ$ No 
    \item Have you ever taken any measurements of air pollutants in your home or office? \\
        $\circ$ Yes $\circ$ No $\circ$ If yes, please specify the health condition.
    \item What do you think are the possible pollution sources in your household? (e.g., Gas stove, Fridge, Incense sticks, etc.)
    \item When do you think your house is more polluted? \\
        $\circ$ Morning (06:00-12:00) $\circ$ Afternoon (12:00-18:00) $\circ$ Evening (18:00-00:00) $\circ$ Night (00:00-06:00)
    \item What do you think are the possible health impacts of air pollutants (e.g., Dizziness, irritation of eyes, etc)
    \item According to you, order the following pollutants with respect to harmfulness (i.e., 1 - least harmful, 5 - most harmful)
    \begin{itemize}
        \item Small dust particles (PM2.5) $\circ$ 1 $\circ$ 2 $\circ$ 3 $\circ$ 4 $\circ$ 5
        \item Carbon dioxides (CO2) $\circ$ 1 $\circ$ 2 $\circ$ 3 $\circ$ 4 $\circ$ 5
        \item Ethanol (C2H5OH) $\circ$ 1 $\circ$ 2 $\circ$ 3 $\circ$ 4 $\circ$ 5
        \item Volatile Organic Compounds (VOC) $\circ$ 1 $\circ$ 2 $\circ$ 3 $\circ$ 4 $\circ$ 5
        \item Nitrogen dioxide (NO2)$\circ$ 1 $\circ$ 2 $\circ$ 3 $\circ$ 4 $\circ$ 5
    \end{itemize}

    \item What would you do in the following scenarios? Select from the options: $\circ$ Exhaust fan on $\circ$ Ceiling fan on $\circ$ Window ventilator on $\circ$ Split AC on $\circ$ Open window $\circ$ Clean the area.
    \begin{itemize}
        \item Kitchen is full of smoke
        \item Food leftover in dining from yesterday
        \item Sweeping dusts in bedroom
        \item Family gathering
        \item Smoking in room
    \end{itemize}
\end{enumerate}

\subsubsection{Perception on Pollution and Countermeasures}
Responses are selected from options: $\circ$ Strongly Disagree  $\circ$ Disagree  $\circ$ Neutral  $\circ$ Agree $\circ$ Strongly Agree.

\begin{enumerate}
    \item[Q1] Exhaust fan helps ventilating the pollutants in kitchen
    \item[Q2] Pollutants are not affected by ceiling fans
    \item[Q3] Sometimes carbon dioxide concentration in the Bedroom is more compared to the Kitchen
    \item[Q4] Split AC ventilates carbon dioxide and other pollutants from the room
    \item[Q5] Indoor gathering increases the carbon dioxide concentration of the room
    \item[Q6] Anyone can feel when pollutants are accumulating in their home
    \item[Q7] Carbon dioxide lowers our ability to concentrate on a task
    \item[Q8] There is less awareness about indoor air pollution among the general public
    \item[Q9] Indoor is more polluted than outdoor
    \item[Q10] You can reduce air pollutants if you can see where they are concentrated in your room
\end{enumerate}

\subsection{Post Session Survey Questionnaire}
\label{appendix:post_study}
GEQ scores are computed as the average value of their items.
For in-game module -- Competence: Q2 and Q9, Sensory and Imaginative Immersion: Q1 and Q4, Flow: Q5 and Q10, Tension: Q6 and Q8, Challenge: Q12 and Q13, Negative affect: Q3 and Q7, Positive affect: Q11 and Q14. For post-game module -- Positive Experience: Q1, Q5, Q7, Q8, Q12, Q16, Negative Experience: Q2, Q4, Q6, Q11, Q14, Q15, Tiredness: Q10, Q13, Returning to Reality: Q3, Q9, Q17. Responses are selected from options: $\circ$ Not at all  $\circ$ Slightly  $\circ$ Moderately  $\circ$ Fairly $\circ$ Extremely.

\subsubsection{In-game Experience Questionnaire}

\begin{enumerate}
    \item[Q1] I was interested in the game's story
    \item[Q2] I felt successful
    \item[Q3] I felt bored
    \item[Q4] I found it impressive
    \item[Q5] I forgot everything around me
    \item[Q6] I felt frustrated
    \item[Q7] I found it tiresome
    \item[Q8] I felt irritable
    \item[Q9] I felt skilful
    \item[Q10] I felt completely absorbed
    \item[Q11] I felt content
    \item[Q12] I felt challenged
    \item[Q13] I had to put a lot of effort into it
    \item[Q14] I felt good
\end{enumerate}

\subsubsection{Post-game Experience Questionnaire}

\begin{enumerate}
    \item[Q1] I felt revived
    \item[Q2] I felt bad
    \item[Q3] I found it hard to get back to reality
    \item[Q4] I felt guilty
    \item[Q5] It felt like a victory
    \item[Q6] I found it a waste of time
    \item[Q7] I felt energised
    \item[Q8] I felt satisfied
    \item[Q9] I felt disoriented
    \item[Q10] I felt exhausted
    \item[Q11] I felt that I could have done more useful things
    \item[Q12] I felt powerful
    \item[Q13] I felt weary
    \item[Q14] I felt regret
    \item[Q15] I felt ashamed
    \item[Q16] I felt proud
    \item[Q17] I had a sense that I had returned from a journey
\end{enumerate}

\subsubsection{Post-experiment Feedback}
\begin{enumerate}
    \item Did your understanding of indoor air pollution improve after playing this game? \\
    $\circ$ Not at all $\circ$ Slightly $\circ$ Moderately $\circ$ Fairly $\circ$ Extremely
    
    \item How often do you play mobile or AR\//VR games? \\
    $\circ$ Never $\circ$ Daily $\circ$ Weekly $\circ$ Monthly $\circ$ Rarely

    \item I feel more familiar with the AR application in this session.\\
    $\circ$ Strongly Disagree $\circ$ Disagree $\circ$ Somewhat Disagree $\circ$ Neutral $\circ$ Somewhat Agree $\circ$ Agree $\circ$ Strongly Agree

    \item I want to use this app in my home to understand where the CO2 bubbles are.\\
    $\circ$ Strongly Disagree $\circ$ Disagree $\circ$ Somewhat Disagree $\circ$ Neutral $\circ$ Somewhat Agree $\circ$ Agree $\circ$ Strongly Agree
    
    \item Would you recommend this game to others in your friend circle?
    $\circ$ Yes $\circ$ No

    \item I want other pollutants (i.e, small dust particles, ethanol, etc.) to be integrated with this app.\\
    $\circ$ Strongly Disagree $\circ$ Disagree $\circ$ Somewhat Disagree $\circ$ Neutral $\circ$ Somewhat Agree $\circ$ Agree $\circ$ Strongly Agree
        
    \item Which game features would you like to see in the future\\
    $\circ$ More players in multiplayer mode $\circ$ AR headset integration $\circ$ AI recommendations to help reduce pollutants
    
    \item Any other suggestions to improve this Game
    
\end{enumerate}

\subsection{Post-Study System Usability Survey}
\label{appendix:pssuq}
PSSUQ consists of four usability scores. The scores are computed as the average value of their items -- Overall: Q1 to Q16, System Usefulness (SYSUSE): Q1 to Q6, Information Quality (INFOQUAL): Q7 to Q12, Interface Quality (INTERQUAL): Q13 to Q16. Responses are selected from a 7-point Likert scale options: $\circ$ Strongly Disagree $\circ$ Disagree $\circ$ Somewhat Disagree $\circ$ Neutral $\circ$ Somewhat Agree $\circ$ Agree $\circ$ Strongly Agree.

\begin{enumerate}
    \item[Q1] Overall, I am satisfied with how easy it is to play this game
    \item[Q2] It was simple to use AR app
    \item[Q3] I was able to see the reducing CO2 concentration using the AR app
    \item[Q4] I felt comfortable using this AR app
    \item[Q5] It was easy to learn to use this AR app
    \item[Q6] I believe I could reduce pollutants surrounding me using this AR app
    \item[Q7] If there is any technical error during my session, the app gave error messages that clearly told me how to fix problems
    \item[Q8] Whenever I made a mistake using the AR app, I could recover easily and quickly. (Wrong tagging the equipment, multiple bubbles)
    \item[Q9] The information (such as information sheet, instructions by RA) provided with this AR app was clear
    \item[Q10] It was easy to find the CO2 concentration
    \item[Q11] The information sheet was effective in helping me complete the tasks and scenarios
    \item[Q12] The organisation of information on the app screen was clear
    \item[Q13] The interface of this app was pleasant
    \item[Q14] I liked using the interface of this app
    \item[Q15] This app has all the functions and capabilities I expect it to have
    \item[Q16] Overall, I am satisfied with this app
\end{enumerate}

\subsection{Semi-structured Interview Topics}
\label{appendix:interview}

\subsubsection{Gameplay Experience}
\begin{itemize}
    \item How did you find the overall experience of interacting with the game?
    \item Which parts did you enjoy the most and least?
    \item How did you feel about the design of the interface in terms of clarity, responsiveness, or understanding?
    \item What improvements would you suggest for the interface?
\end{itemize}
    
\subsubsection{Impact of In-game Interactions on Perception of Air Pollutants}
\begin{itemize}
    \item Before playing the game, how aware were you of air pollutants in your environment?
    \item Did the game change your awareness or understanding? If yes, how?
    \item Did the visualizations make the pollutants more relatable or noticeable to you?
\end{itemize}
    
\subsubsection{Suggestions to Improve the Platform}
\begin{itemize}
    \item What features or changes would you suggest to enhance the game's impact on your understanding of air quality?
    \item Are there any tools, information that you felt were missing?
\end{itemize}

\end{document}